\documentclass[twocolumn]{aastex6}
\usepackage{multirow}
\usepackage{amsmath}

\shorttitle{Detection of Planetary and Stellar Companions}
\shortauthors{Stephen R. Kane et al.}

\begin{document}

\title{Detection of Planetary and Stellar Companions to Neighboring
  Stars via a Combination of Radial Velocity and Direct Imaging
  Techniques}

\author{
  Stephen R. Kane\altaffilmark{1},
  Paul A. Dalba\altaffilmark{1},
  Zhexing Li\altaffilmark{1},
  Elliott P. Horch\altaffilmark{2},
  Lea A. Hirsch\altaffilmark{3},
  Jonathan Horner\altaffilmark{4},
  Robert A. Wittenmyer\altaffilmark{4},
  Steve B. Howell\altaffilmark{5},
  Mark E. Everett\altaffilmark{6},
  R. Paul Butler\altaffilmark{7},
  Christopher G. Tinney\altaffilmark{8},
  Brad D. Carter\altaffilmark{4},
  Duncan J. Wright\altaffilmark{4},
  Hugh R.A. Jones\altaffilmark{9},
  Jeremy Bailey\altaffilmark{8},
  Simon J. O'Toole\altaffilmark{10,11}
}
\altaffiltext{1}{Department of Earth and Planetary Sciences,
  University of California, Riverside, CA 92521, USA}
\altaffiltext{2}{Department of Physics, Southern Connecticut State
  University, New Haven, CT 06515, USA}
\altaffiltext{3}{Kavli Institute for Particle Astrophysics and
  Cosmology, Stanford University, Stanford, CA 94305, USA}
\altaffiltext{4}{Centre for Astrophysics, University of Southern
  Queensland, Toowoomba, QLD 4350, Australia}
\altaffiltext{5}{NASA Ames Research Center, Moffett Field, CA 94035,
  USA}
\altaffiltext{6}{National Optical Astronomy Observatory, 950 N. Cherry
  Ave, Tucson, AZ 85719, USA}
\altaffiltext{7}{Department of Terrestrial Magnetism, Carnegie
  Institution of Washington, NW, Washington, DC, 20015-1305, USA}
\altaffiltext{8}{School of Physics and Australian Centre for
  Astrobiology, University of New South Wales, Sydney 2052, Australia}
\altaffiltext{9}{Centre for Astrophysics Research, University of
  Hertfordshire, Hatfield, Herts AL10 9AB, UK}
\altaffiltext{10}{Australian Astronomical Observatory, North Ryde, NSW
  2113, Australia}
\altaffiltext{11}{Australian Astronomical Optics, Faculty of Science
  and Engineering, Macquarie University, North Ryde, NSW 2113,
  Australia}
\email{skane@ucr.edu}


\begin{abstract}

The sensitivities of radial velocity (RV) surveys for exoplanet
detection are extending to increasingly long orbital periods, where
companions with periods of several years are now being regularly
discovered. Companions with orbital periods that exceed the duration
of the survey manifest in the data as an incomplete orbit or linear
trend, a feature that can either present as the sole detectable
companion to the host star, or as an additional signal overlain on the
signatures of previously discovered companion(s). A diagnostic that
can confirm or constrain scenarios in which the trend is caused by an
unseen stellar, rather than planetary, companion is the use of
high-contrast imaging observations. Here, we present RV data from the
Anglo-Australian Planet Search (AAPS) for twenty stars that show
evidence of orbiting companions. Of these, six companions have
resolved orbits, with three that lie in the planetary regime. Two of
these (HD~92987b and HD~221420b) are new discoveries. Follow-up
observations using the Differential Speckle Survey Instrument (DSSI)
on the Gemini South telescope revealed that five of the twenty
monitored companions are likely stellar in nature. We use the
sensitivity of the AAPS and DSSI data to place constraints on the mass
of the companions for the remaining systems. Our analysis shows that a
planetary-mass companion provides the most likely self-consistent
explanation of the data for many of the remaining systems.

\end{abstract}



\section{Introduction}
\label{intro}

Radial velocity (RV) surveys for exoplanets have now been operating
over sufficiently long temporal baselines that their sensitivity
extends out to semi-major axes of several AU around other stars
\citep{wright2008,wittenmyer2011a,wittenmyer2016c}. This means that
the requirement of measuring a complete planetary orbital phase is
fulfilled for direct Jupiter analogs. Beyond this orbital regime, the
RV data cover a fraction of the total orbital phase, and the
sensitivity of experiments to Uranus/Neptune analogs \citep{kane2011d}
and the occurrence rate of long-period giant planets is more
adequately covered by exoplanet surveys using the microlensing
\citep{cassan2012,mroz2017b,penny2019} and imaging
\citep{meshkat2017,kopparapu2018,stone2018} techniques. Long-period
planets also present a compelling advantage due to thw synergy between
RV and astrometric observations, wherein such planets impart a
significant reciprocal astrometric motion on their host star
\citep{eisner2002}. With the advent of the extremely high astrometric
precision offered by observations carried out by the {\it Gaia}
spacecraft \citep{prusti2016,brown2018}, such motion should be readily
detected. As the RV method measures the line-of-sight motion of a
star, and the astrometric method measures motion at right angles to
the line of sight, combining data from these two methods will allow
the orbits and masses of currently unseen companions to be precisely
calculated, a result of great interest to the exoplanet community.

Detectable companions to host stars moving on long-period orbits
typically manifest in the form of curvatures or linear trends in RV
data. Such trends reveal which stars should continue to be monitored,
but also present significant challenges to providing a complete
characterization of the companion's orbit as well as limiting
constraints on the companion mass \citep{eisner2002,wright2009b}. A
major problem with such incomplete orbital coverage is that the
observed trends could be the result of a much larger, possibly
stellar, mass companion moving on a very long period orbit. There are
numerous surveys that aim to resolve the potential presence of stellar
companions through the combination of RV data exhibiting linear trends
and high-resolution imaging data that should be able to directly
detect such companions
\citep{kane2014c,crepp2016,wittrock2016,wittrock2017,crepp2018,kane2019a}.
High-resolution imaging has also played an important role in the
validation of small transiting exoplanets orbiting relatively faint
host stars \citep{everett2015}, such as those detected by the {\it
  Kepler} mission \citep{howell2011,quintana2014a} and the case of the
TRAPPIST-1 system \citep{howell2016b}.

One of the longest running RV surveys is that carried out by the
Anglo-Australian Planet Search (AAPS), which detected their first
planet (orbiting HD~179949) at the turn of the millennium
\citep{tinney2001}. Since then, the survey has focused on improving
their sensitivity to giant planets beyond the snow line through
continued observations in order to aid with the development of planet
formation theories for solar system analogs
\citep{wittenmyer2016c}. As a result of the long temporal baseline and
precision achieved by AAPS, the survey is an ideal source of suitable
targets for direct imaging surveys \citep{kane2018c}, and for studies
investigating the occurrence rate of giant planets in the Habitable
Zone (HZ) of their host stars \citep{hill2018}. However, the parameter
space of star--planet separation is exceedingly large, and a full
investigation of long-period planets beyond $\sim$10~AU using the RV
technique would require the ongoing monitoring of target stars for
timescales of decades to come, albeit at relatively low cadence.

In this paper we present new results from a survey that aims to study
the cause of RV signatures detected for nearby stars. In
Section~\ref{rv} we provide RV data from the AAPS for twenty stars
that show evidence of a companion, and present the orbital solutions
for 6 companions including two new planet discoveries. In
Section~\ref{imaging} we further provide the results of a follow-up
observing campaign using the Differential Speckle Survey Instrument
(DSSI) operating on the Gemini South telescope. These data reveal that
5 of the 20 target stars have stellar companions that potentially
explain the signatures observed in the RV data, described in detail in
Section~\ref{stars}. We quantify the mass limits on planetary
companions for the remaining 15 stars in Section~\ref{planets}, and
provide a description of potential additional observations together
with our concluding remarks in Section~\ref{conclusions}.


\section{Radial Velocity Variations}
\label{rv}

The AAPS is one of the ``legacy'' RV surveys, having gathered data for
17 years between 1998 and 2015. The AAPS used the UCLES
high-resolution spectrograph \citep{diego1990} on the 3.9m
Anglo-Australian Telescope (AAT) to deliver a consistent velocity
precision of 2--3~m\,s$^{-1}$. An iodine absorption cell provided
wavelength calibration from 5000 to 6200\,\AA. The spectrograph
point-spread function and wavelength calibration were derived from the
iodine absorption lines embedded on the spectrum by the cell
\citep{valenti1995b,butler1996}. The use of the same iodine
calibration cell and same target list for 17 years have made the AAPS
a pioneer in the detection of long-period planets, in particular
``Jupiter analogs'': giant planets with orbital periods $P \gtrsim
10$~years
\citep[e.g.][]{butler2006,jones2010,wittenmyer2014b,wittenmyer2016c}.

\begin{deluxetable*}{lccccccccc}
  \tablewidth{0pc}
  \tablecaption{\label{sumtab} Summary of AAPS observations and
    stellar parameters.}
  \tablehead{
    \colhead{Star} &
    \colhead{$\Delta$RV (m\,s$^{-1}$)} &
    \colhead{$dv/dt$ (m\,s$^{-1}$\,day$^{-1}$)} &
    \colhead{$N_\mathrm{obs}$} &
    \colhead{$V\,^{\dagger}$} &
    \colhead{$d\,^{\dagger}$ (pcs)} &
    \colhead{$M_\star\,^{\dagger}$ ($M_\odot$)}&
    \colhead{$T_\mathrm{eff}\,^{\dagger}$ (K)} &
    \colhead{$\log g\,^{\dagger}$} &
    \colhead{[Fe/H]$\,^{\dagger}$}
  }
  \startdata
   \sidehead{\bf Keplerian orbits}
~~~~HD 45701  &  2416.41 & --                 &  35 & 6.45 & 31.8 & $1.40\pm0.12$ & 5886 & 4.28 &  0.16 \\
~~~~HD 92987  &   308.31 & --                 &  53 & 7.03 & 44.0 & $1.05\pm0.11$ & 5774 & 4.06 &  0.03 \\
~~~~HD 145825 &  2180.14 & --                 &  17 & 6.55 & 21.9 & $1.08\pm0.09$ & 5803 & 4.49 &  0.03 \\
~~~~HD 212330 &  5838.34 & --                 &  33 & 5.31 & 20.5 & $1.40\pm0.11$ & 5739 & 4.15 &  0.01 \\
~~~~HD 219077 &   369.95 & --                 &  72 & 6.12 & 29.2 & $1.51\pm0.13$ & 5364 & 4.05 & -0.10 \\
~~~~HD 221420 &   104.69 & --                 &  88 & 5.82 & 31.8 & $1.67\pm0.11$ & 5830 & 4.08 &  0.29 \\
   \sidehead{\bf Trends}
~~~~HD 51929  &   494.32 &  $0.1047\pm0.0060$ &  16 & 7.39 & 37.6 & $1.30\pm0.11$ & 5805 & 4.43 & -0.48 \\
~~~~HD 52447  &   664.60 & $-0.1700\pm0.0074$ &  24 & 8.38 & 78.6 & $1.14\pm0.16$ & 6051 & 4.23 &  0.23 \\
~~~~HD 80913  &   143.61 &  $0.0241\pm0.0013$ &  35 & 7.49 & 64.1 & $1.27\pm0.13$ & 5983 & 4.05 & -0.60 \\
~~~~HD 100623 &    39.92 &  $0.0093\pm0.0004$ & 104 & 5.96 &  9.5 & $0.96\pm0.13$ & 5189 & 4.68 & -0.32 \\
~~~~HD 108309 &    23.66 &  $0.0018\pm0.0004$ &  69 & 6.25 & 26.7 & $1.26\pm0.11$ & 5778 & 4.26 &  0.09 \\
~~~~HD 117939 &    56.31 &  $0.0094\pm0.0007$ &  35 & 7.29 & 30.2 & $1.07\pm0.10$ & 5671 & 4.46 & -0.17 \\
~~~~HD 161050 &  1158.85 &  $0.1932\pm0.0009$ &  31 & 7.16 & 49.5 & $1.22\pm0.13$ & 5980 & 4.14 & -0.06 \\
~~~~HD 166553 &   205.70 &  $0.0365\pm0.0039$ &  43 & 7.30 & 42.4 & $0.84\pm0.16$ & 5960 & 4.17 &  0.03 \\
~~~~HD 191408 &    67.12 &  $0.0081\pm0.0003$ & 187 & 5.30 &  6.1 & $0.77\pm0.06$ & 4922 & 4.58 & -0.33 \\
~~~~HD 199509 &   431.16 & $-0.0758\pm0.0021$ &  33 & 6.98 & 24.2 & $1.05\pm0.09$ & 5770 & 4.55 & -0.27 \\
~~~~HD 207700 &    60.80 & $-0.0082\pm0.0006$ &  36 & 7.43 & 39.8 & $1.40\pm0.14$ & 5680 & 4.39 &  0.09 \\
~~~~HD 212708 &   229.62 & $-0.0338\pm0.0008$ &  38 & 7.48 & 35.8 & $1.06\pm0.10$ & 5689 & 4.39 &  0.24 \\
~~~~HD 214953 &    38.15 &  $0.0031\pm0.0004$ &  83 & 6.30 & 23.6 & $0.81\pm0.05$ & 6049 & 4.29 &  0.03 \\
~~~~HD 217958 &   112.34 &  $0.0171\pm0.0011$ &  37 & 8.05 & 53.7 & $1.15\pm0.18$ & 5962 & 4.40 &  0.26
\enddata
\tablenotetext{\dagger}{\citet{valenti2005}}
\end{deluxetable*}

\begin{figure*}
  \begin{center}
    \begin{tabular}{cc}
      \includegraphics[clip,width=8.2cm]{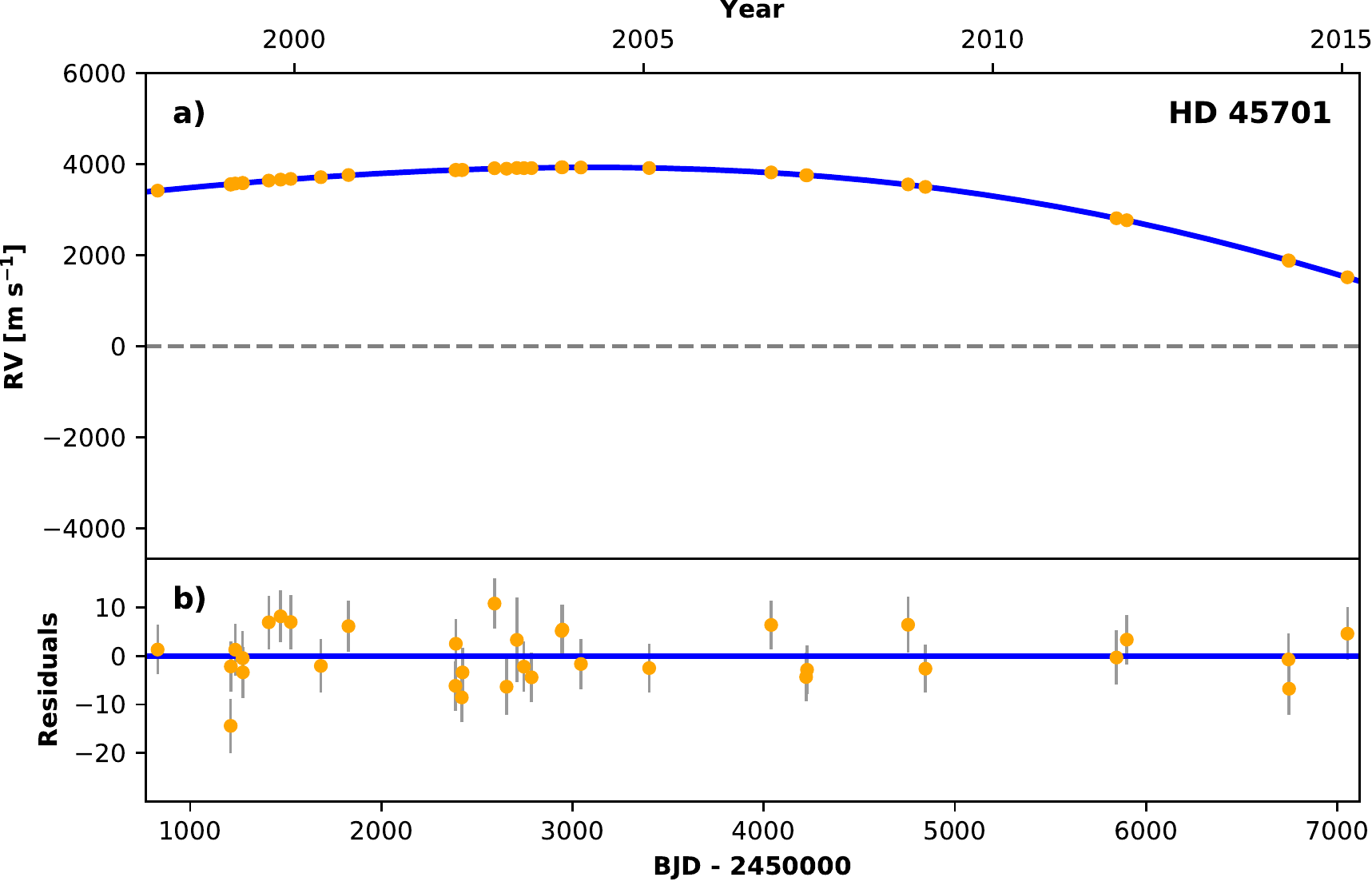} &
      \includegraphics[clip,width=8.2cm]{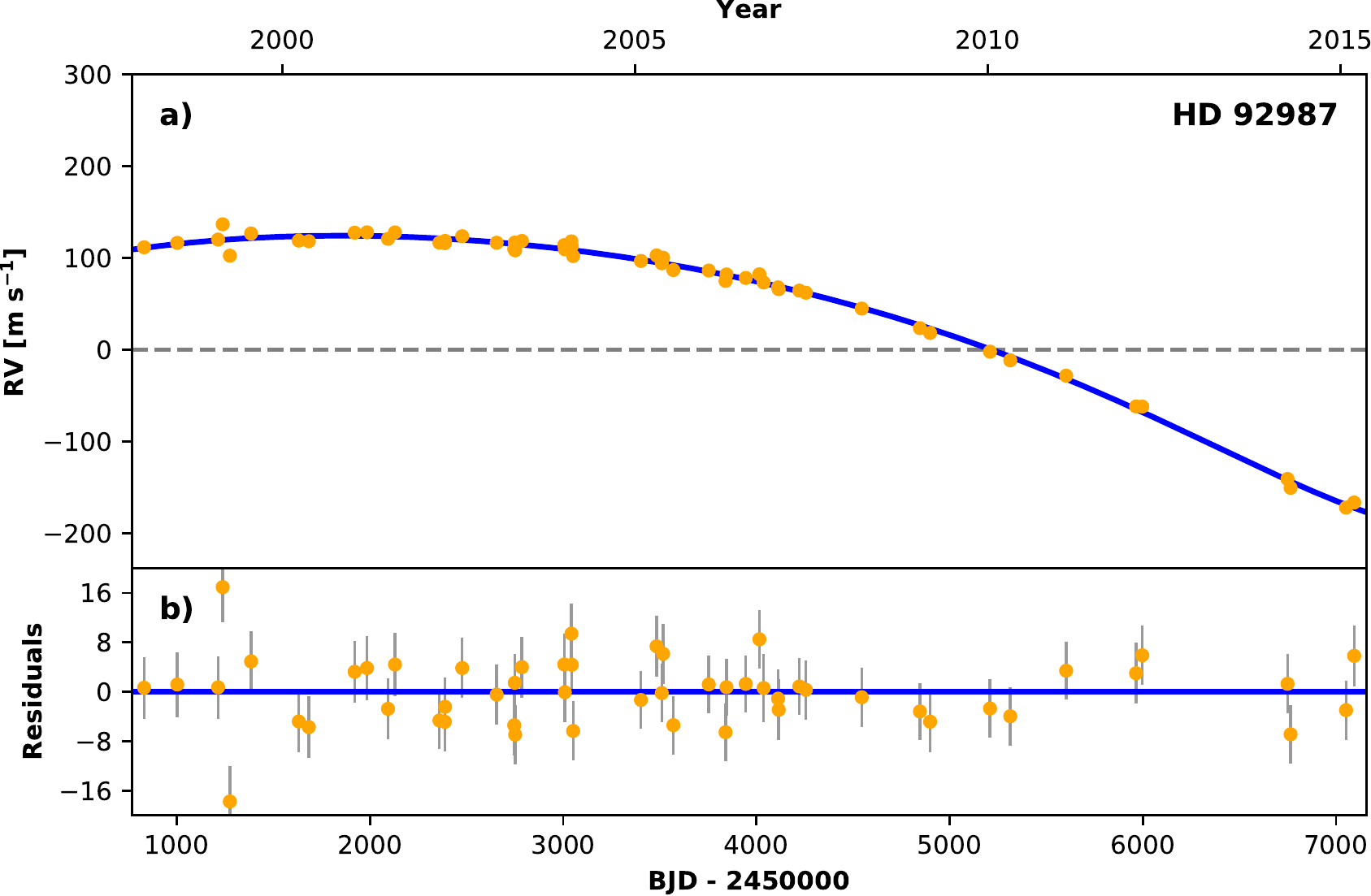} \\
      \includegraphics[clip,width=8.2cm]{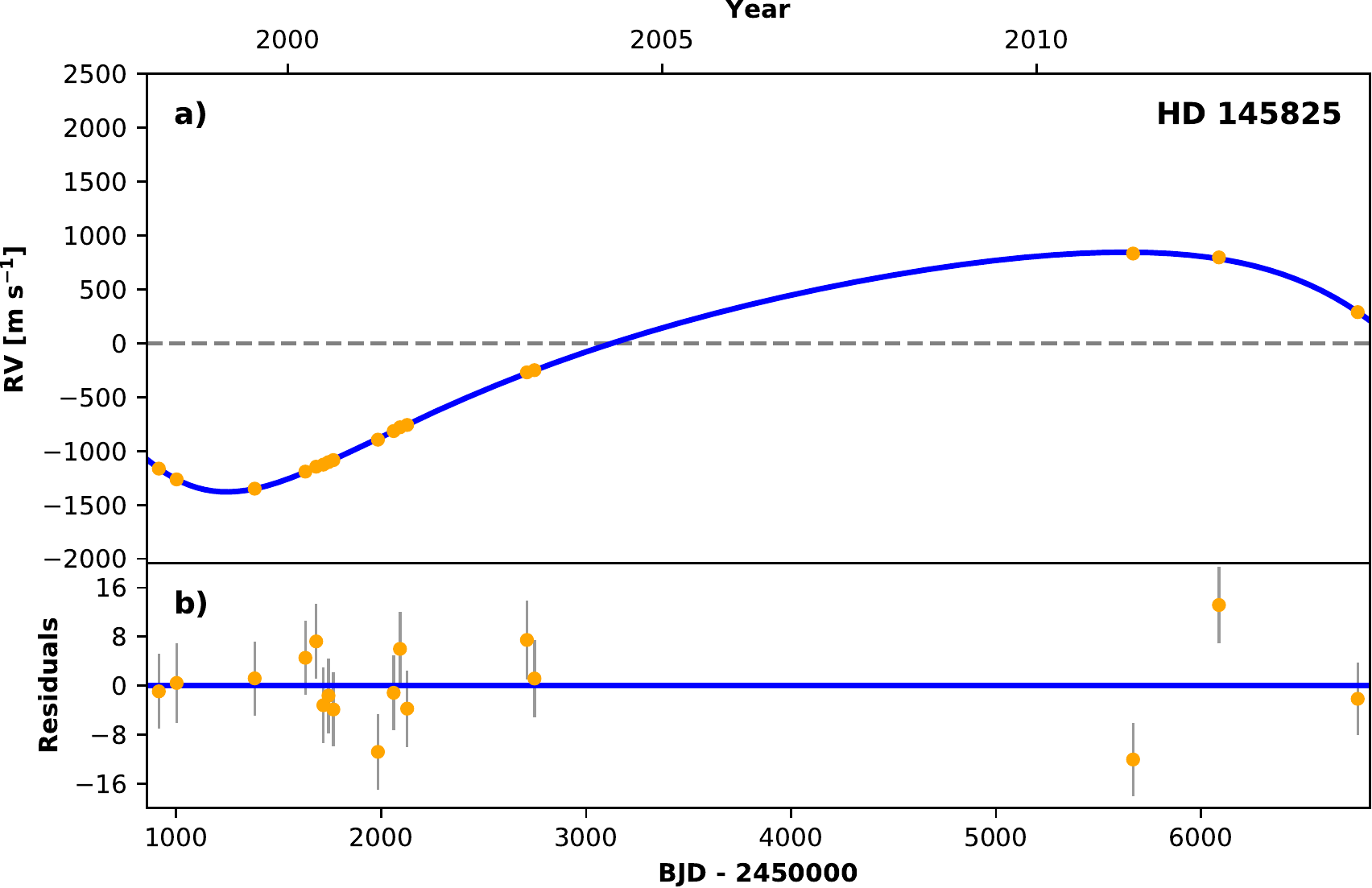} &
      \includegraphics[clip,width=8.2cm]{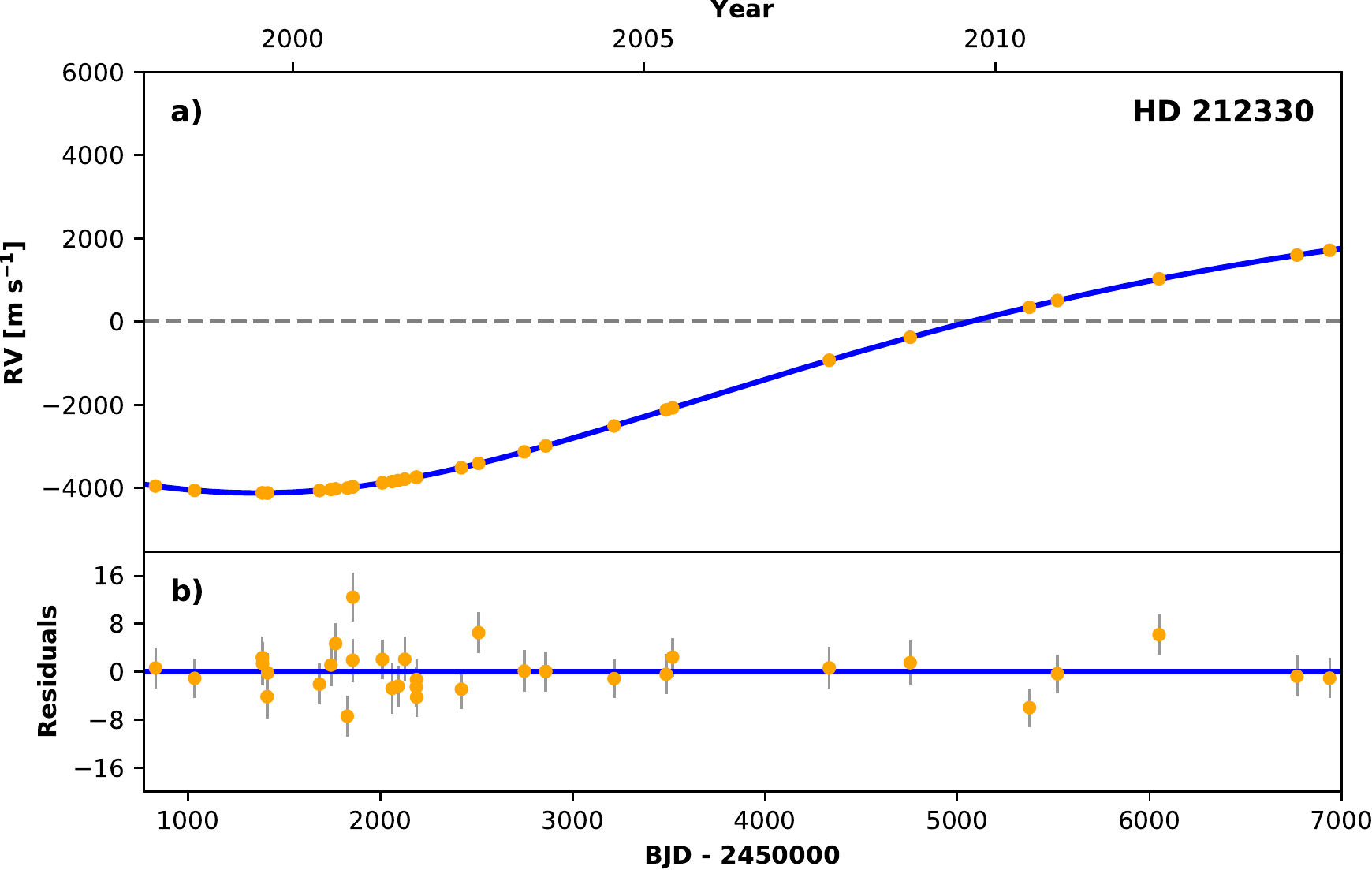} \\
      \includegraphics[clip,width=8.2cm]{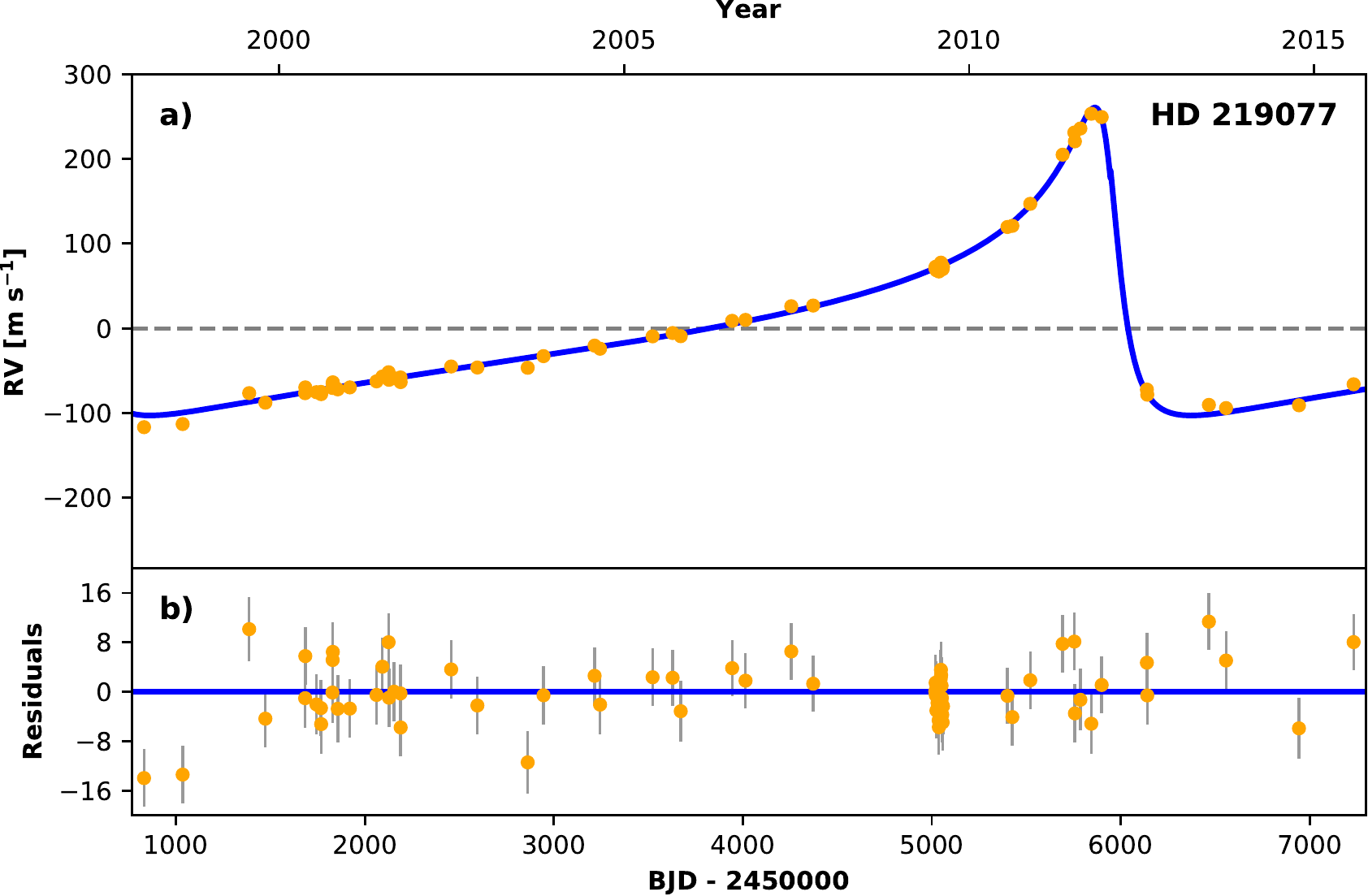} &
      \includegraphics[clip,width=8.2cm]{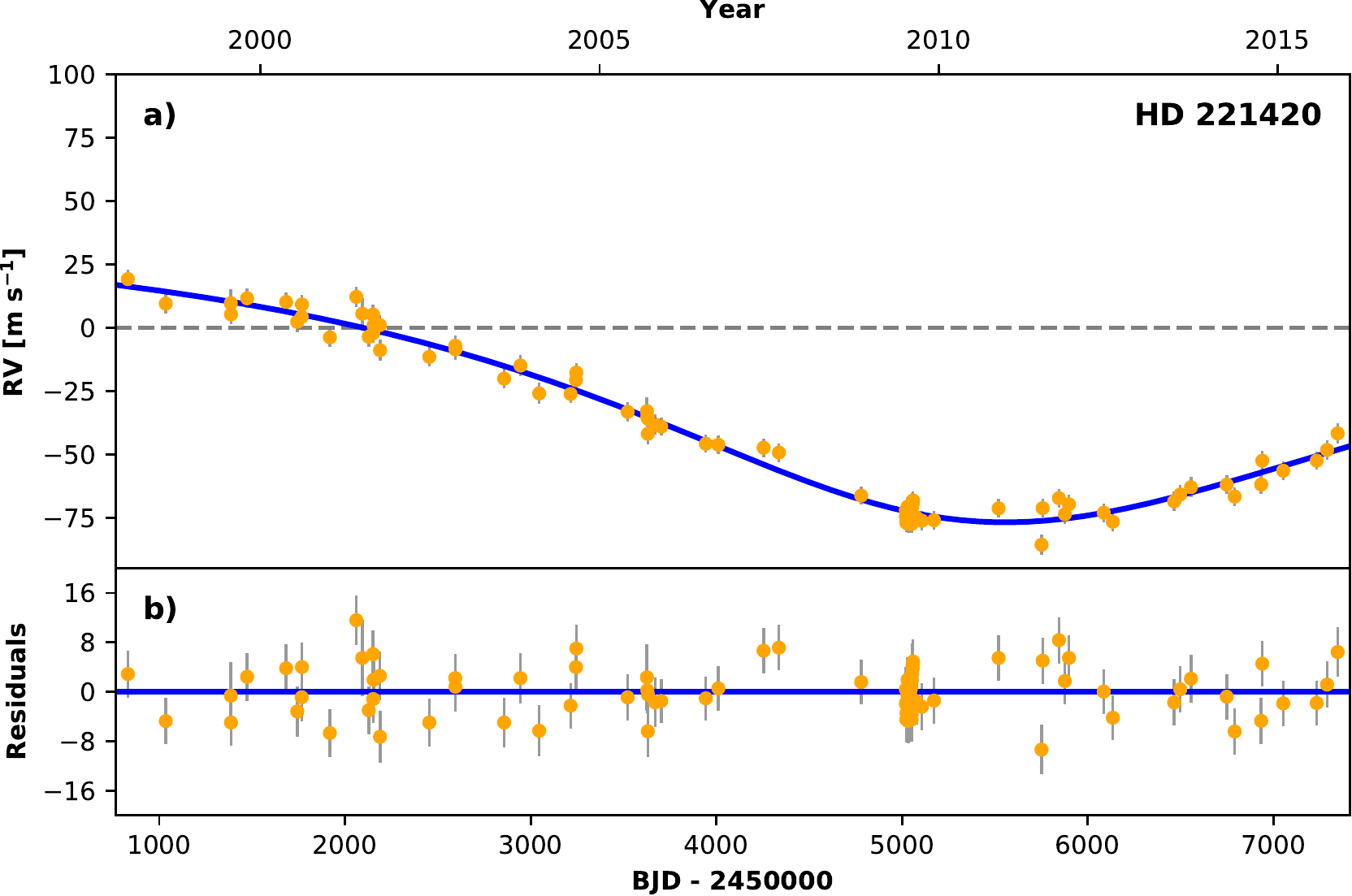}
    \end{tabular}
  \end{center}
  \caption{RV data and Keplerian orbital fits for the six targets
    described in Table~\ref{sumtab} and \ref{keptab}.}
  \label{rvplots1}
\end{figure*}

\begin{figure*}
  \begin{center}
    \begin{tabular}{ccc}
      \includegraphics[width=5.6cm]{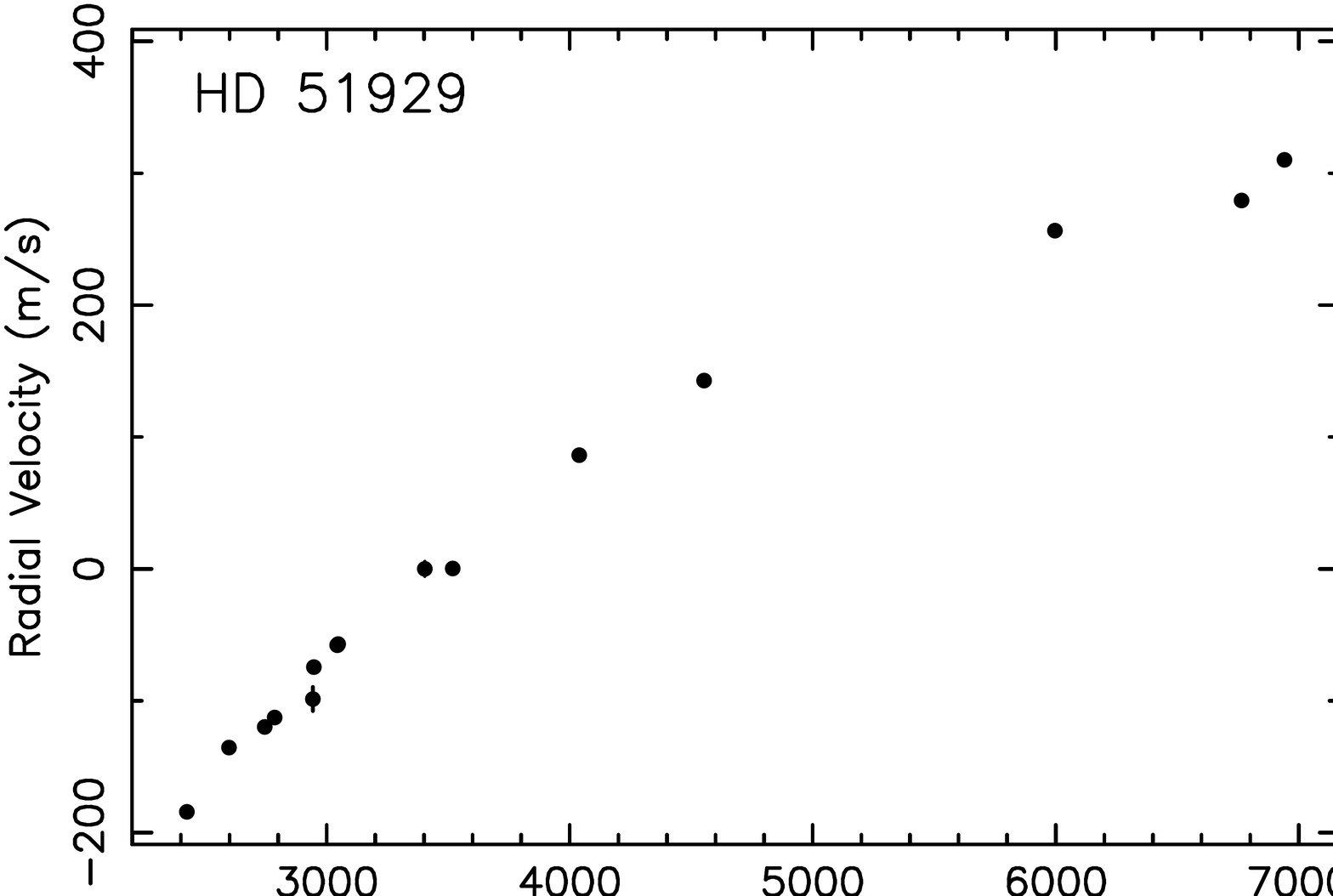} &
      \includegraphics[width=5.6cm]{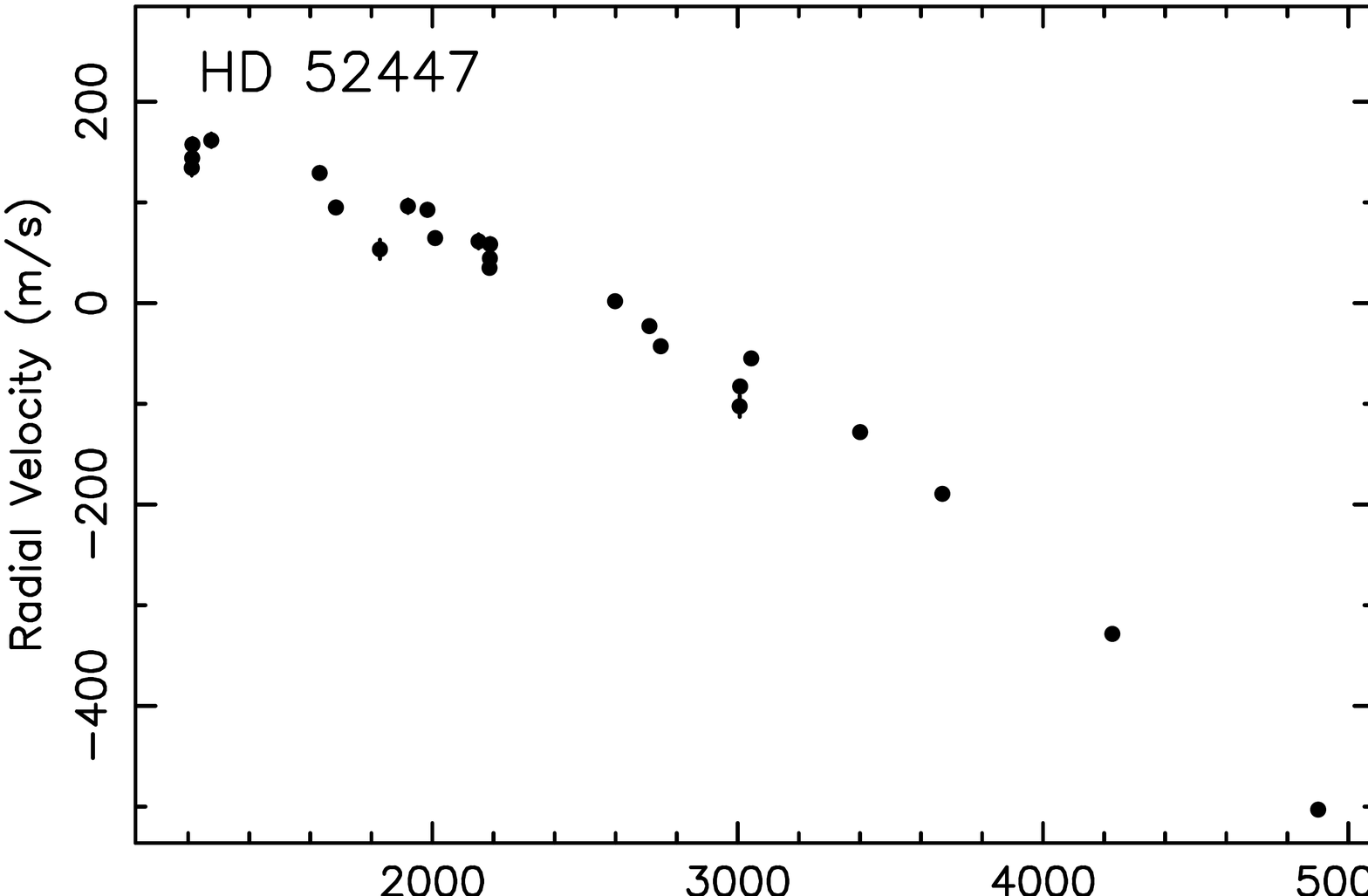} &
      \includegraphics[width=5.6cm]{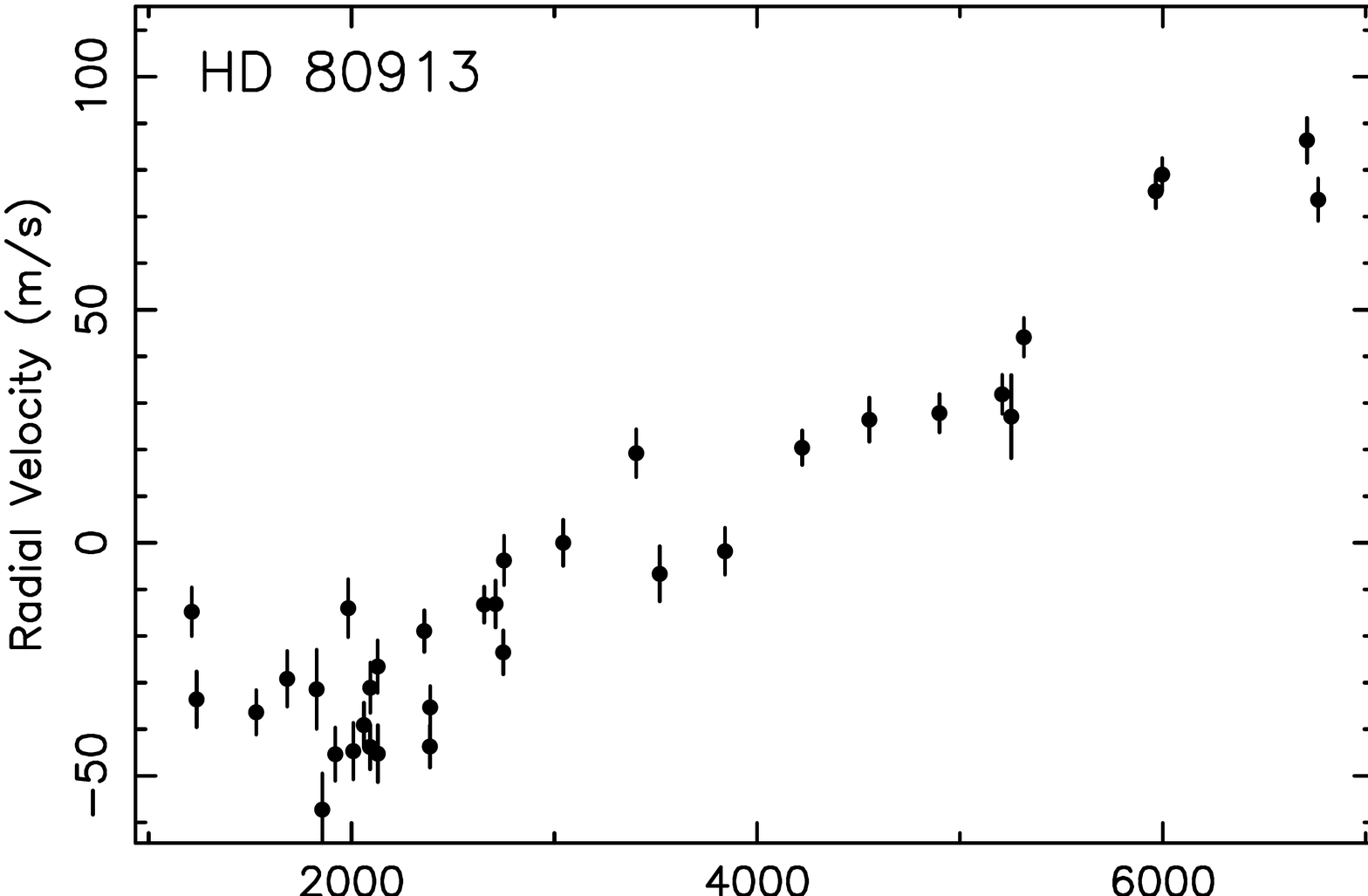} \\
      \includegraphics[width=5.6cm]{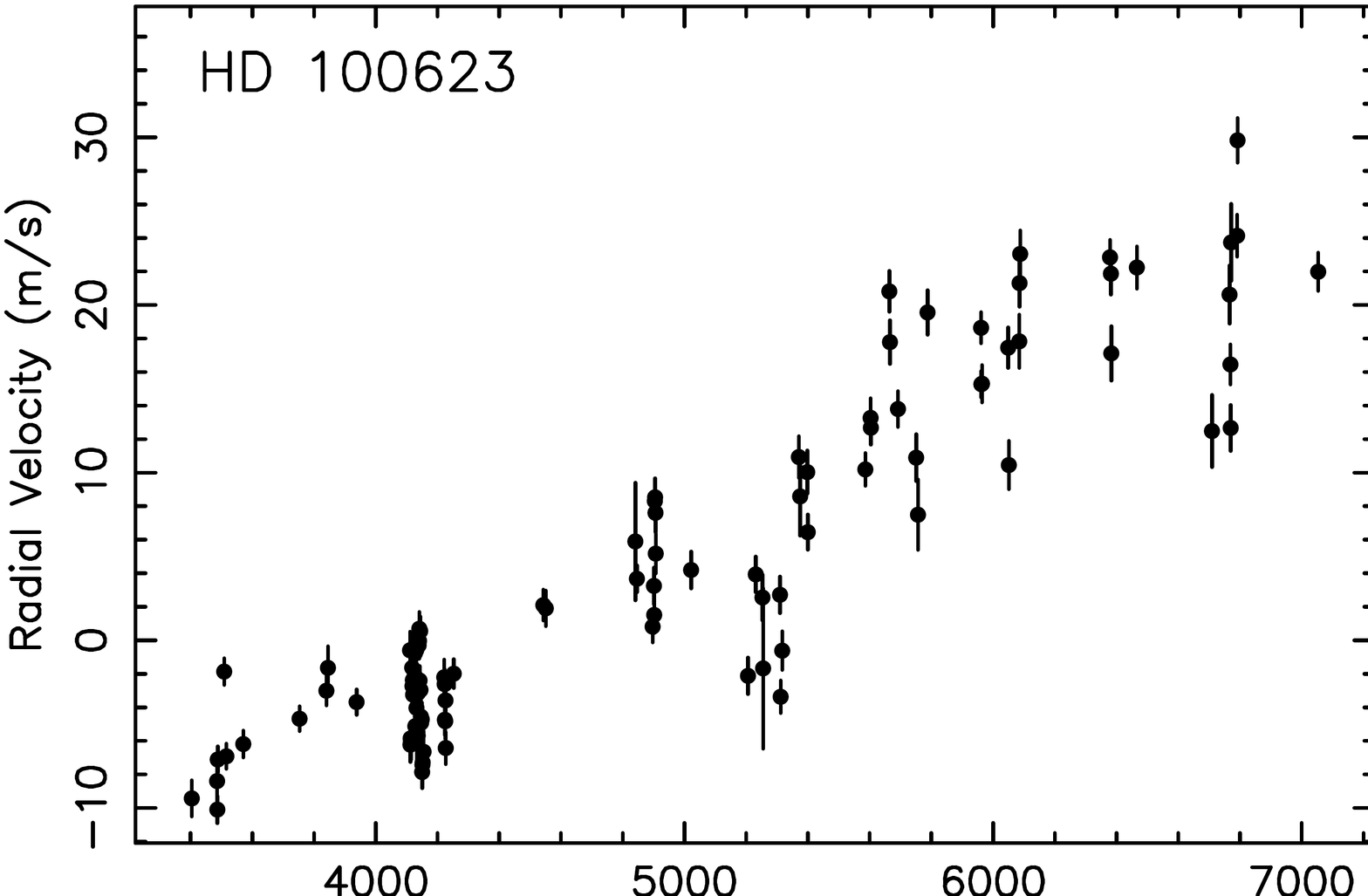} &
      \includegraphics[width=5.6cm]{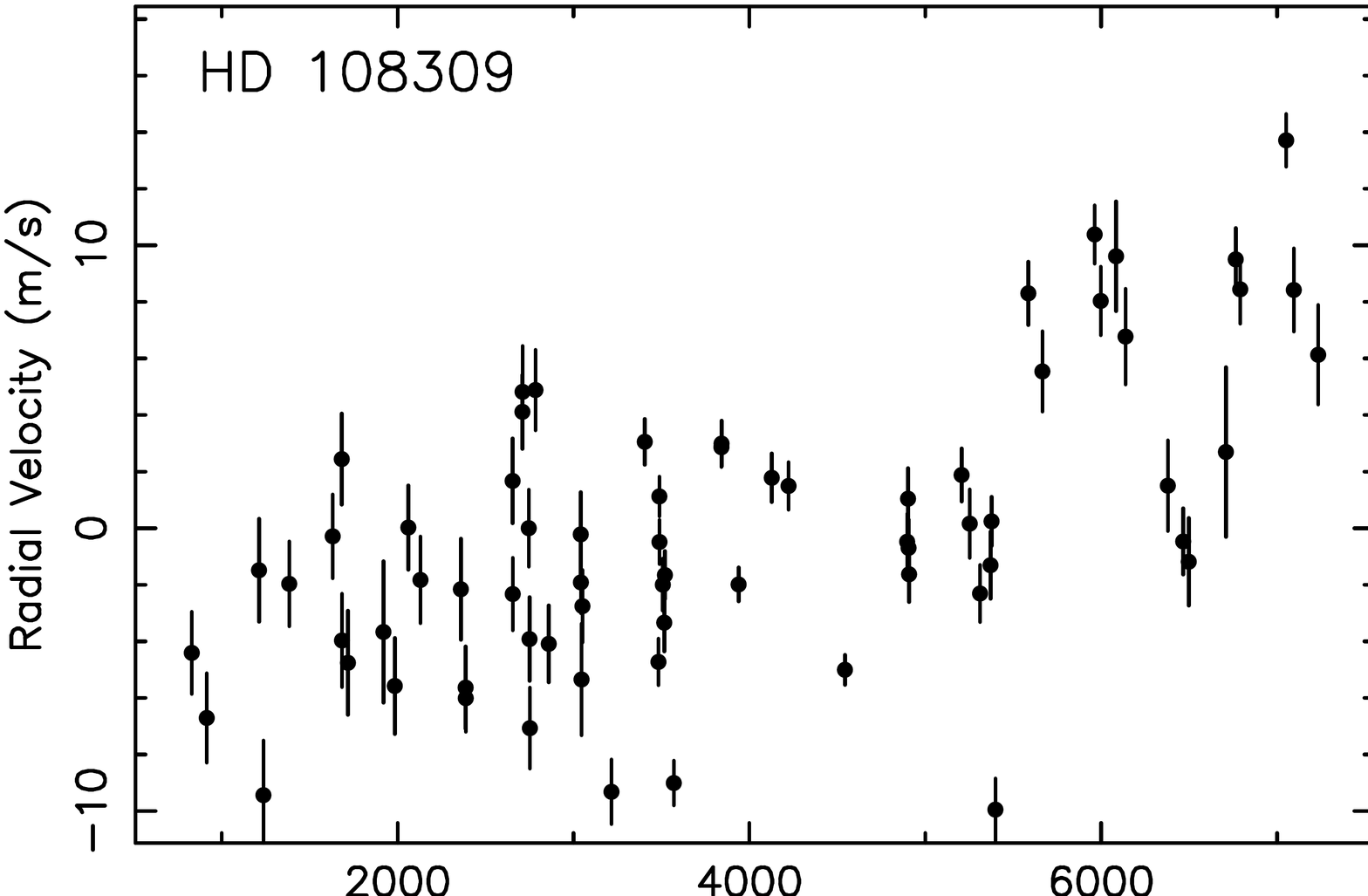} &
      \includegraphics[width=5.6cm]{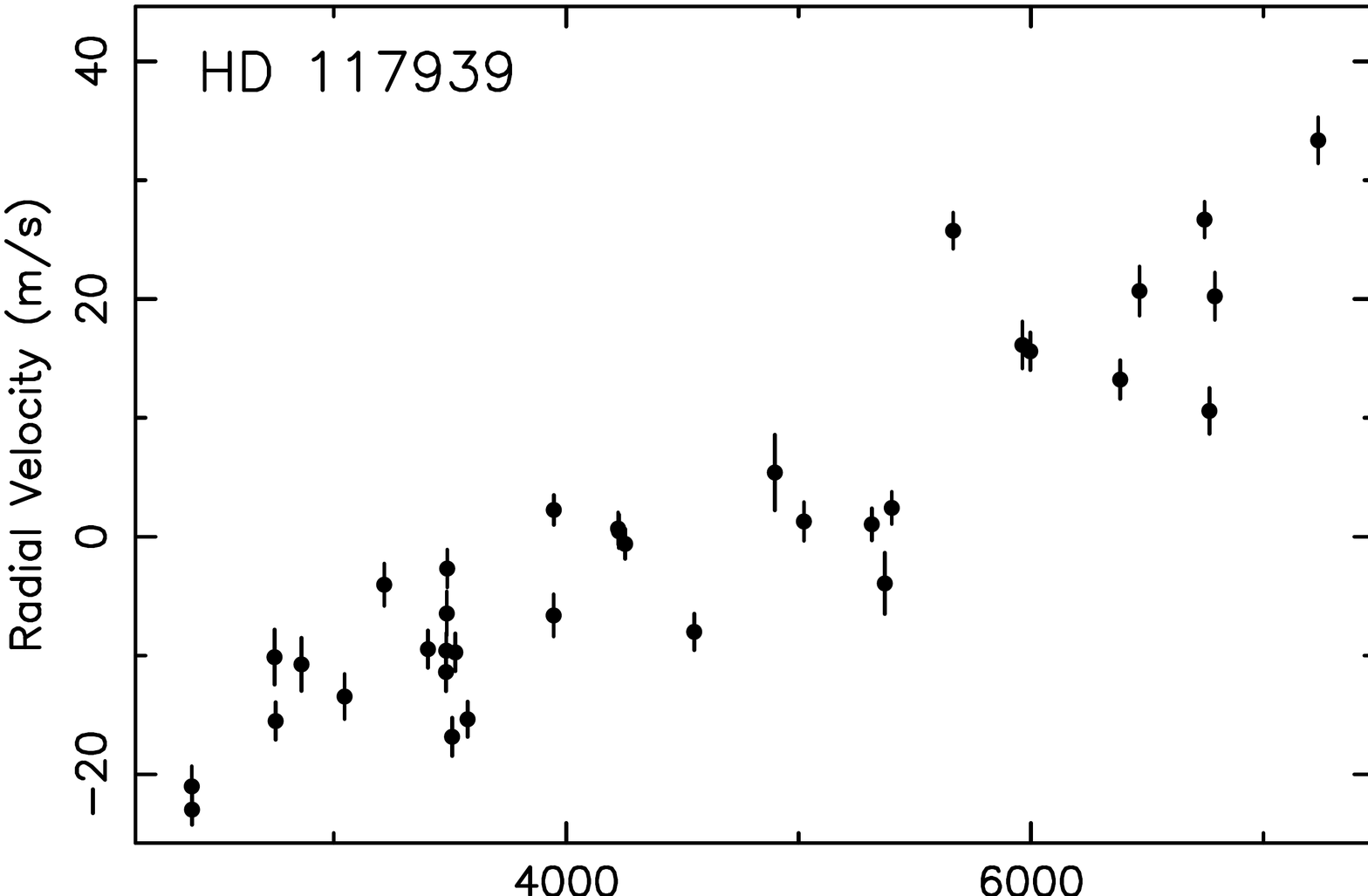} \\
      \includegraphics[width=5.6cm]{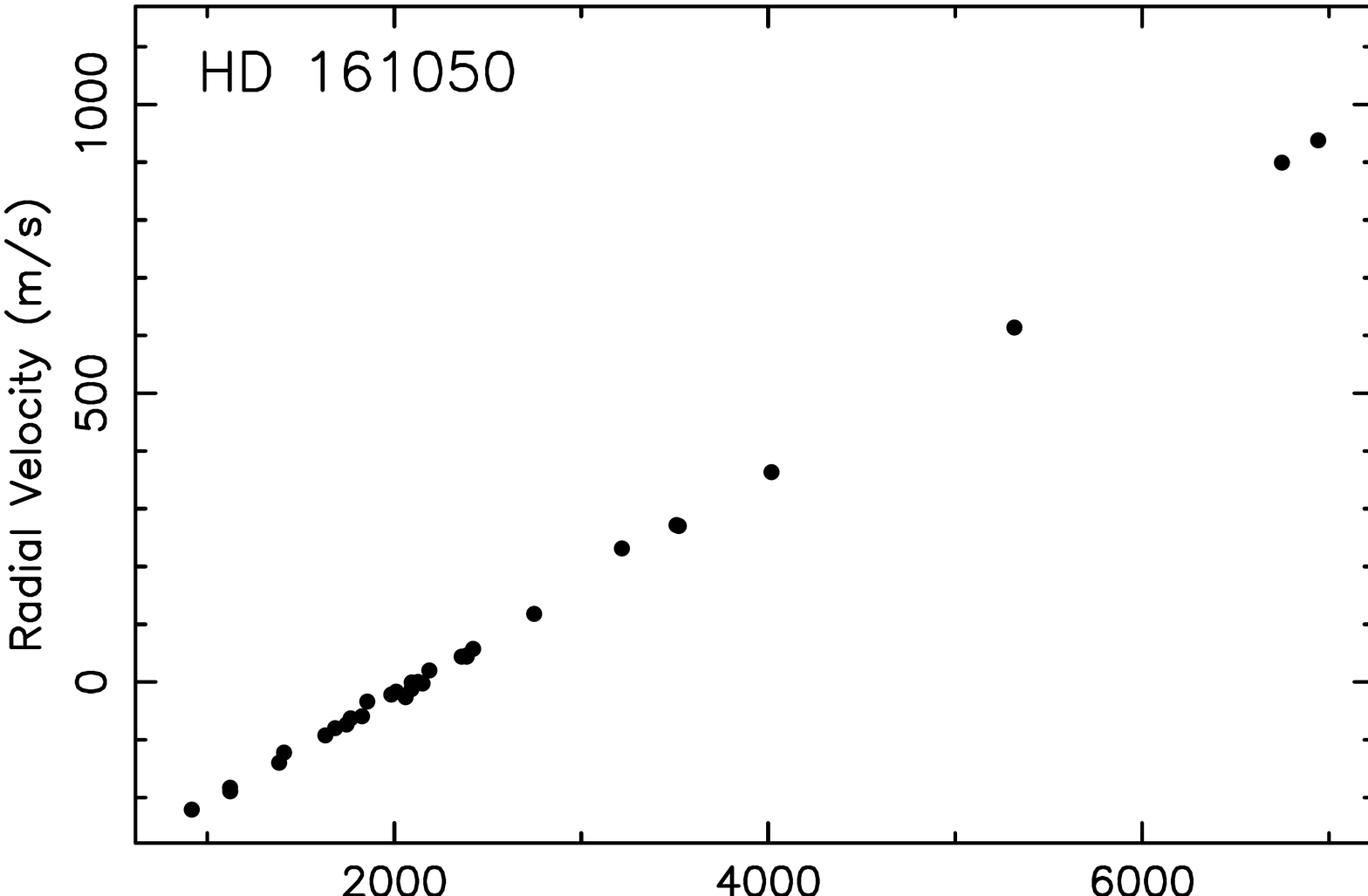} &
      \includegraphics[width=5.6cm]{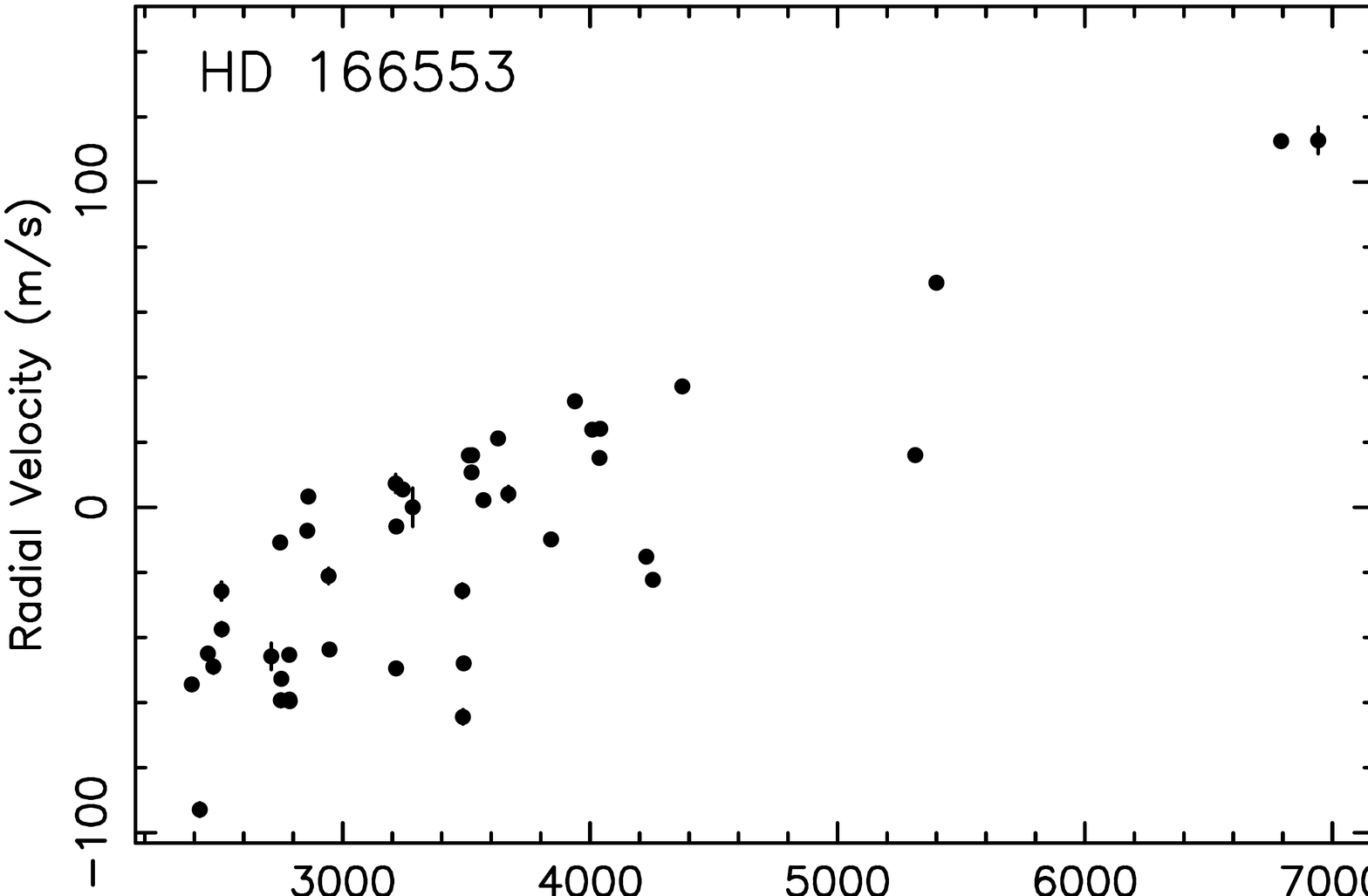} &
      \includegraphics[width=5.6cm]{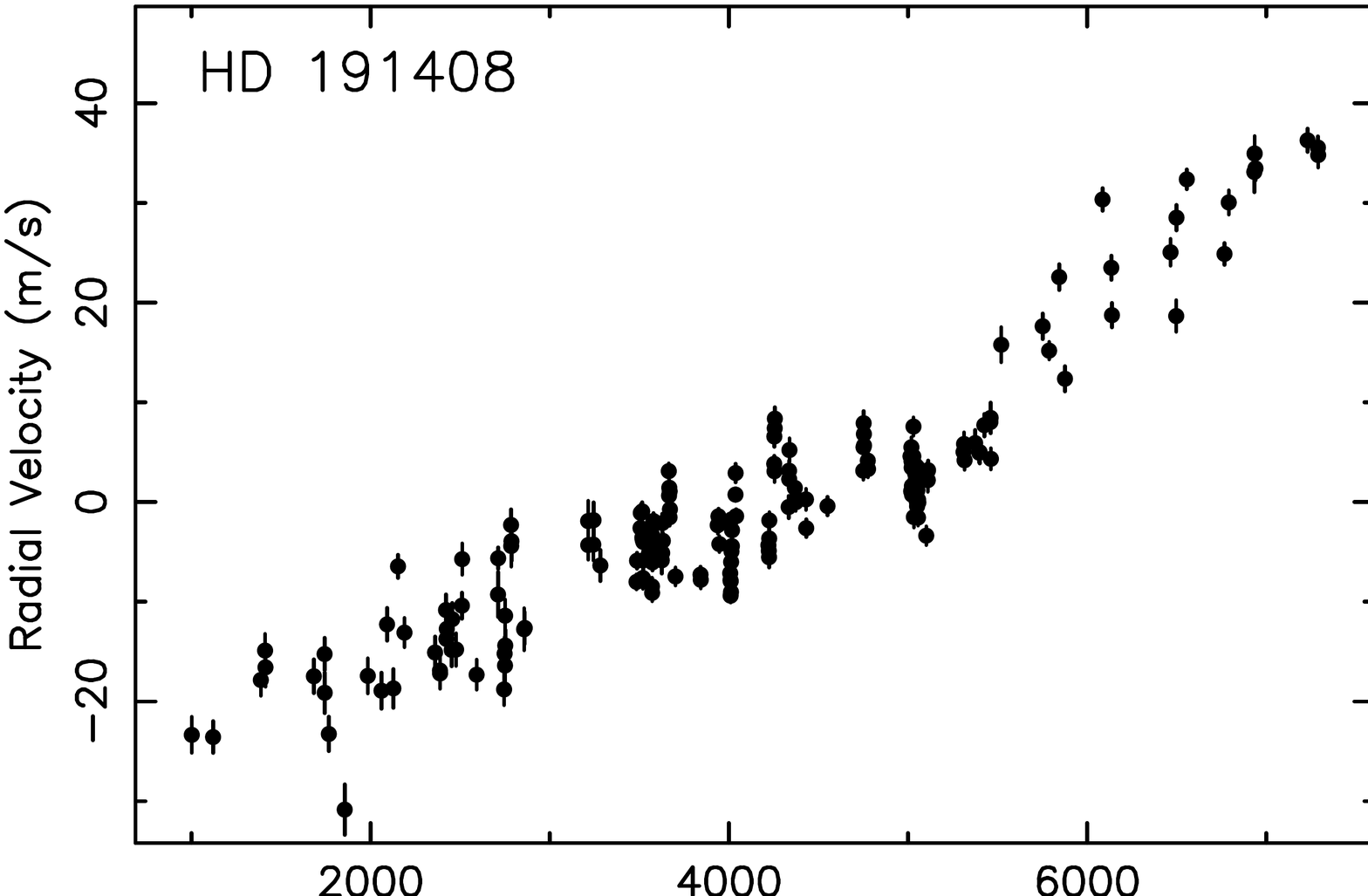} \\
      \includegraphics[width=5.6cm]{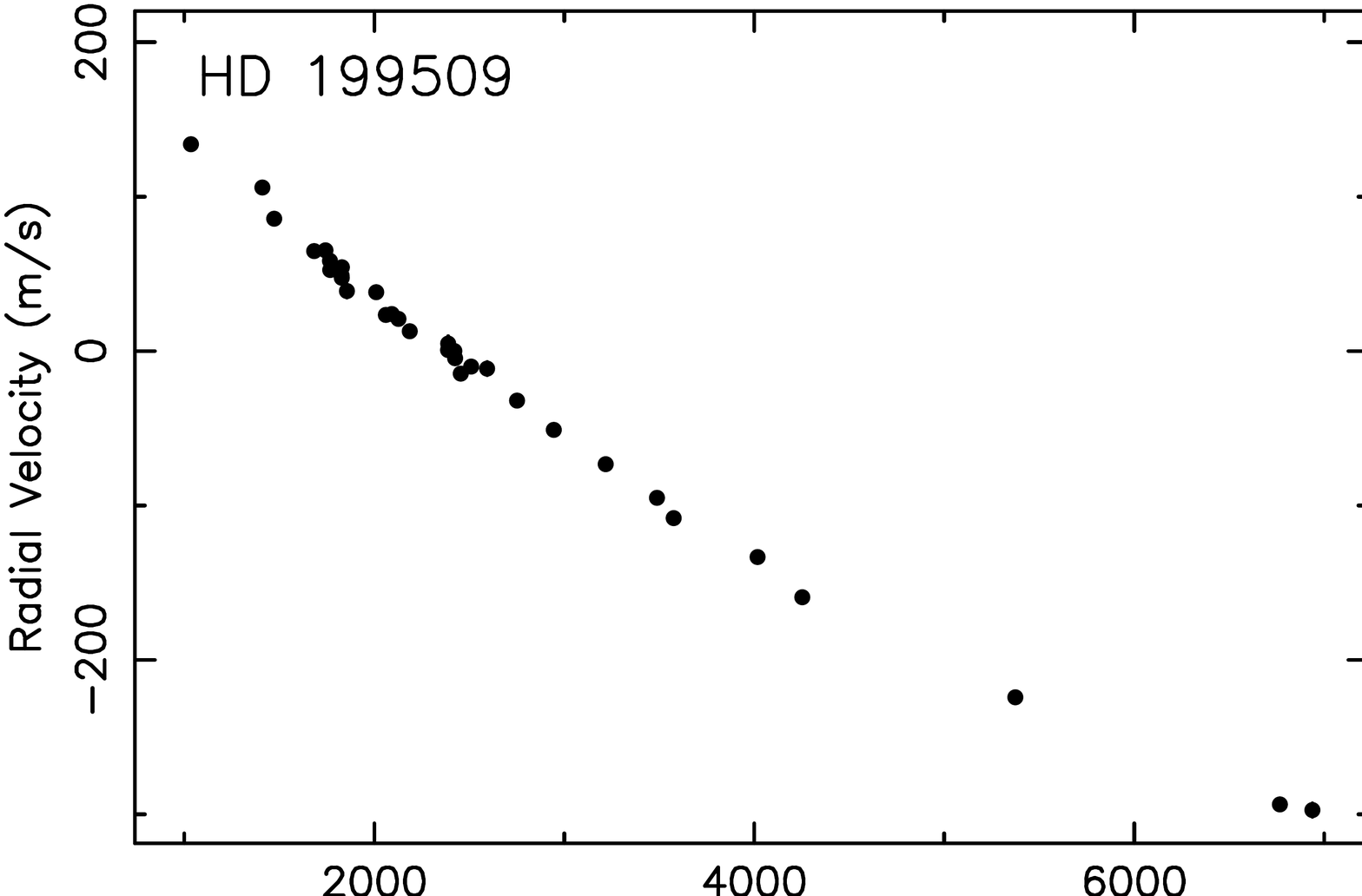} &
      \includegraphics[width=5.6cm]{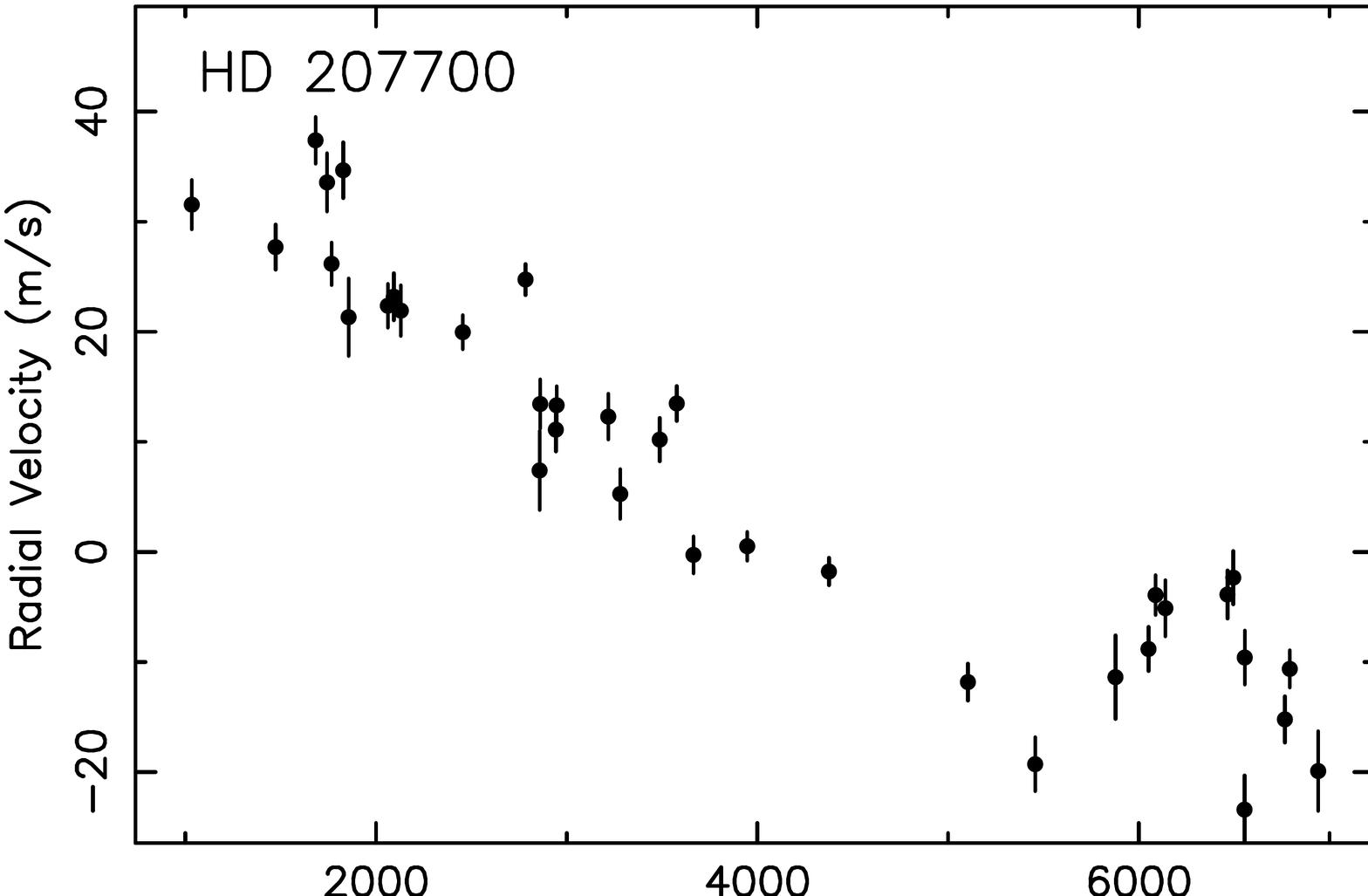} &
      \includegraphics[width=5.6cm]{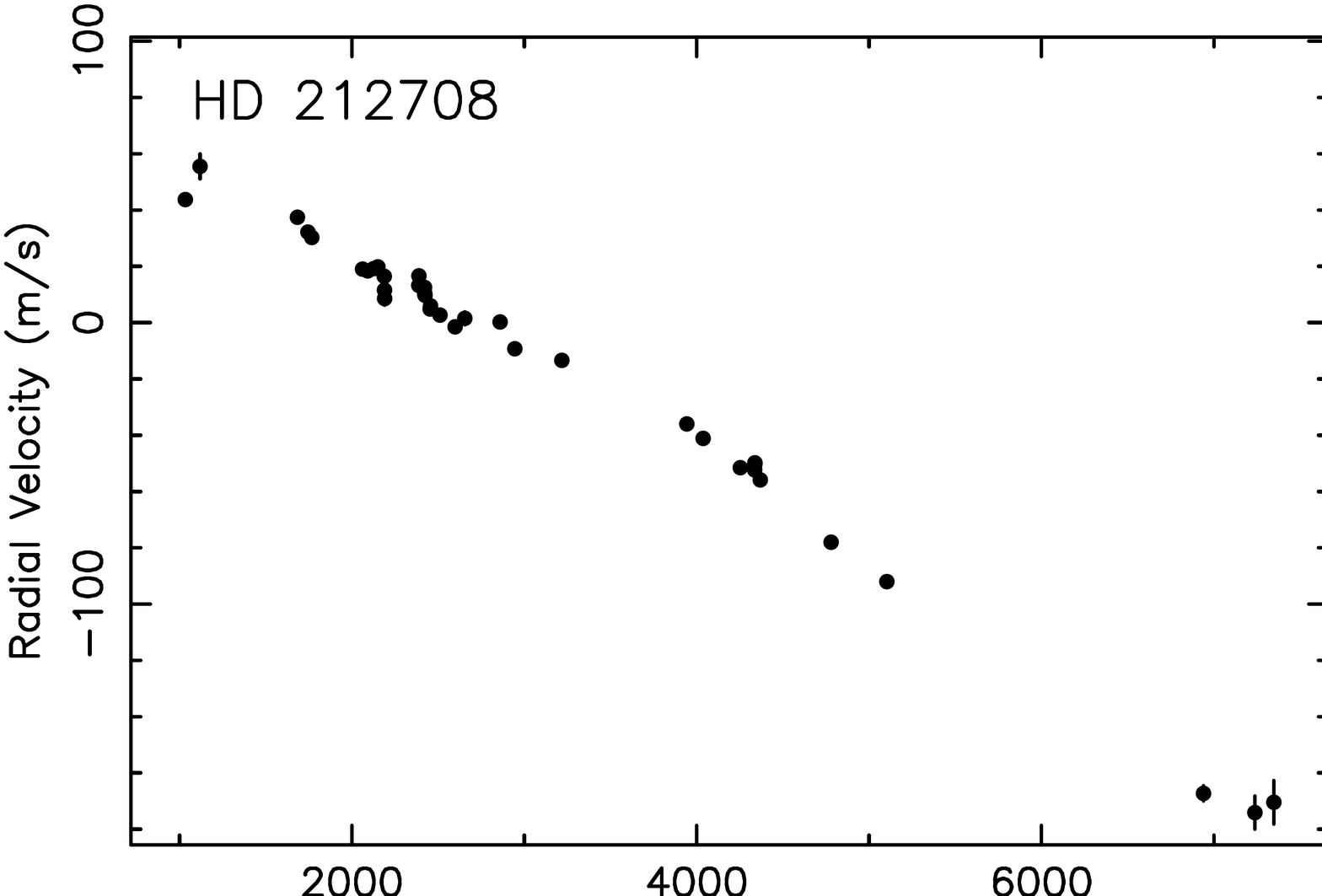} \\
      \includegraphics[width=5.6cm]{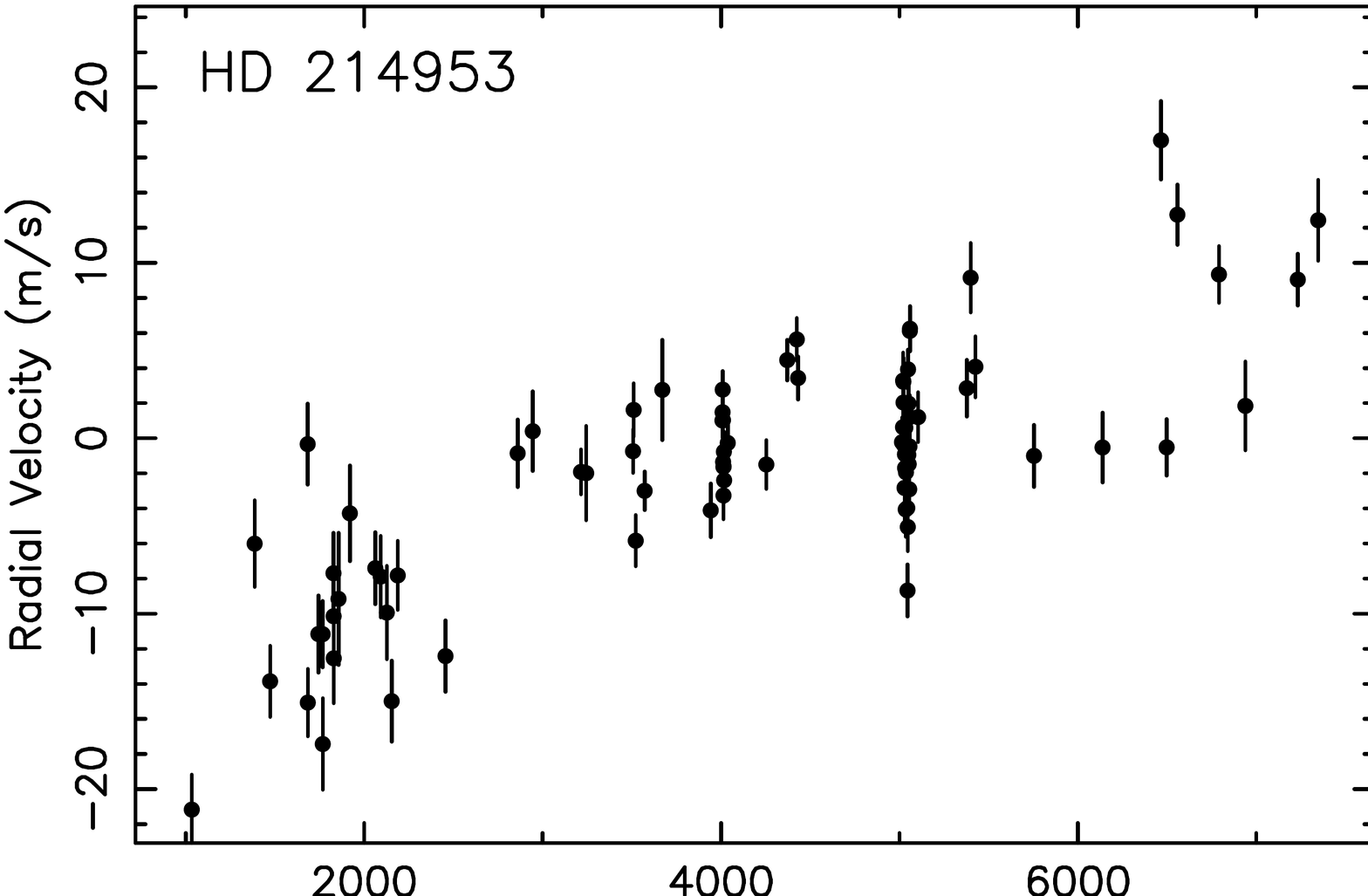} &
      \includegraphics[width=5.6cm]{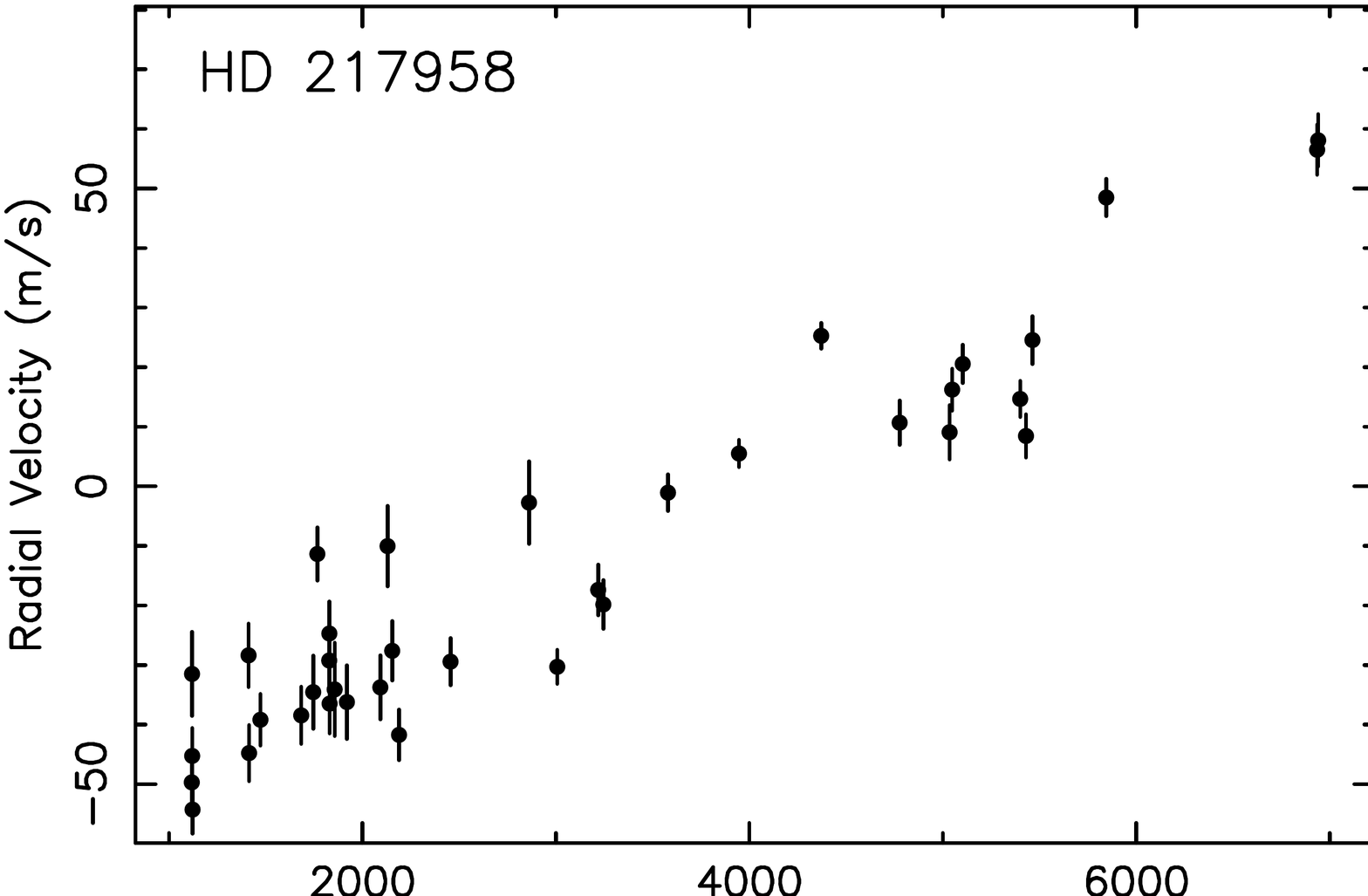} &
    \end{tabular}
  \end{center}
  \caption{RV data for the fourteen targets exhibiting approximately
    linear trends, described in Table~\ref{sumtab}.}
  \label{rvplots2}
\end{figure*}

In this work, we considered a total of twenty targets, for which a
variety of RV signals had been measured by the AAPS. A summary of the
targets, including the host star properties, RV variability (range),
and RV trend ($dv/dt$) is shown in Table~\ref{sumtab}. The host star
properties were all extracted from the same source, namely the
Spectroscopic Properties of Cool Stars (SPOCS) catalog compiled by
\citet{valenti2005}, in order to provide a self-consistent sample of
stellar information. As specified by \citet{valenti2005}, the
uncertainties in the stellar parameters are 44~K in $T_\mathrm{eff}$,
0.06 dex in $\log g$, and 0.03 dex in [Fe/H]. In the majority of cases
(fourteen out of twenty), the temporal baseline of the RV datasets is
insufficient to reasonably constrain the parameter space Keplerian
orbital solutions. We therefore broadly divide our sample into those
with Keplerian orbital solutions and those that are best represented
with a linear trend, shown at the top and bottom of Table~\ref{sumtab}
respectively. For the Keplerian orbital solution sample, we require
that the data contain evidence of a quadrature (``turn around'') point
that will enable a sufficient constraint to the orbital period. For
the six companions with sufficient data for a Keplerian orbit fit, the
orbital solutions are shown in Table~\ref{keptab}, and the data with
fits and residuals are shown in the panels of
Figure~\ref{rvplots1}. The RV data were fit using the RadVel package
\citep{fulton2018a}\footnote{https://radvel.readthedocs.io/en/latest/},
modified to allow for massive companions outside of the planetary
regime \citep{kane2019a}. For the fourteen companions detected as a
roughly linear trend, the trends are quantified in Table~\ref{sumtab}
in units of m\,s$^{-1}$\,day$^{-1}$, and the data for these targets
are shown in the panels of Figure~\ref{rvplots2}.

\begin{deluxetable*}{lcccccc}
    \tablecaption{\label{keptab} Keplerian orbital parameters derived
      from the fits to RV data}
    \tablecolumns{7}
    \tablewidth{0pt}
    \tablehead{ 
        \colhead{Parameter} & 
        \colhead{HD 45701} &
        \colhead{HD 92987} &
        \colhead{HD 145825} &
        \colhead{HD 212330} &
        \colhead{HD 219077} &
        \colhead{HD 221420} 
        }
    \startdata 
       $P$ (days) & $24859^{+210}_{-200}$    & $10790^{+850}_{-800}$   & $6667\pm31$    & $16681^{+440}_{-410}$    & $5513^{+50}_{-45}$      & $22482^{+4200}_{-4100}$ \\ 
       $P$ (years) & $68.06^{+0.57}_{-0.55}$ & $29.54^{+2.33}_{-2.19}$ & $18.25\pm0.08$ & $45.67^{+1.20}_{-1.12}$ & $15.09^{+0.14}_{-0.12}$ & $61.55^{+11.50}_{-11.23}$ \\ 
       $T_c$ (BJD) & $2458258\pm 110$ & $2455674^{+160}_{-110}$ & $2457105^{+13}_{-12}$ & $2448380^{+51}_{-52}$ & $2455992\pm 5$ & $2453143^{+140}_{-180}$ \\ 
       $e$  & $0.170\pm0.004$ & $0.25\pm0.03$ & $0.343\pm0.009$ & $0.212\pm0.008$ & $0.768\pm 0.004$ & $0.42^{+0.05}_{-0.07}$ \\ 
       $\omega$ (deg) & $103.9\pm4.2$ & $198.4^{+6.9}_{-8.0}$ & $134.8^{+1.2}_{-1.2}$ & $174.8^{+1.4}_{-1.5}$ & $55.6\pm 0.7$ & $164.4^{+6.9}_{-6.3}$ \\ 
       $K$ (m s$^{-1}$) & $4092^{+200}_{-190}$ & $162.0^{+14.0}_{-8.8}$ & $1111.4^{+4.3}_{-4.0}$ & $3400^{+28}_{-27}$  & $181.9\pm 1.7$ & $54.7^{+4.2}_{-3.6}$ \\ 
       $M_p \sin{i}$ ($M_J$) & $1034^{+87}_{-81}$& $17.9^{+2.4}_{-1.9}$ & $108.0^{+6.1}_{-6.2}$ & $673^{+33}_{-33}$ & $13.40^{+0.76}_{-0.78}$ & $9.7^{+1.1}_{-1.0}$ \\ 
       $a$ (AU) & $22.28^{+0.57}_{-0.58}$ & $9.75^{+0.61}_{-0.59}$ & $7.33^{+0.20}_{-0.21}$ & $16.21^{+0.48}_{-0.47}$ & $7.03^{+0.20}_{-0.21}$ & $18.5^{+2.3}_{-2.3}$ \\ 
       rms (m s$^{-1}$) & 5.37 & 5.36 & 6.20 & 3.70 & 4.76 & 3.93 \\ 
       $\chi^2_{\rm red}$  & 1.26 & 1.25 & 1.71 & 1.34 & 1.13 & 1.12 
    \enddata
    \label{rvparamtab}
\end{deluxetable*}

For the six companions described in Table~\ref{keptab}, three have
minimum masses in the stellar regime and three have minimum masses in
the planetary regime. Of those in the stellar category, the companion
to HD~145825 was previously published by \citet{jenkins2010} using a
subset of the data presented in this work. Their analysis produced an
estimated minimum mass based on an incomplete orbit of $M =
44.5$~$M_J$, also presenting imaging data that were unable to reveal
the suspected stellar companion. Our additional RV data, acquired in
the years following that publication, result in a closed orbital
solution, and a new minimum mass for the companion of $M = 108$~$M_J$,
or $M = 0.103$~$M_\odot$, which is comfortably in the stellar
regime. Of those in the planetary category, the highly eccentric
planet HD~219077b was previously discovered by
\citet{marmier2013}. Our Keplerian orbital parameters of the planet
presented in Table~\ref{keptab} are in close agreement with those
provided by \citet{marmier2013}, and the lower planet mass of
10.39~$M_J$ is attributable to the lower stellar mass 1.05~$M_\odot$
used by \citet{marmier2013} in their calculations. The two other
planets presented here, HD~92987b and HD~221420b, are new discoveries
with masses of 17.9~$M_J$ and 9.7~$M_J$ respectively.


\section{Imaging Observations}
\label{imaging}

In order to attempt to confirm or constrain the presence of potential
stellar companions to these RV trend stars, we used DSSI whilst it was
at the Gemini South Telescope as a visitor instrument. Most of the
data were taken during a run that occurred 2017 June 6 through 12,
with a few objects added during a second run that occurred 2018
October 26 through 2018 November 5. The instrument itself was
originally described in \citet{horch2009}, and a subsequent upgrade to
its current configuration, which uses two electron-multiplying CCDs,
is detailed in \citet{horch2011}.  The instrument magnifies the image
received at the telescope focal plane so that individual speckles are
critically sampled, and it also sends the light in the collimated beam
through a dichroic beamsplitter. Once re-imaged onto the two
detectors, this allows for two different speckle patterns to be
simultaneously recorded in two filters. For all of the observations
discussed here, these were a 692-nm filter with a bandpass of 40 nm,
and an 880-nm filter with a bandpass of 50 nm.

\citet{horch2011} also described the reduction process for companion
detection when observing with DSSI, which we review briefly here. A
sequence of short-exposure (60 ms) images is taken of the target,
resulting in two fits data cubes, one for each filter. The standard
file contains 1000 images; for fainter targets, more fits cubes are
recorded in sequence, and the results are combined in the analysis
phase. To have an estimate of the speckle transfer function at the
time of the observation, a bright point source that is close to the
science target in the sky position is observed either right before or
right after each science target. Beacause these stars are bright, even
a single 1000-frame sequence yields a very high signal-to-noise ratio
with which to complete a deconvolution in the analysis phase.

The analysis begins with the calculation of autocorrelation functions
and image bispectra from the raw data frames. These are then used to
generate diffraction-limited reconstructions of each star in the
Fourier plane using the method of \citet{lohmann1983}. Upon
Fourier-transforming the autocorrelation, the spatial frequency power
spectrum is obtained. If a similar calculation is performed on the
point-source observation, then its power spectrum can be used to
deconvolve the speckle transfer function from the science target's
power spectrum via simple division. By taking the square root of the
result, the diffraction-limited modulus of the object's Fourier
transform is obtained. The bispectrum contains within it an estimate
of the derivative of the phase of the object's Fourier
transform. Reconstructing the phase from the bispectrum is completed
with the relaxation algorithm of \citet{meng1990}. By combining with
the modulus, a diffraction-limited estimate of the full,
complex-valued transform is obtained. This result is low-pass filtered
with a Gaussian function, and inverse-transformed to arrive at the
final reconstructed image.

Companion detection is performed by studying the reconstructed image
and examining the statistics of local maxima and minima in the image
as a function of separation from the primary star
\citep{horch2011}. We draw a sequence of concentric annuli centered on
the primary star, and determine the average value of local maxima and
local minima, and their standard deviations. This allows us to
estimate for each annulus what the 5$\sigma$ value above the noise is
as a function of separation. Values are associated with the center
radius of each annulus, which is chosen at 0.1-arcsecond intervals. To
produce a continuous curve, we make a cubic-spline interpolation
between the final values, and assume that, at the diffraction limit,
our sensitivity to companions goes to a $\Delta m$ of 0. If a peak in
the image exceeds the 5$\sigma$ value for its separation, then it is
studied as a possible companion. Generally, if a similar peak occurs
in images of both filters, this is judged to be confirmation of the
detection of a stellar component. Occasionally, very red components
are only detected in the redder filter; these detections must be
viewed as less certain until follow-up observations can confirm the
result.


\section{Detection of Stellar Companions}
\label{stars}

\begin{figure*}
  \begin{center}
    \begin{tabular}{cccc}
      \includegraphics[width=3.5cm]{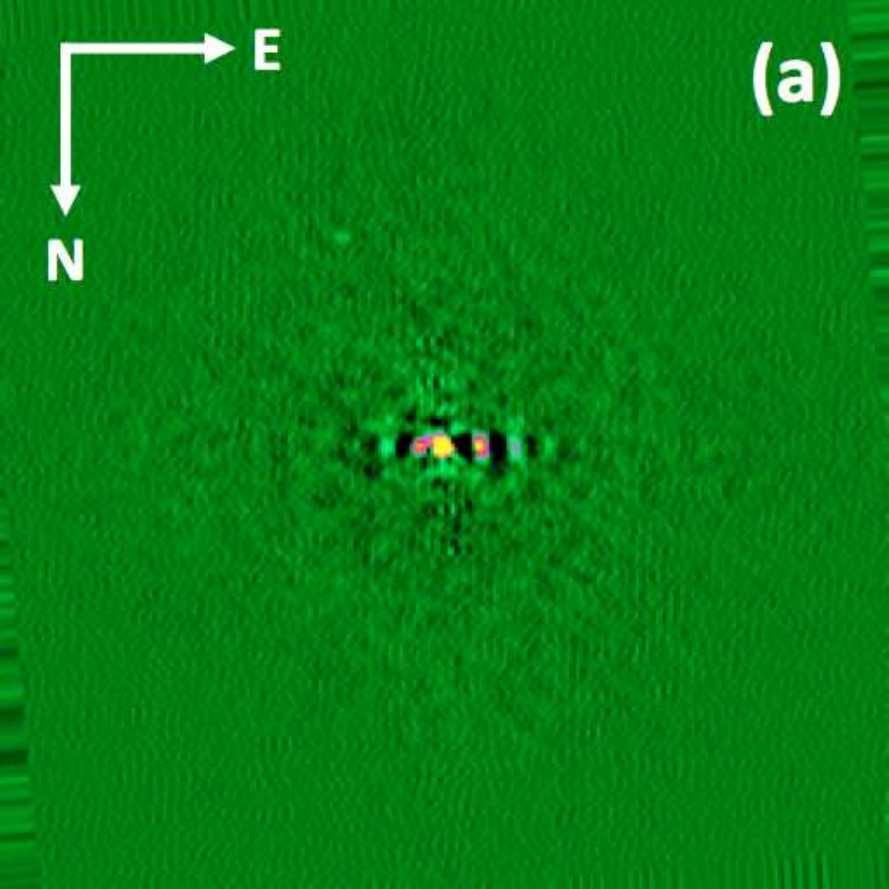} &
      \includegraphics[width=5.0cm]{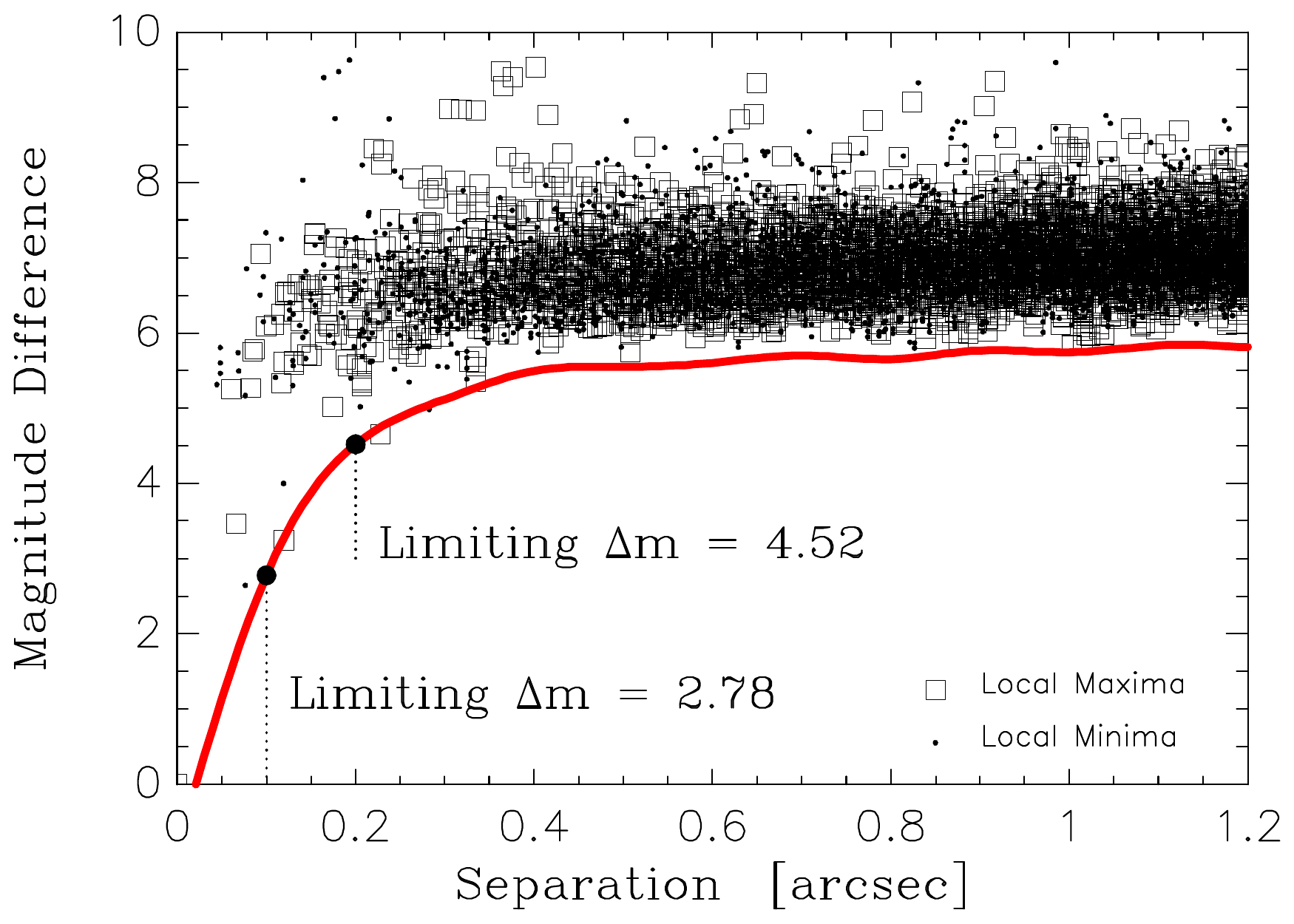} &
      \includegraphics[width=3.5cm]{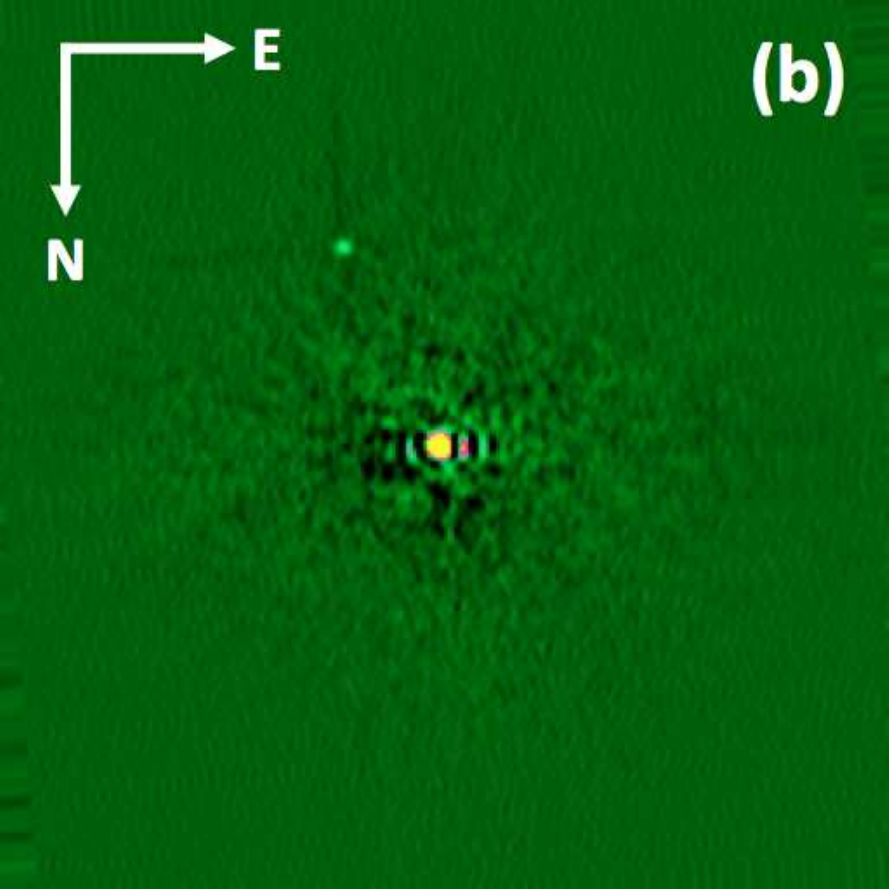} &
      \includegraphics[width=5.0cm]{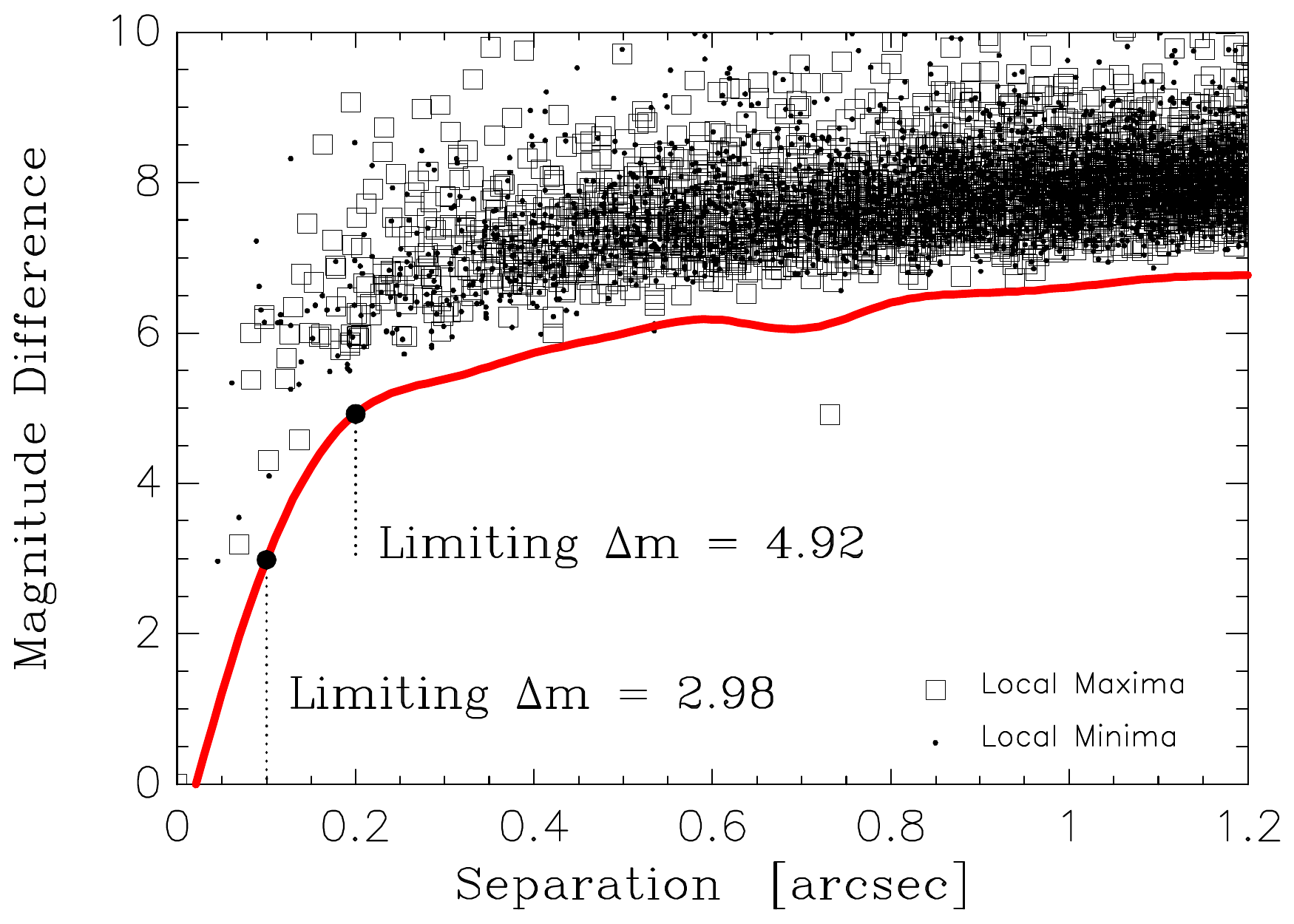} \\
      \includegraphics[width=3.5cm]{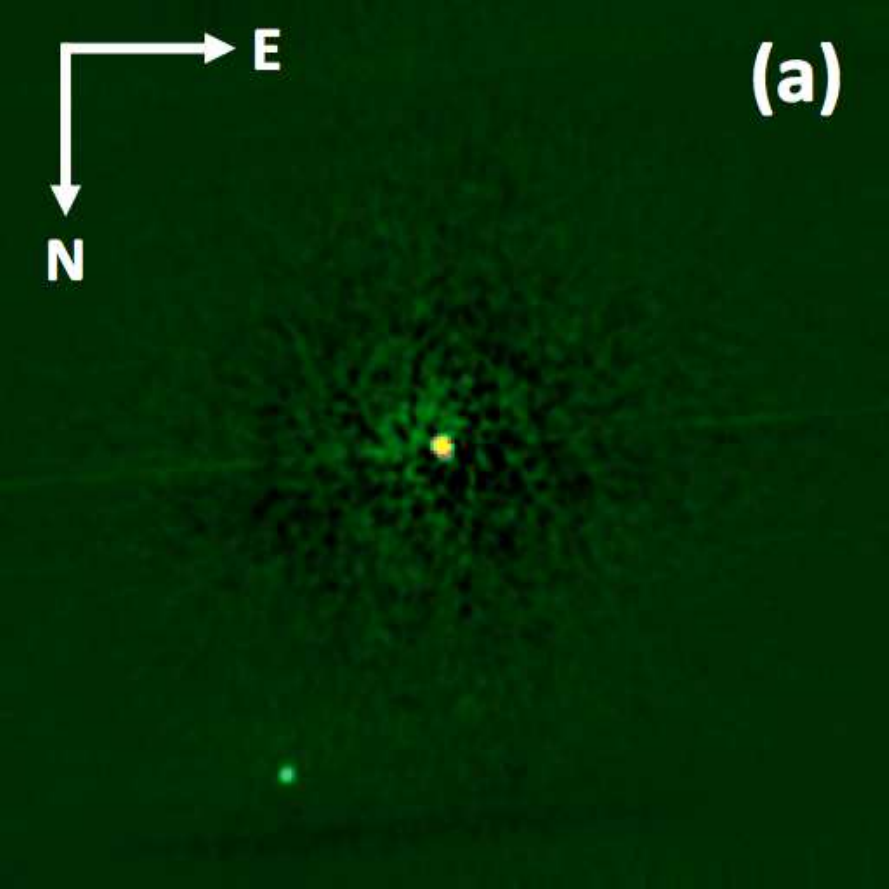} &
      \includegraphics[width=5.0cm]{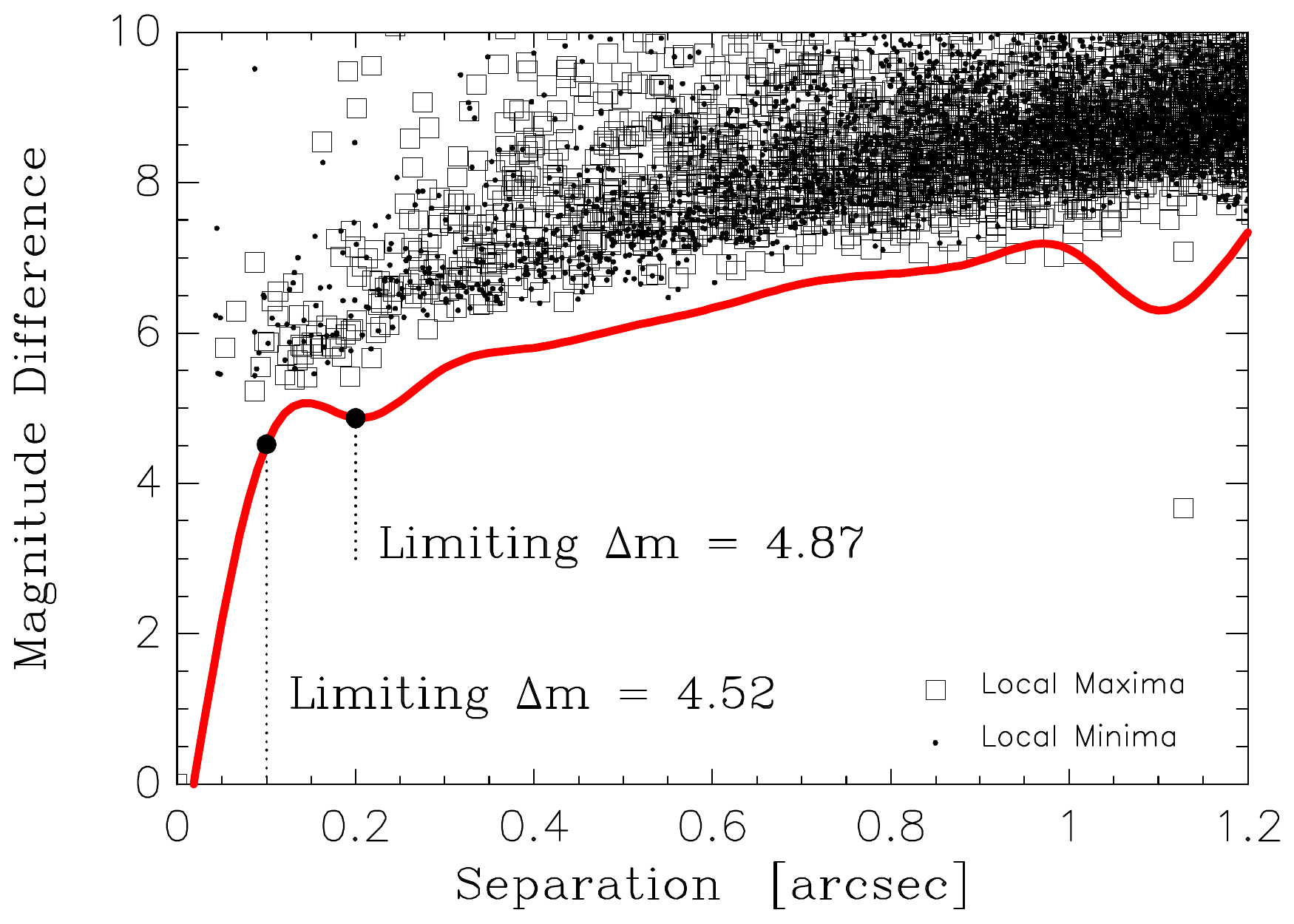} &
      \includegraphics[width=3.5cm]{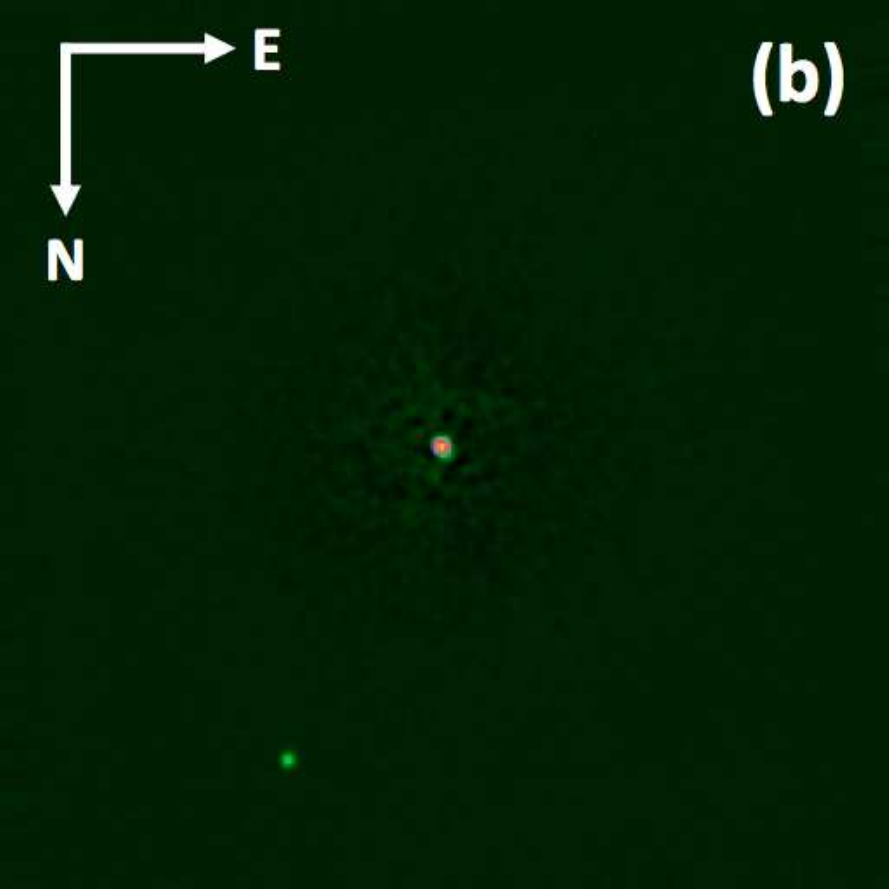} &
      \includegraphics[width=5.0cm]{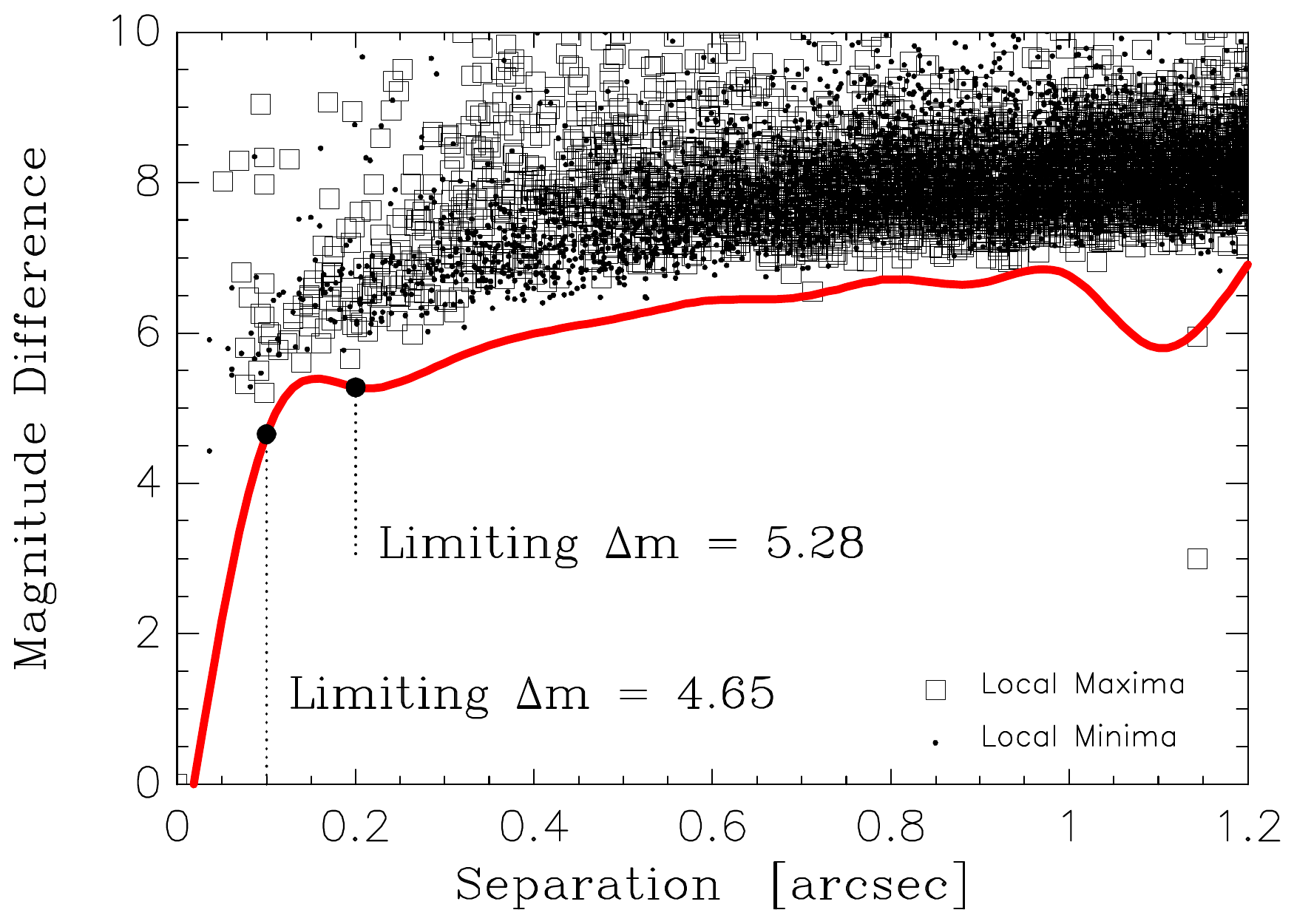} \\
      \includegraphics[width=3.5cm]{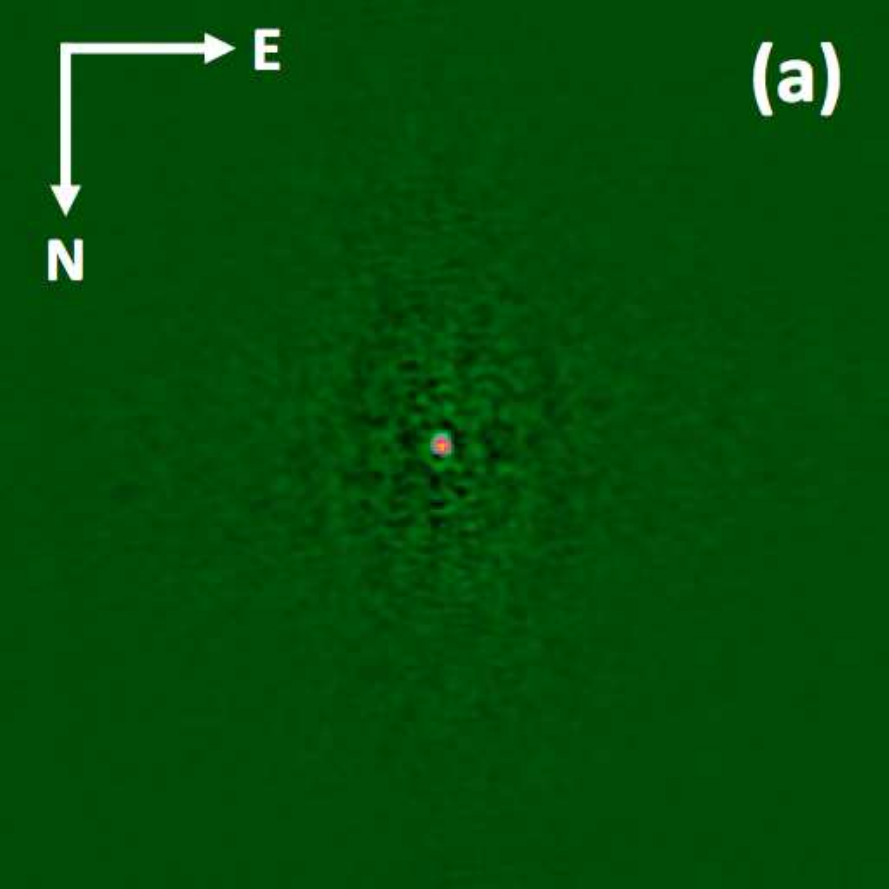} &
      \includegraphics[width=5.0cm]{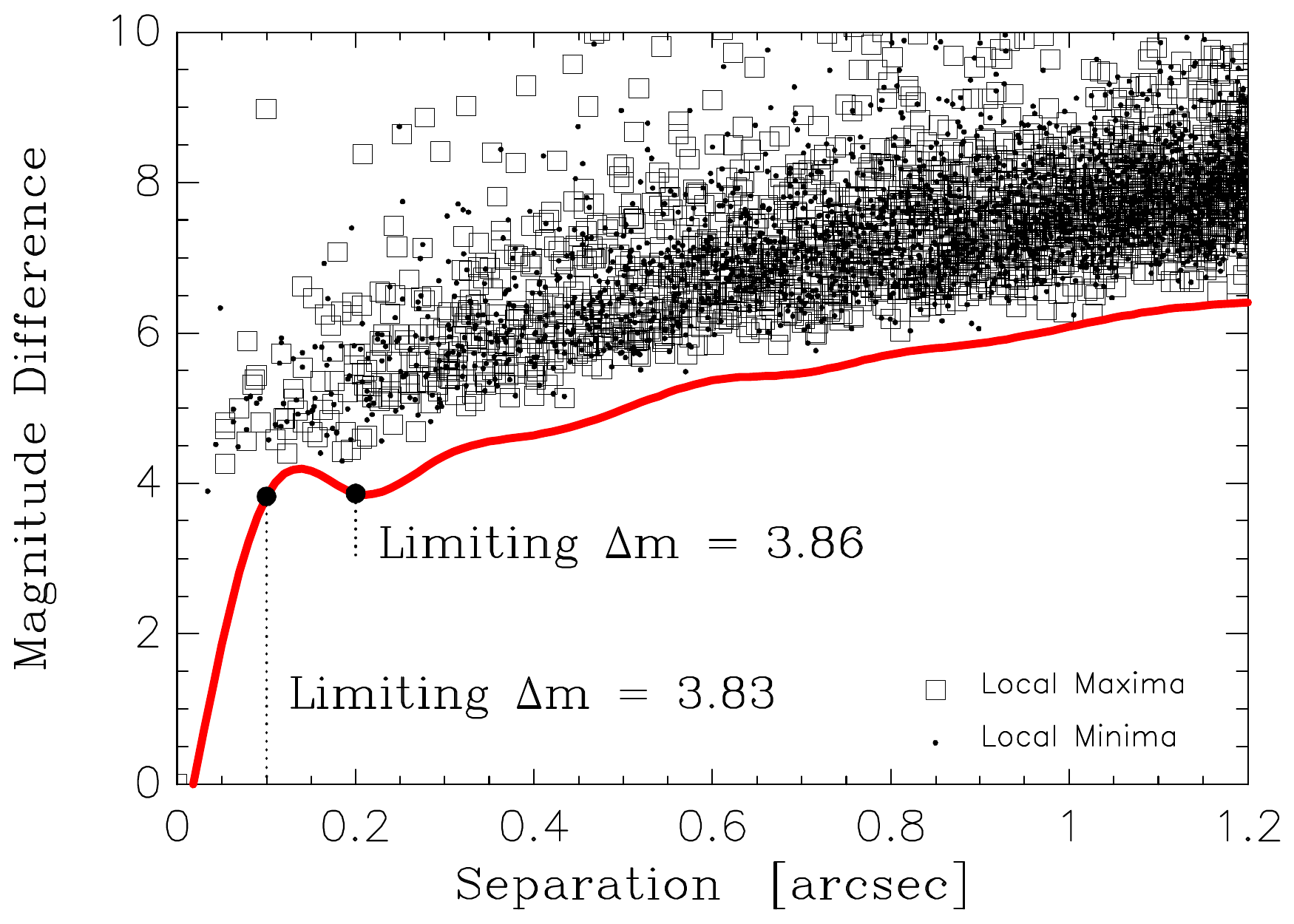} &
      \includegraphics[width=3.5cm]{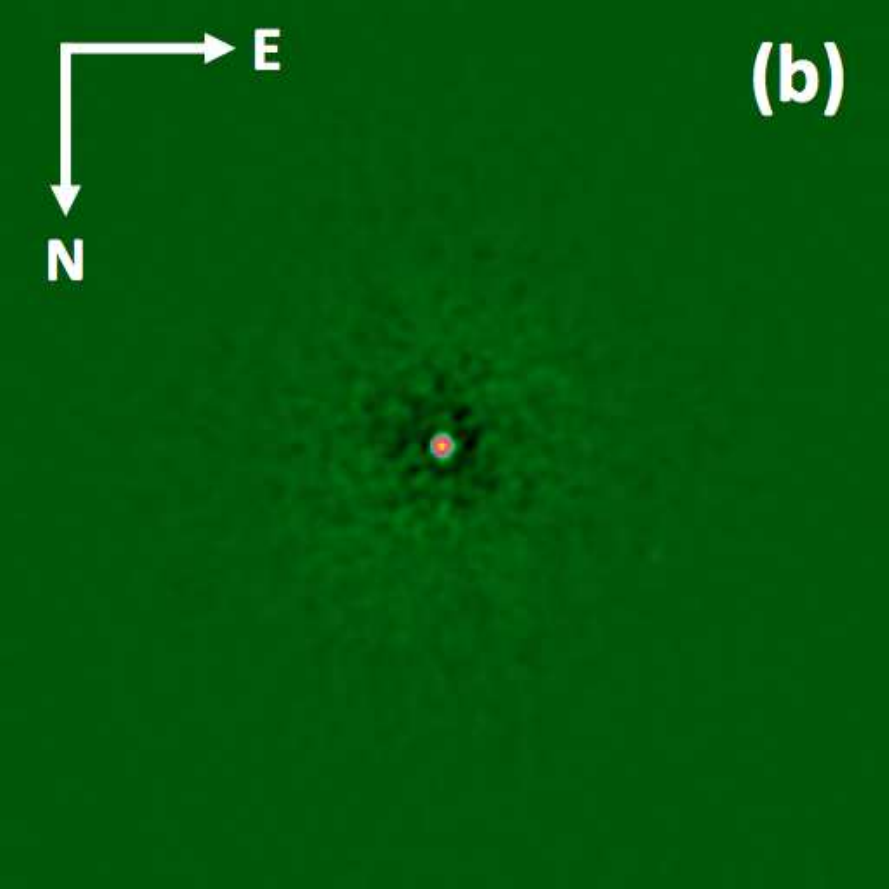} &
      \includegraphics[width=5.0cm]{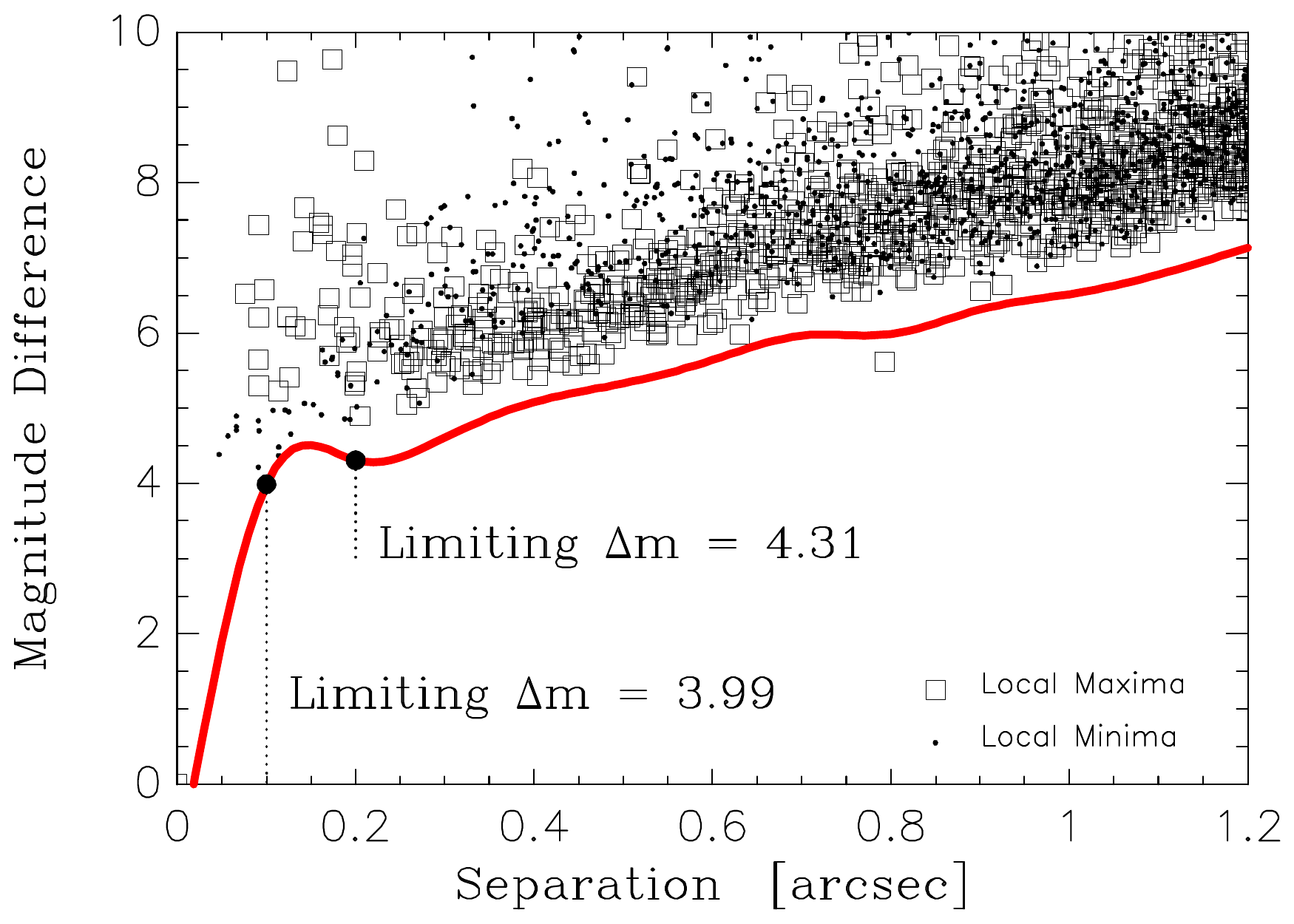} \\
      \includegraphics[width=3.5cm]{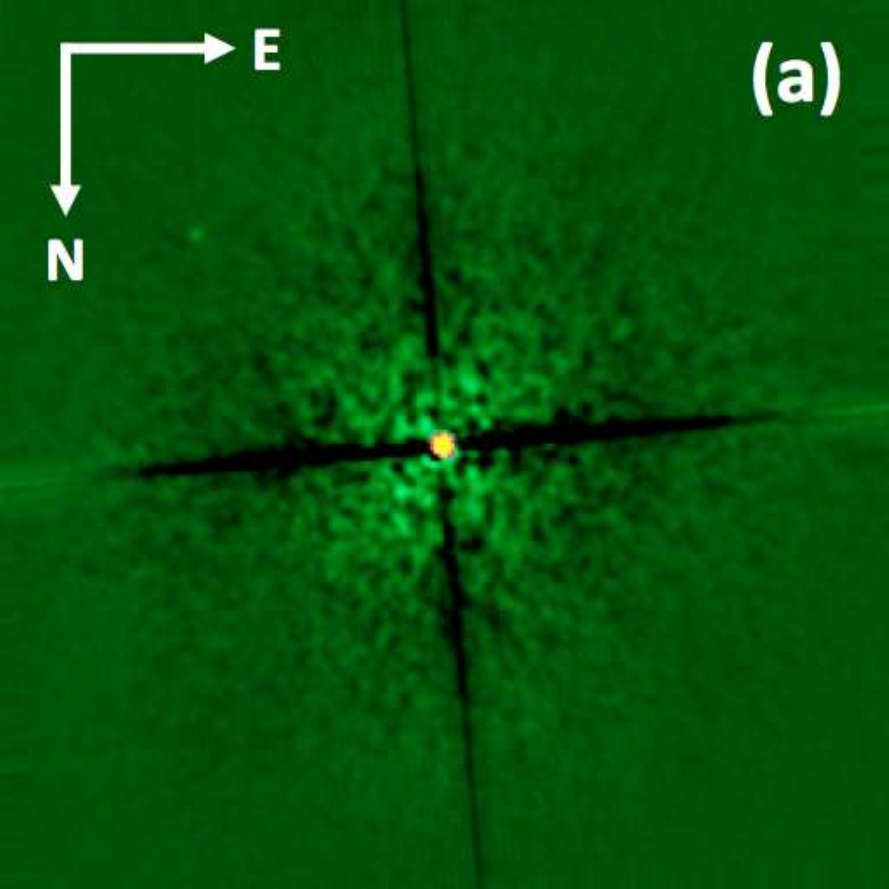} &
      \includegraphics[width=5.0cm]{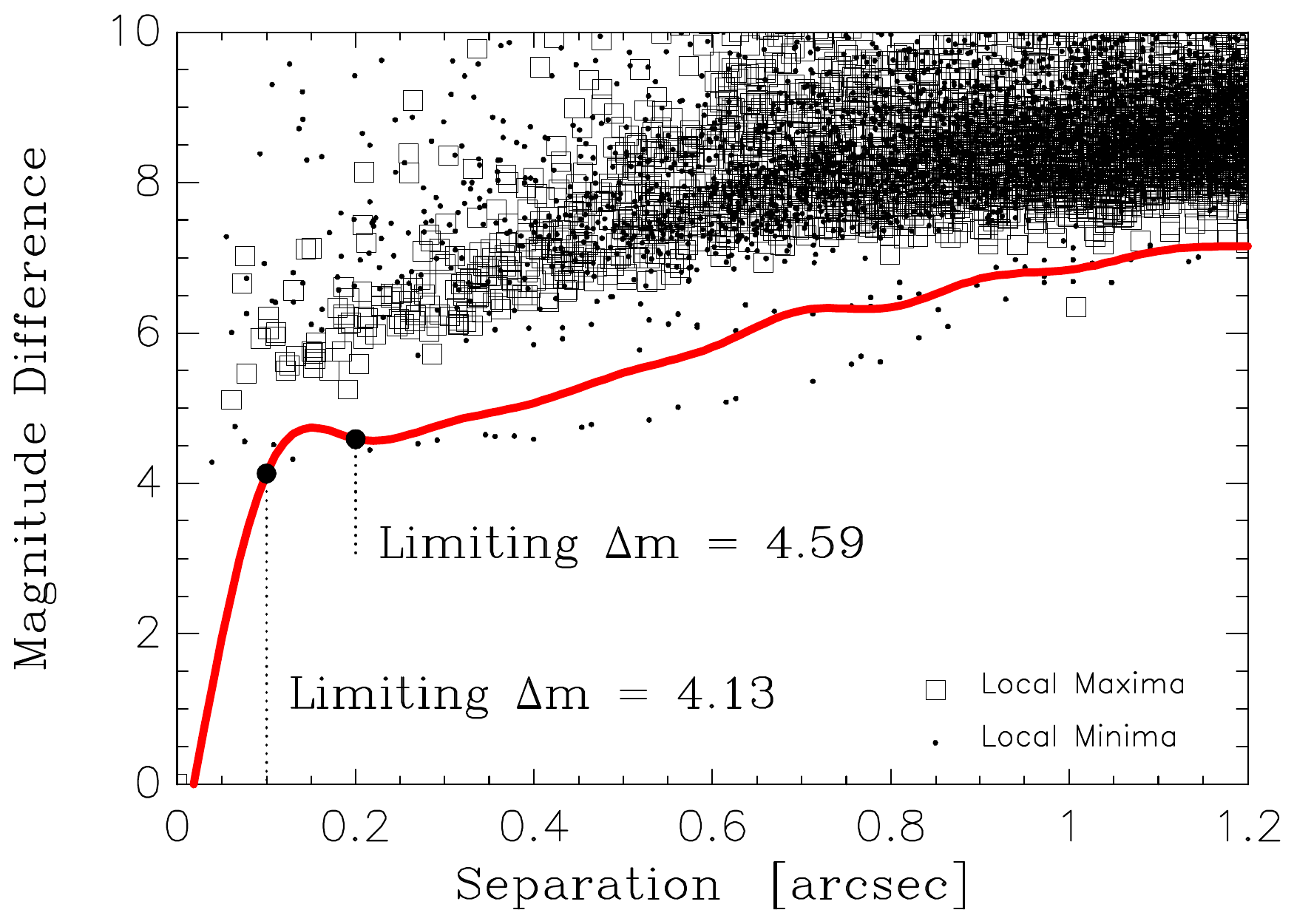} &
      \includegraphics[width=3.5cm]{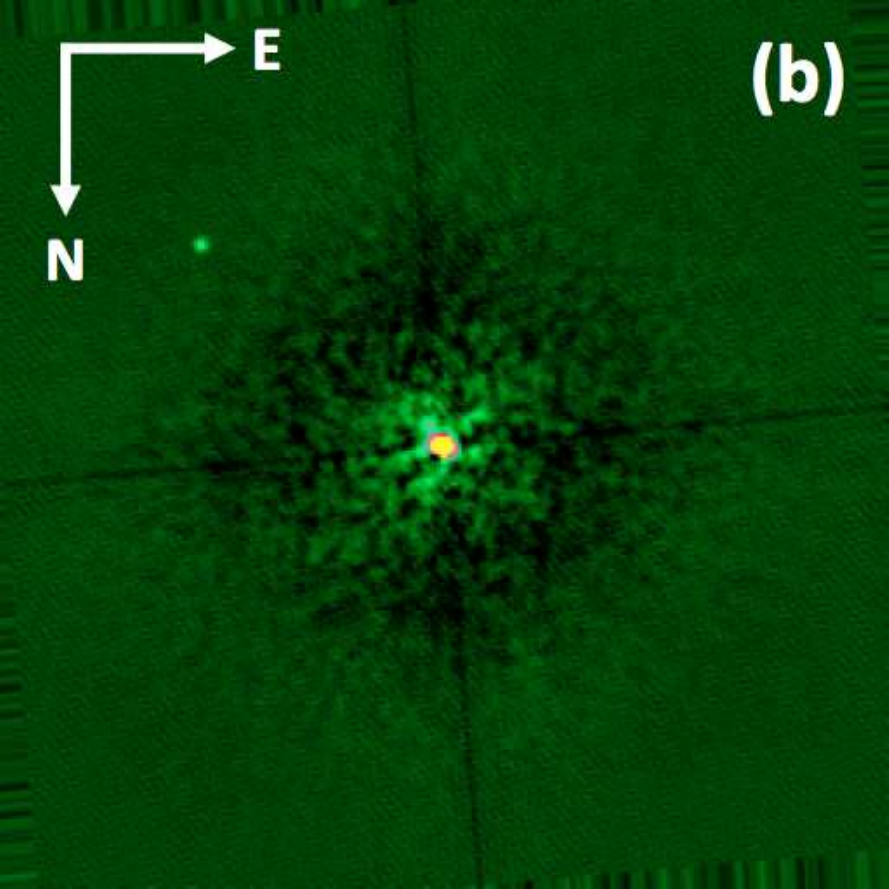} &
      \includegraphics[width=5.0cm]{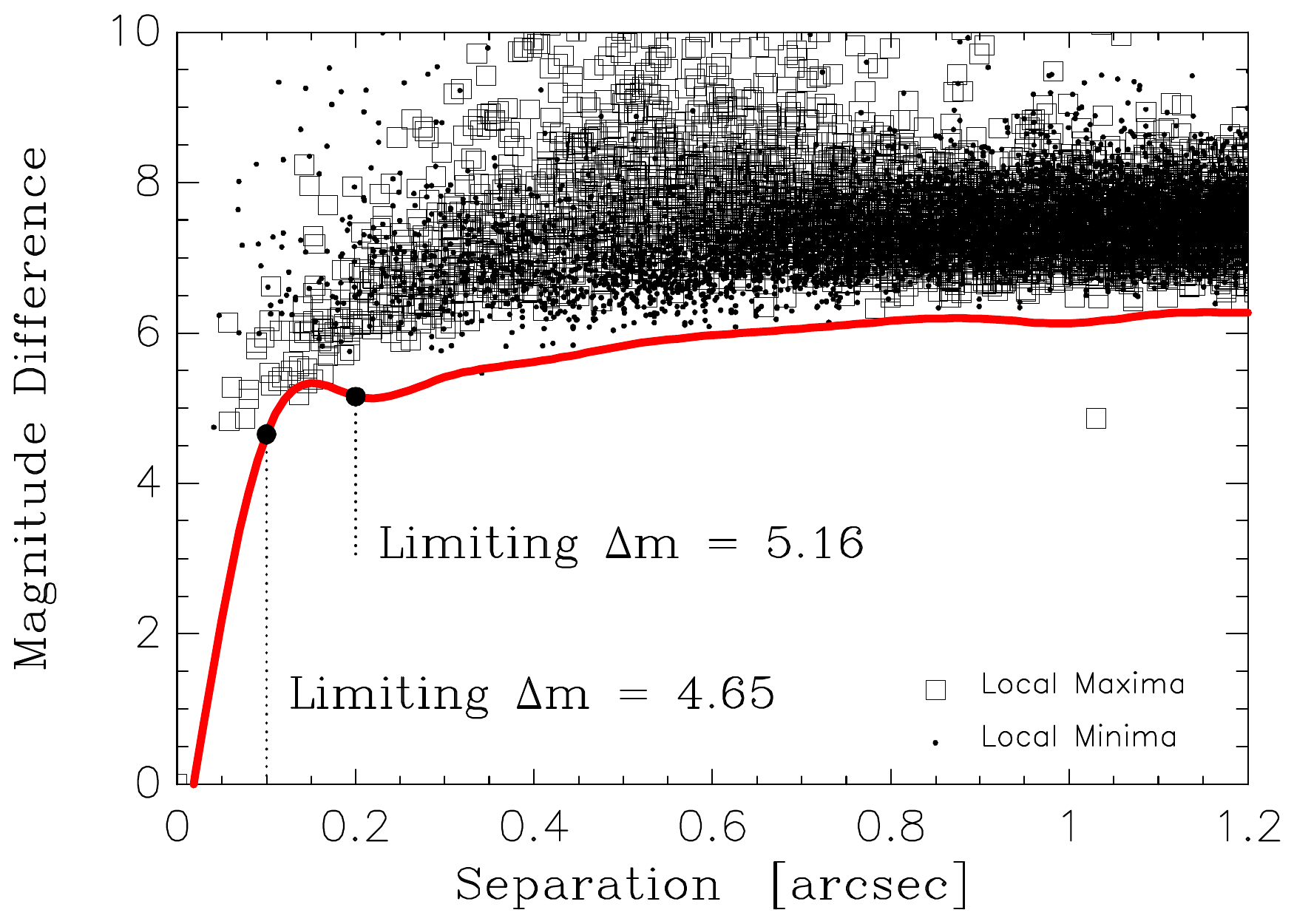} \\
      \includegraphics[width=3.5cm]{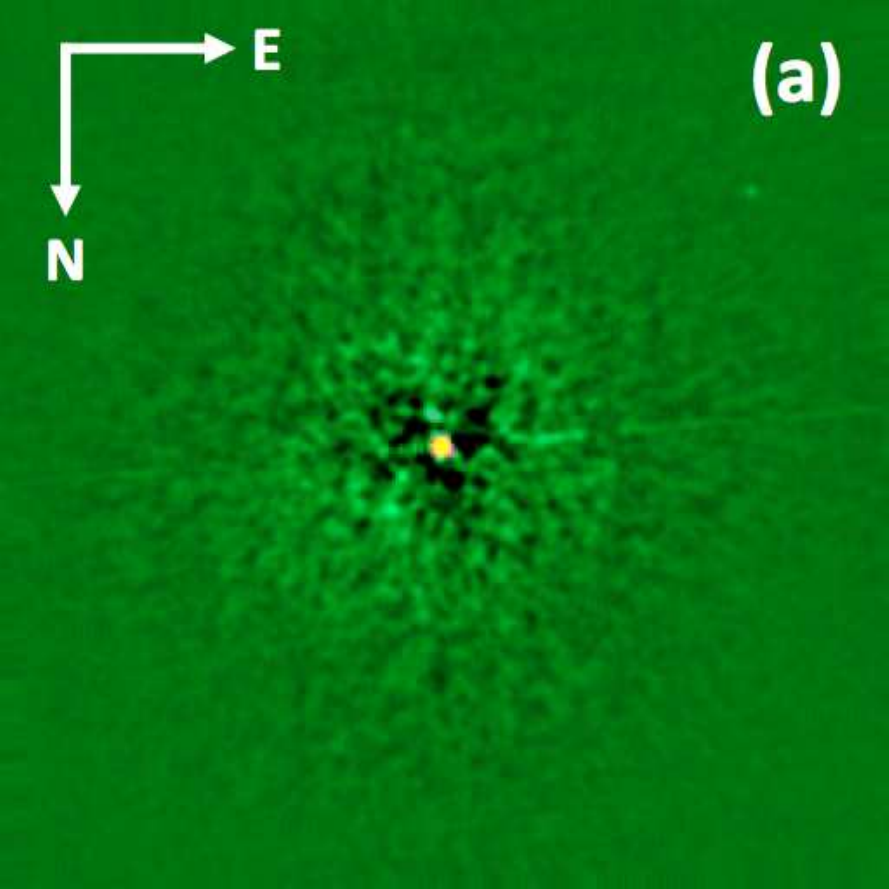} &
      \includegraphics[width=5.0cm]{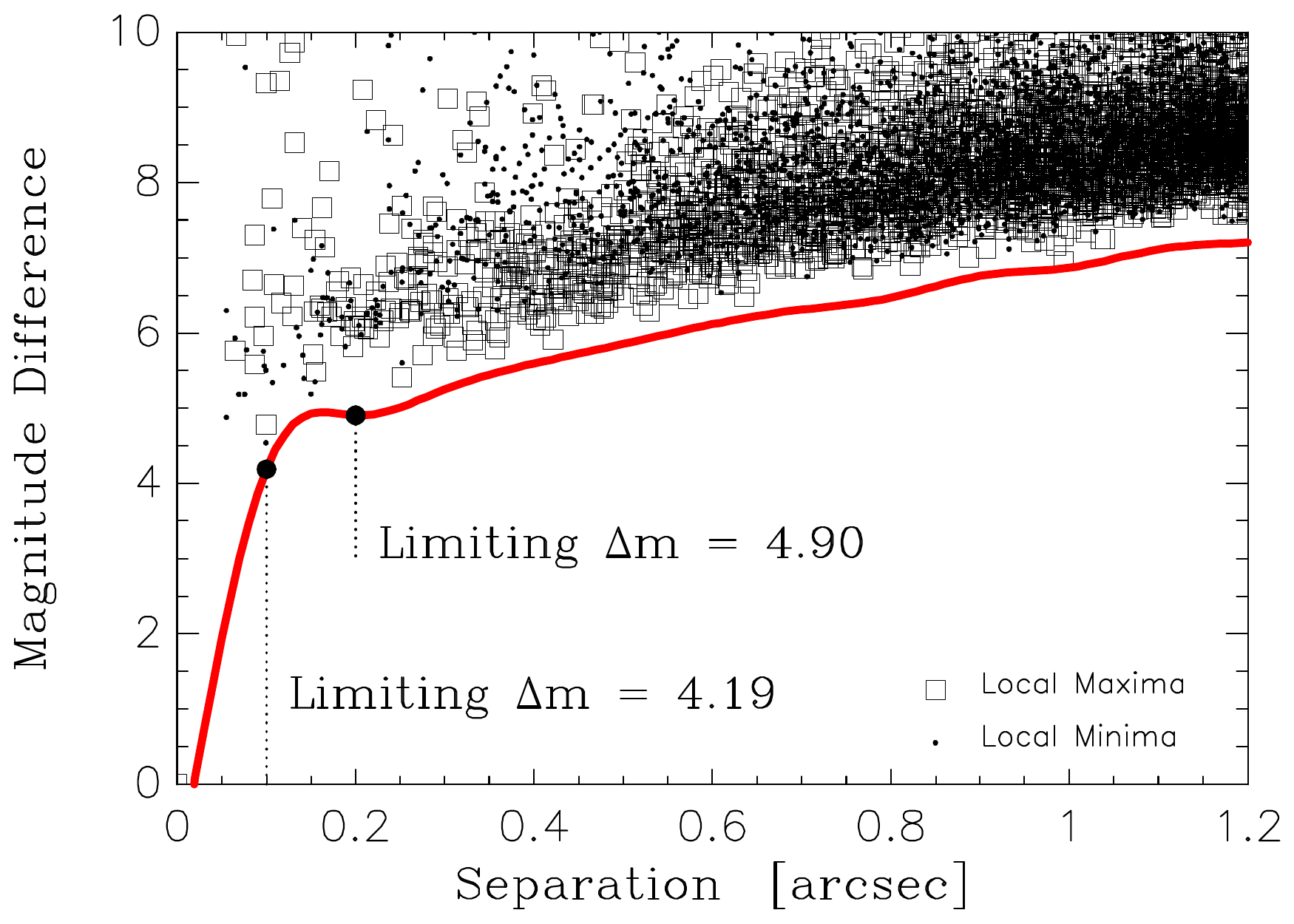} &
      \includegraphics[width=3.5cm]{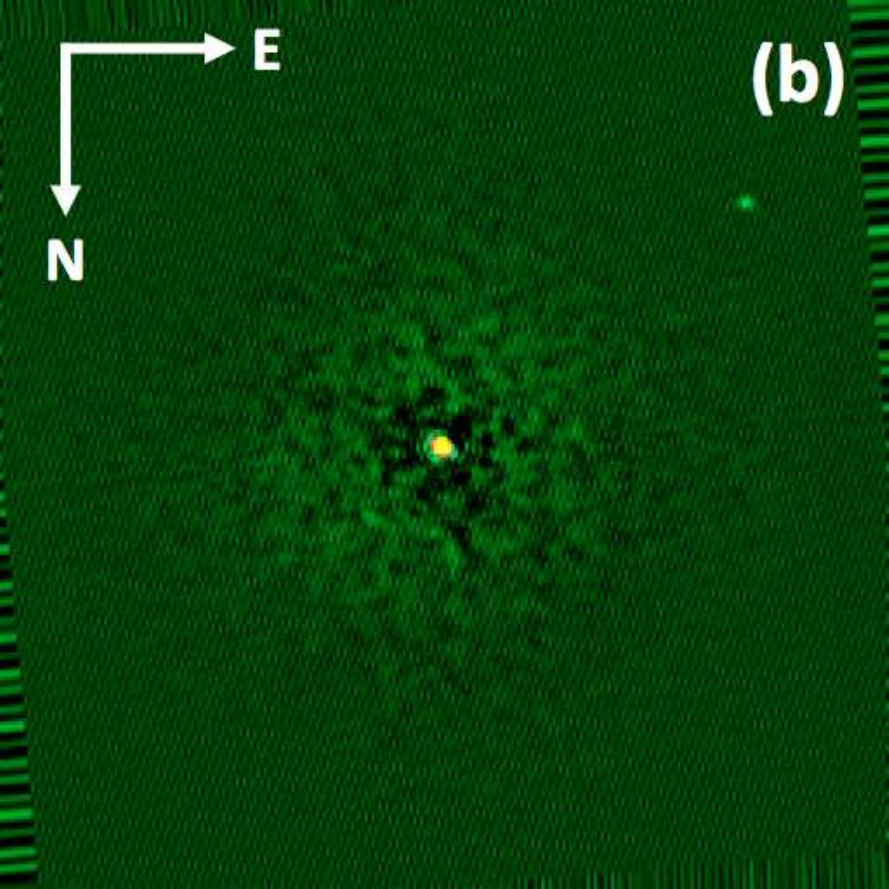} &
      \includegraphics[width=5.0cm]{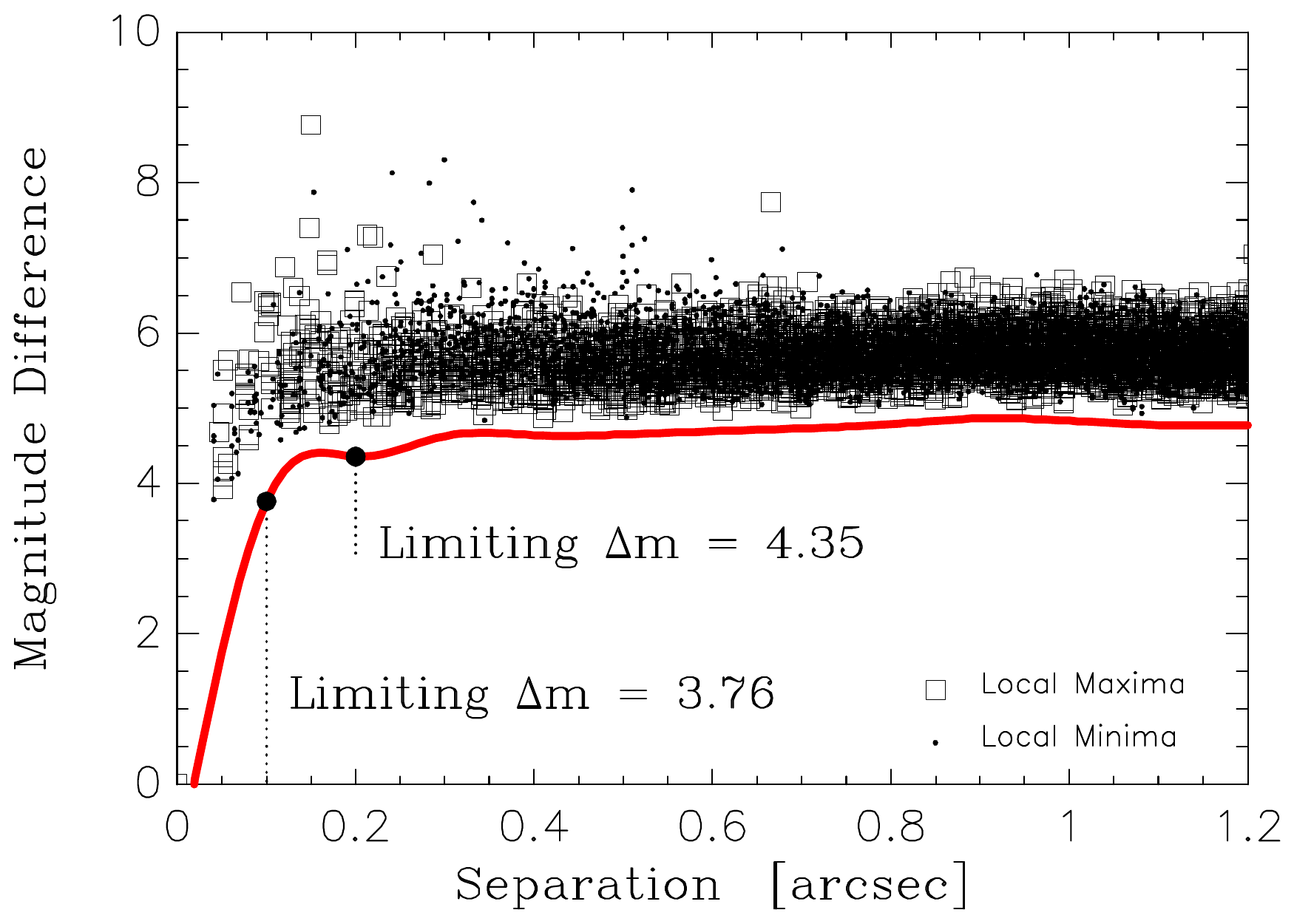}
    \end{tabular}
  \end{center}
  \caption{DSSI images and detection limits for (from top to bottom)
    HD~51929, HD~166553, HD~212330, HD~212708, and HD~217958. The data
    shown are for the 692~nm (a) and 880~nm (b) filters, and the
    field-of-view for the images is 2.8$\times$2.8$\arcsec$. The
    detection limit plots include data for the local maxima (squares)
    and minima (points) and the solid curves are cubic spline
    interpolations of the 5$\sigma$ detection limit. All five of these
    targets show evidence of companions, described in
    Section~\ref{imaging} and Table~\ref{stellartab}. The
    uncertainties in PA, sep, and $\Delta m$ are 0.2$\degr$,
    0.0025$\arcsec$, 0.15~mag respectively (see Section~\ref{stars}).}
  \label{dssiplots}
\end{figure*}

Using the methodology described in Section~\ref{imaging}, we examined
the DSSI data for all twenty targets. Of those, five showed
significant evidence of a possible stellar companion to the host
star. Specifically, stellar companion evidence was found for the
following targets: HD~51929, HD~166553, HD~212330, HD~212708, and
HD~217958. The DSSI images for each of these five targets are shown in
Figure~\ref{dssiplots}, for both the 692~nm (a) and 880~nm (b)
passbands. Each of the images have a field-of-view of
2.8$\times$2.8$\arcsec$ and are oriented such that north is down and
east is to the right. The contrasts of the images were selected to
enhance the visibility of the detected companions, though this was
challenging for the companion to HD~212330. To the right of each panel
are the limiting magnitude plots constructed from each of the images
(see Section~\ref{imaging}), where the data represent local maxima and
minima and the solid curve is a cubic spline interpolation of the
5$\sigma$ detection limit. We derived the properties of the stellar
companions using the methodology described by \citet{horch2004}. The
details regarding the DSSI derived properties of the stellar companion
are shown in Table~\ref{stellartab}, including the position angle
(PA), the separation from the host star (sep), and the difference in
magnitude from the host star ($\Delta m$) in both DSSI bandpasses. We
estimated the uncertainty of these properties based on previous
similar observations using DSSI \citep{horch2012b}: 0.2$\degr$,
0.0025$\arcsec$, and 0.15~mag for PA, sep, and $\Delta m$
respectively.

\begin{deluxetable}{lcccc}
  \tablecaption{\label{stellartab} DSSI derived companion properties.}
  \tablehead{
    \colhead{Host Star} &
    \colhead{Filter (nm)} &
    \colhead{PA ($\degr$)} &
    \colhead{Sep ($\arcsec$)} &
    \colhead{$\Delta m$}
  }
  \startdata
  HD~51929  & 692 & 205.7 & 0.7221 & 5.83 \\
  HD~51929  & 880 & 205.4 & 0.7247 & 4.50 \\
  HD~166553 & 692 & 334.7 & 1.1319 & 3.55 \\
  HD~166553 & 880 & 334.7 & 1.1354 & 2.91 \\
  HD~212330 & 692 &  62.2 & 0.7655 & 6.65 \\
  HD~212330 & 880 &  61.7 & 0.7653 & 5.50 \\
  HD~212708 & 692 & 229.2 & 1.0103 & 6.48 \\
  HD~212708 & 880 & 229.2 & 1.0122 & 5.27 \\
  HD~217958 & 692 & 129.5 & 1.2570 & 6.51 \\
  HD~217958 & 880 & 129.6 & 1.2509 & 4.20
  \enddata
\end{deluxetable}

\begin{deluxetable*}{lccccl}
  \tablecaption{\label{stellartab2} Additional companion properties.}
  \tablehead{
    \colhead{Host Star} &
    \colhead{Spectral Type} &
    \colhead{$\Delta V$} &
    \colhead{Color Offset ($\sigma$)} &
    \colhead{Bound Probability (\%)$^{\dagger}$} &
    \colhead{Comments}
  }
  \startdata
  HD~51929  & M4 & 6.53 & 2.37 & 29    & \\
  HD~166553 & M0 & 4.17 & 1.18 & 14    & \\
  HD~212330 & M6 & 8.33 & 0.04 & 25    & \\
  HD~212708 & M6 & 7.93 & 0.60 & 14    & \\
  HD~217958 & M5 & 7.48 & 4.75 & $<14$ & Additional observations needed  
  \enddata
  \tablenotetext{\dagger}{The probability of being gravitationally
    bound based only on angular separation (see text).}
\end{deluxetable*}

The question remains as to whether the detected companions are
gravitationally bound to the host star. Analysis of DSSI data for {\it
  Kepler} exoplanet candidate host stars by \citet{horch2014} used a
statistical approach to demonstrate that most of the detected
companions within the DSSI separation range are indeed bound to the
host, with similar results found by \citet{matson2018}. 

We applied the results of \citet{matson2018} from the Gemini telescope
to approximate the probability that the companions were
gravitationally bound based only on angular separation. These
estimates are shown in Table~\ref{stellartab2}. For all targets except
HD~217958, the \citet{matson2018} results suggest that there is a
relatively low probability of being bound. The angular separation of
the companion to HD~217958 fell outside the range of the
\citet{matson2018} investigation, meaning that the angular separation
is insufficient as an indicator that the companion is bound.

The color information provided by the two DSSI filters also allows for
an isochrone analysis to determine if there are significant
differences between the observed and predicted companion
properties. The isochrones were extracted from the Dartmouth Stellar
Evolution Database \citep{dotter2008}. We used the $\Delta m$ values
for each filter shown in Table~\ref{stellartab} to interpolate down
the Dartmouth isochrones from the position of the primary (as
determined by the stellar properties shown in Table~\ref{sumtab}) to
the position of a hypothetical bound companion with the measured
contrast. We performed this translation down the isochrones for each
of the measured $\Delta m$ values, then defined a companion model from
the weighted average of these individual $\Delta m$ models. This
analysis results in a predicted model color for the companion, that we
compare against the measured color of the companion, as detailed by
\citet{hirsch2017}. A color offset of $\leq 3\sigma$ between the
observed and modeled companion color is taken to imply that the object
is gravitationally bound.

\begin{figure}
  \begin{center}
    \includegraphics[width=8.5cm]{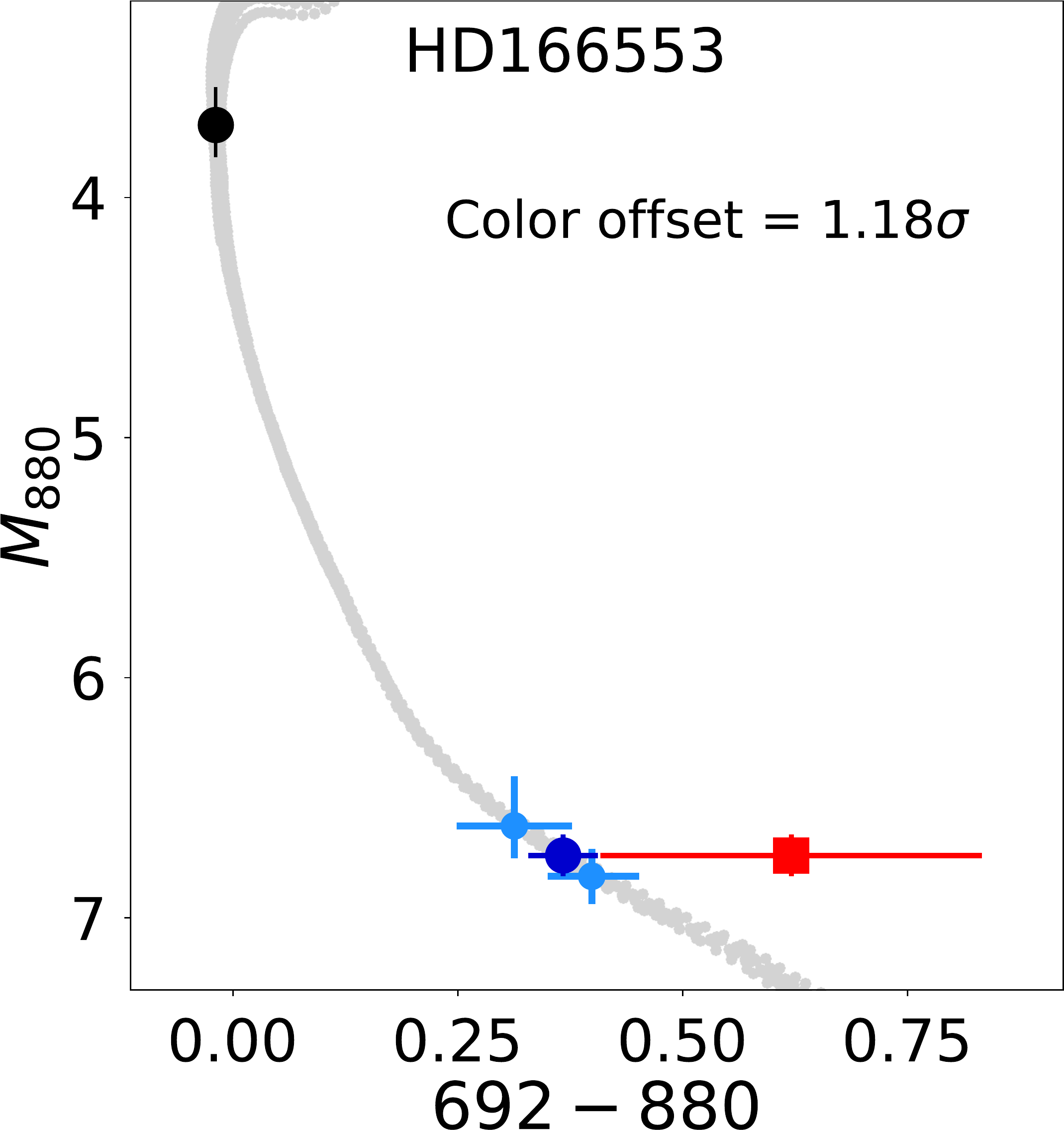}
  \end{center}
  \caption{Isochrone model for the star HD~166553 and the detected
    companion. The light blue points are the color predictions of the
    secondary based on each of the individual $\Delta m$ measurements,
    combined with the properties of the primary star. The weighted
    mean of the light blue points is shown in dark blue. In this case,
    the color offset between the model color (dark blue) and the
    observed color (red) of the HD~166553 companion is 1.18$\sigma$.}
  \label{isochrone}
\end{figure}

The case of isochrone analysis for HD~166553 is shown as an example in
Figure~\ref{isochrone}. The isochrone track for the primary is shown
in light gray and the positions of the secondary for the 692~nm and
880~nm filters are shown in light blue. The weighted average of these
positions, shown in dark blue, was compared to the observed color of
the secondary, shown in red, and then we calculated the color offset
between the model color and observed color in units of the measured
uncertainty. As shown in Figure~\ref{isochrone}, the color offset for
the HD~166553 case is 1.18$\sigma$ which is within the 3$\sigma$
criteria adopted by \citet{hirsch2017}. The color offsets for the
other four cases are listed in Table~\ref{stellartab2}. Only the
companion to HD~217958 does not meet the described criteria, lending
credence to most of the companions being bound. Note that these
results do not mean that the detected object near HD~217958 is not
gravitationally bound; rather they indicate that it does not fulfill
the assumptions of similar age and formation location that the
\citet{hirsch2017} criteria apply and will require additional
observations to verify its bound state.

The measured information regarding the companions and their host stars
may be used to provide a rough estimate of the companion spectral
types. To do this, we utilized the Pickles spectral library
\citep{pickles1998} combined with the measured $\Delta m$ values for
each of the DSSI filters. Using this methodology, we also infer
$\Delta V$ values between the companions and the primaries (see Table
\ref{stellartab2}). When compared with the $V$ magnitudes and
distances of the host stars (see Table~\ref{sumtab}), these are
consistent with the possible companions having spectral types in the
range M4--6, with the exception of the companion to HD~166553 that is
closer to M0. Note that these are only approximate estimates of the
spectral types and should be treated with caution.

Given the measured separations of the detected objects from the target
stars shown in Table~\ref{stellartab}, we performed a search of the
{\it Gaia} second data release (DR2) to check if our detections are
also in that catalog \citep{brown2018}. The only companion present in
DR2 is the companion to HD~166553, but the companion does not have a
measured parallax, and so a distance could not be established. The
$\Delta m$ for the {\it Gaia} passband between the target and
companion is 2.92 and the separation is 1.15$\arcsec$, both of which
are consistent with the DSSI values shown in
Table~\ref{stellartab}. The {\it Gaia} detection represents a second
epoch and indicates a common proper motion between the target and the
companion.

Of further note is that one of the five targets for which a companion
was detected, HD~212330, also has a full Keplerian orbital solution
(see Table~\ref{keptab}). The minimum mass provided by the RV data is
consistent with a stellar companion, and the orbital parameters
combined with the distance to the host star (see Table~\ref{sumtab})
result in a maximum angular separation of $0.96\arcsec \pm
0.03\arcsec$ \citep{kane2013c,kane2018c}. Comparison with the angular
separation of the detected companion shown in Table~\ref{stellartab}
indicates that the companion was imaged close to maximum angular
separation. This system also has the smallest color offset
(0.04$\sigma$) from the isochrone analysis.


\section{Combined RV and Imaging Analysis}
\label{planets}

The imaging observations presented in Section~\ref{stars} yielded
upper limits on the masses of potential stellar companions. Here, we
define the range of possible planet parameters by combining the
imaging and RV observations for the remaining fifteen targets in the
survey.

First, we place lower limits on the mass and separation of a companion
using the RV observations. Treating the RV range ($\Delta$RV,
Table~\ref{sumtab}) as a lower limit for twice the RV semi-amplitude,
we numerically solve the following relation for the minimum mass of a
companion ($M_p$) as a function of semi-major axis ($a$):
\begin{equation}\label{eq:rvsemiamp}
    \frac{\Delta \mathrm{RV}}{2} \le \sqrt{\frac{G}{a(1-e^2)}}
    \frac{M_p \sin{i}}{\sqrt{M_{\star}+M_p}}
\end{equation}
where $G$ is the gravitational constant, $e$ is the companion orbital
eccentricity, and $i$ is the companion orbital inclination. We
accounted for the unknown companion eccentricities and inclinations
through a Monte Carlo approach. As a function of semi-major axis, we
evaluated Equation \ref{eq:rvsemiamp} 1,000 times drawing inclination
from a uniform distribution in $\cos i$ and drawing eccentricity from
a Beta distribution with shape parameters $\alpha=0.867$ and
$\beta=3.03$. This Beta distribution is motivated by empirical trends
in the eccentricities of RV exoplanets \citep{kipping2013b}. Note that
the use of the Beta distribution assumes that shorter-period RV
planets and those presented in this study follow a similar
eccentricity distribution.

Second, we place upper limits on the mass and separation of a
companion using the imaging observations and following the procedure
of \citet{kane2014c}. Briefly, we use the known distance to each
target system and the mass-luminosity relations of \citet{henry1993}
to estimate the apparent $V$-band magnitude of a possible stellar
companion as a function of $M_p$. Comparison to the known apparent
$V$-band magnitude of the host star yields visual $\Delta m$ values
for each target, also as a function of $M_p$. Then, using the Pickles
spectral library \citep{pickles1998} and the transmission curves of
each DSSI imaging filter, we transform visual $\Delta m$ value to
speckle $\Delta m$ values. We compare these to the DSSI limiting
magnitude curves of each target to find $M_p$ as a function of angular
separation, which we convert to semi-major axis using the distance.

\begin{figure*}
  \begin{center}
    \begin{tabular}{ccc}
      \includegraphics[width=5.6cm]{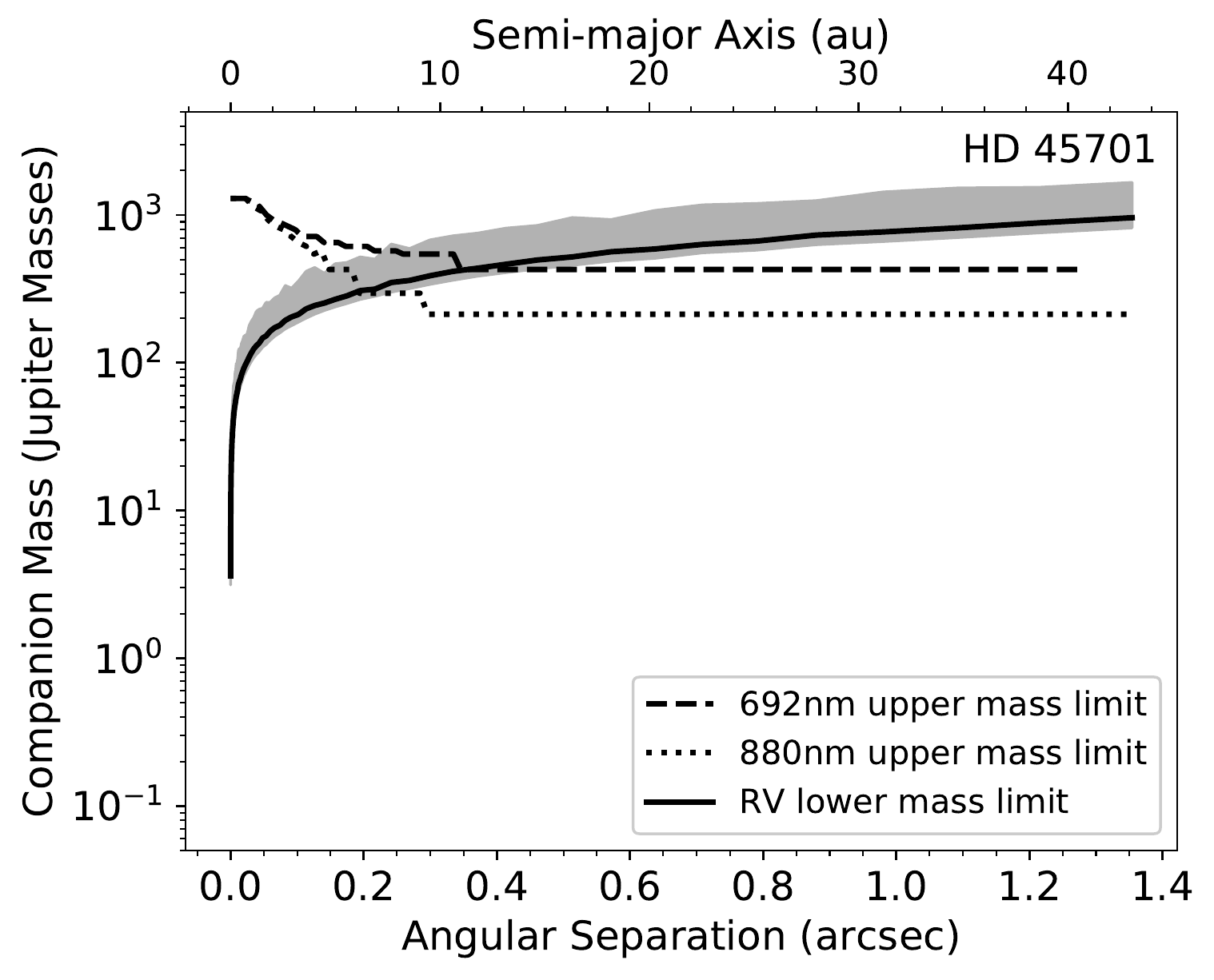} &
      \includegraphics[width=5.6cm]{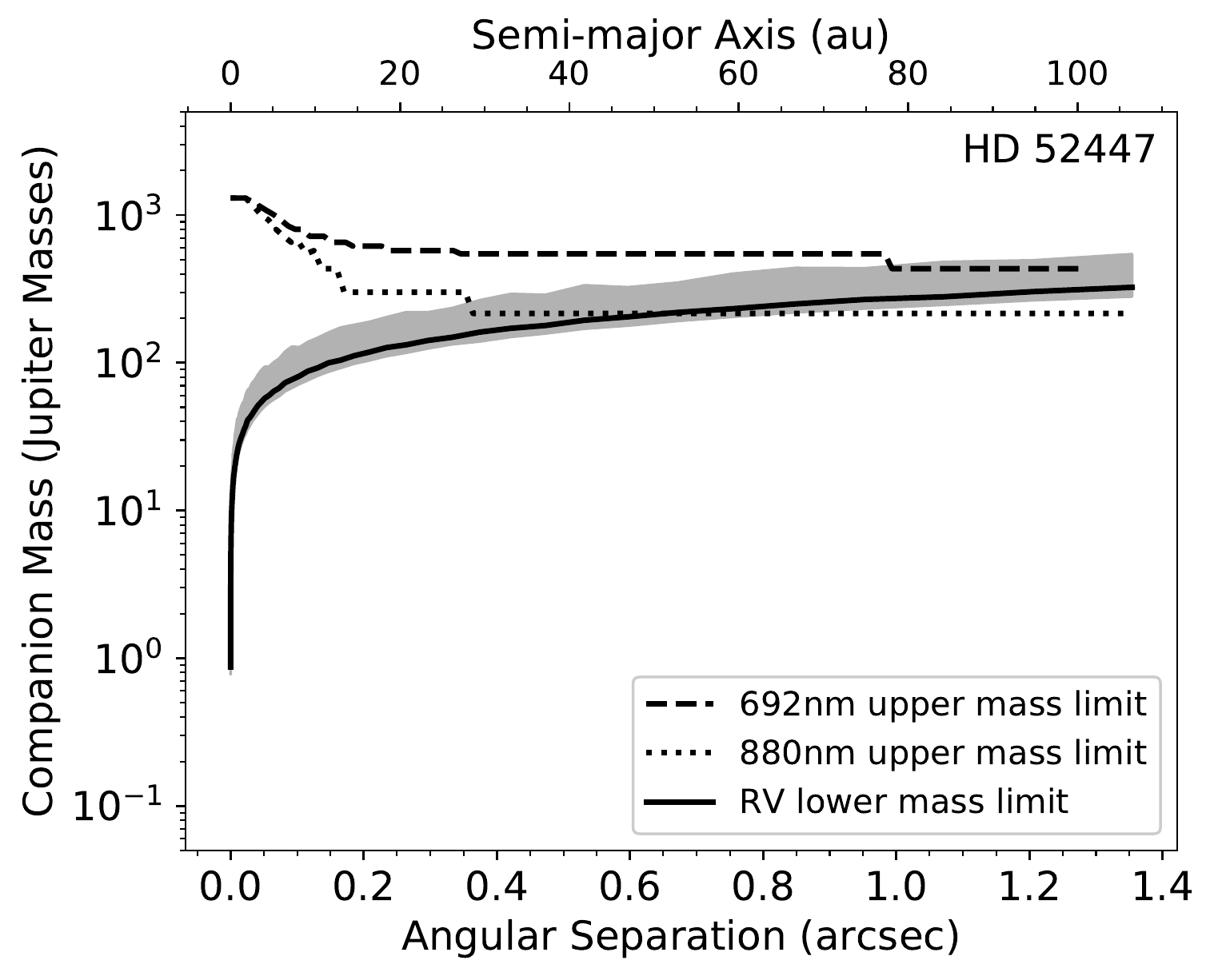} &
      \includegraphics[width=5.6cm]{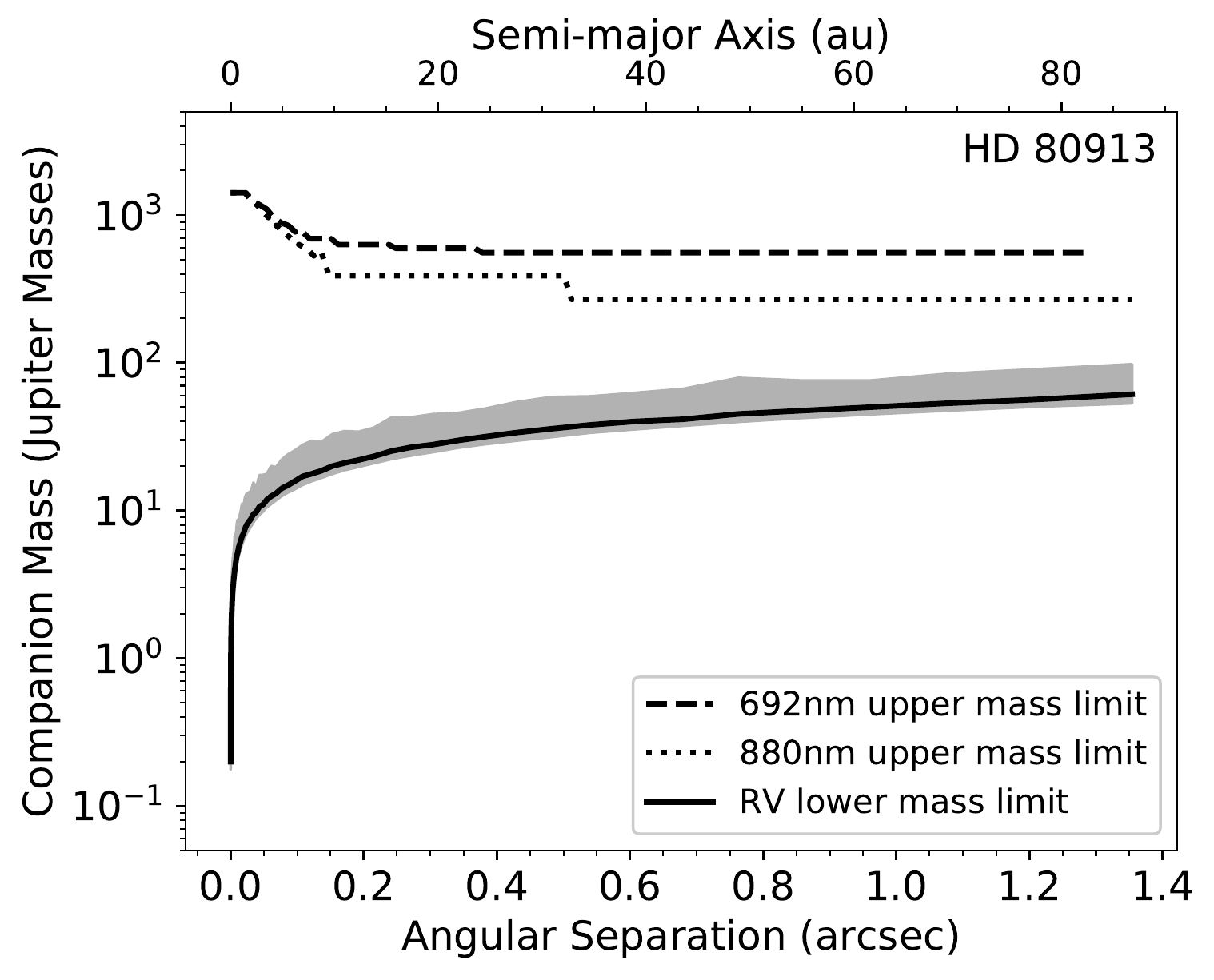} \\
      \includegraphics[width=5.6cm]{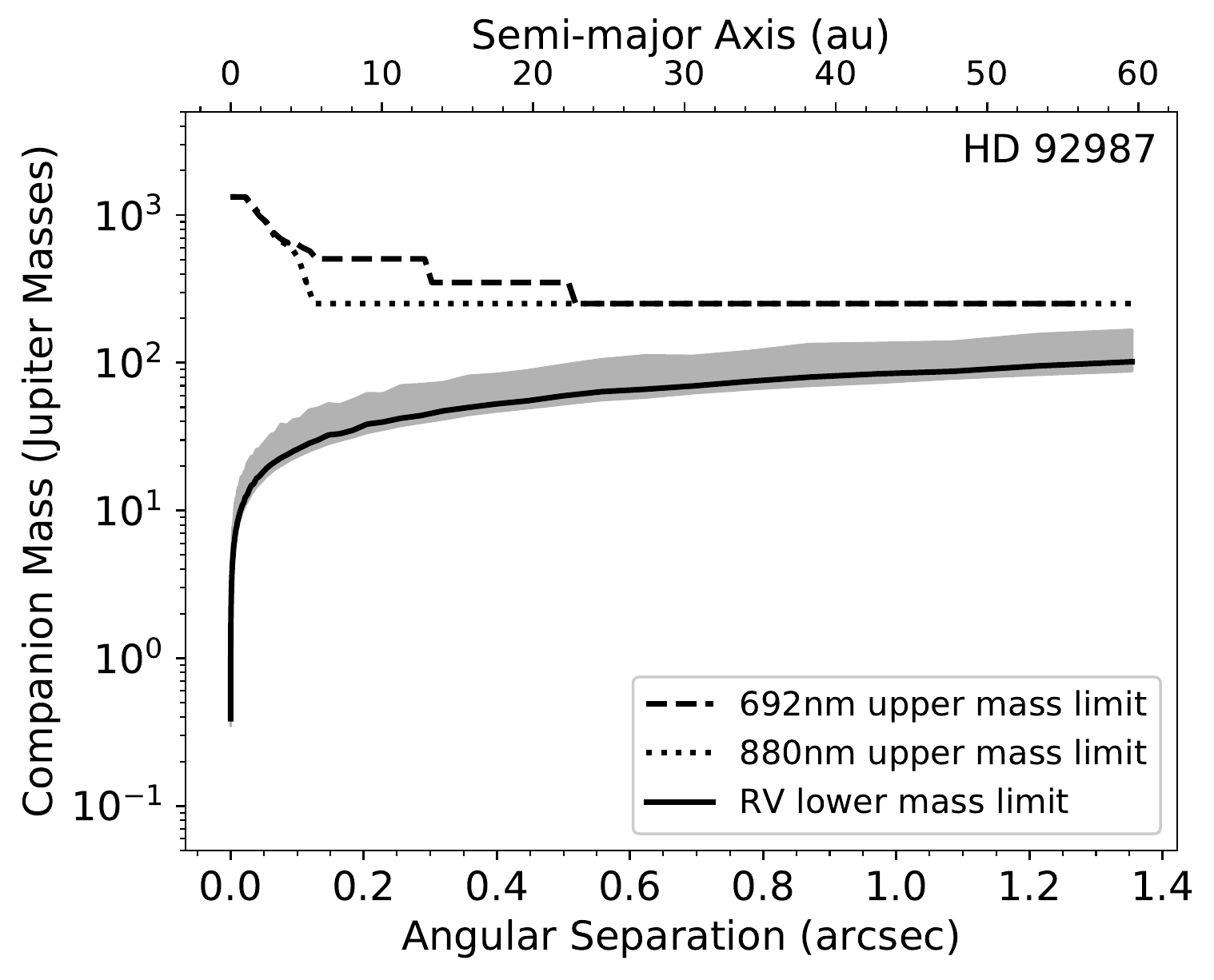} &
      \includegraphics[width=5.6cm]{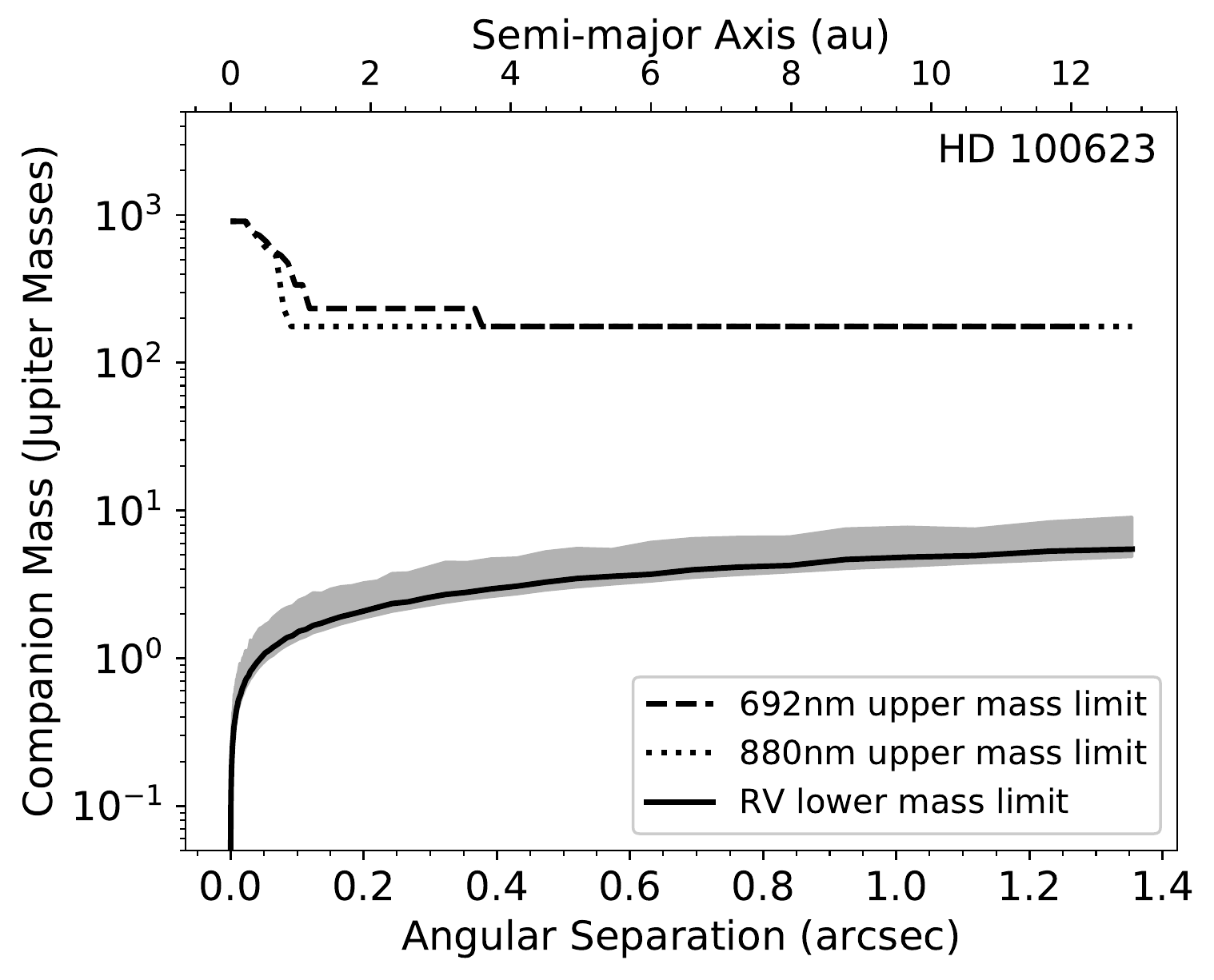} &
      \includegraphics[width=5.6cm]{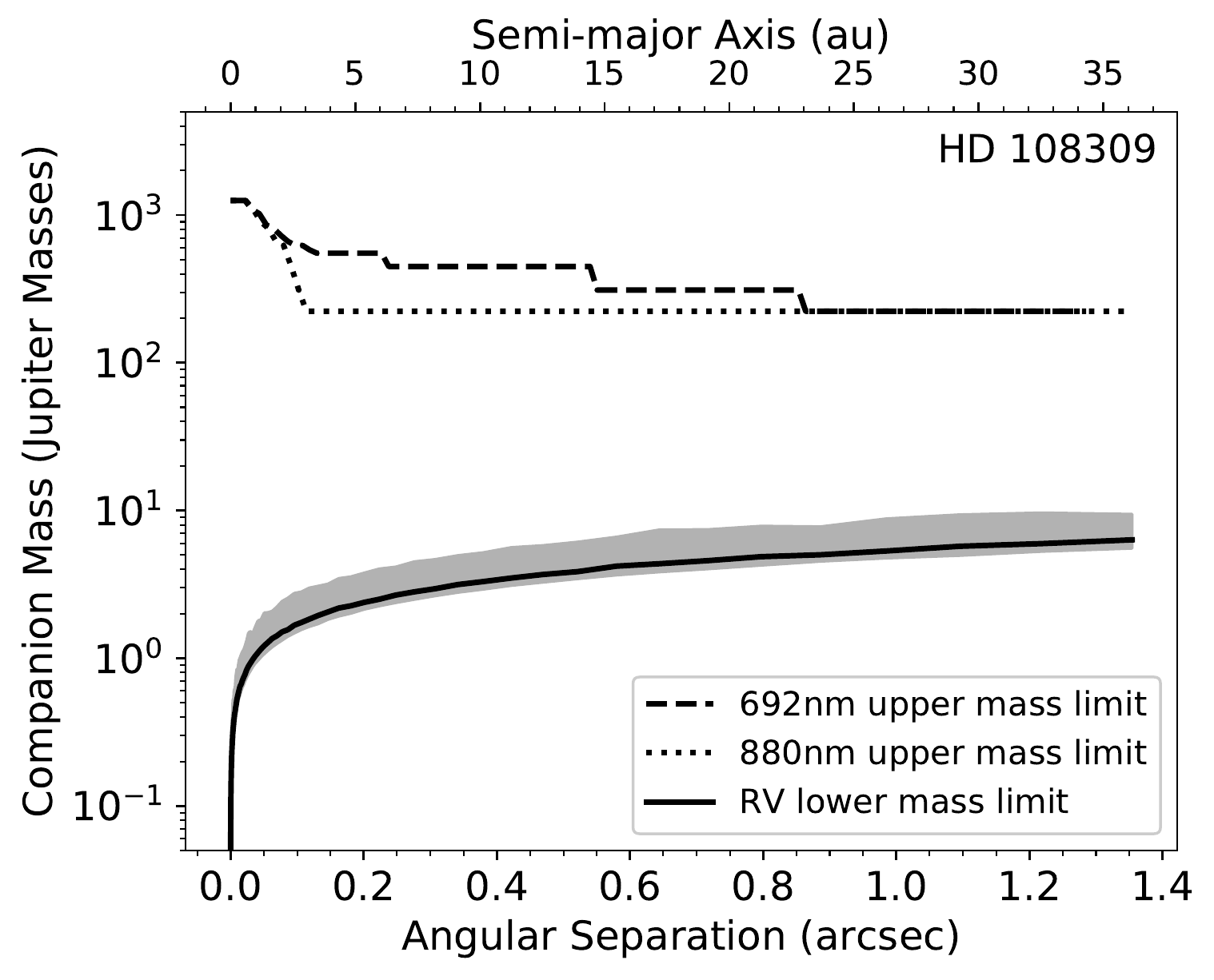} \\
      \includegraphics[width=5.6cm]{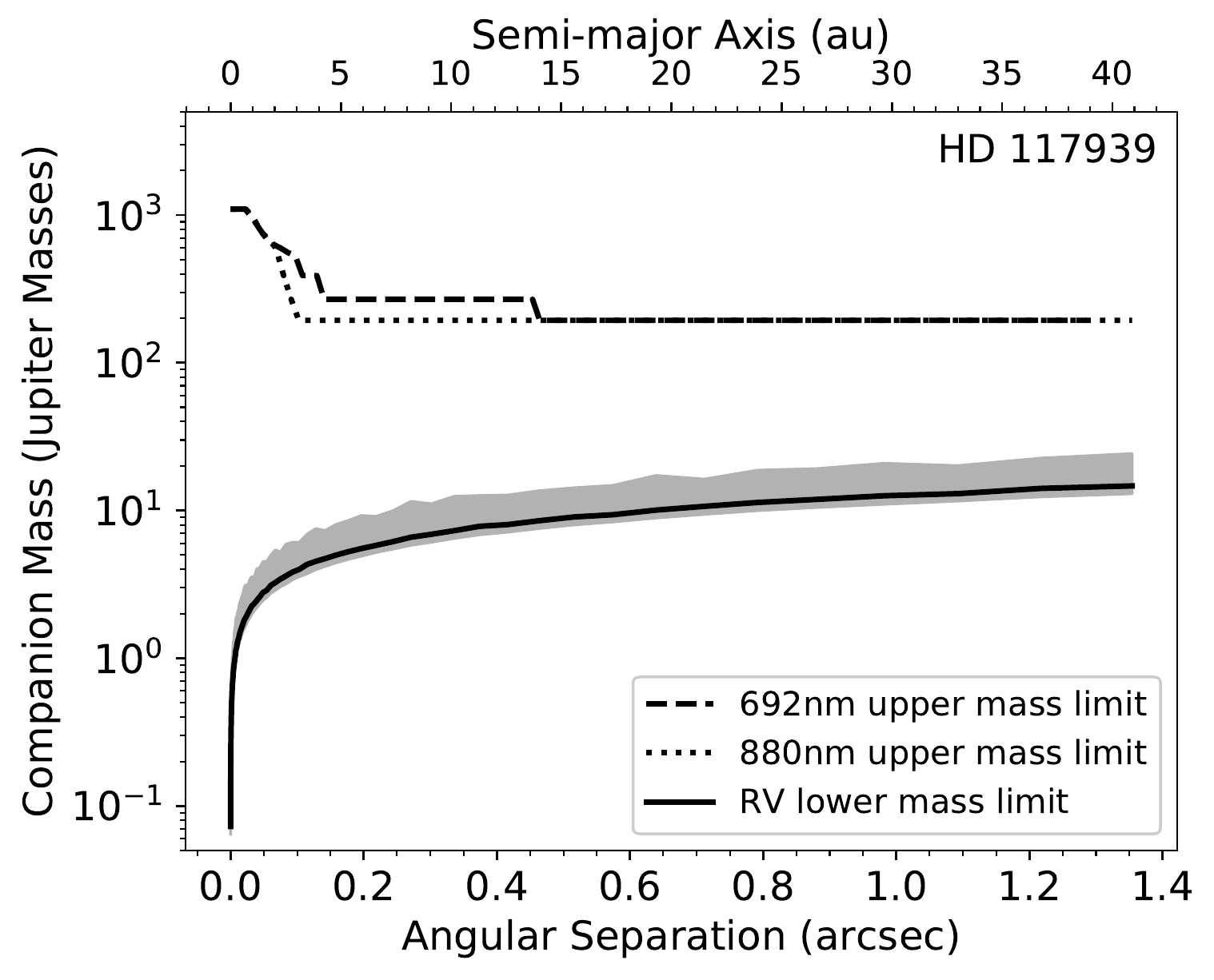} &
      \includegraphics[width=5.6cm]{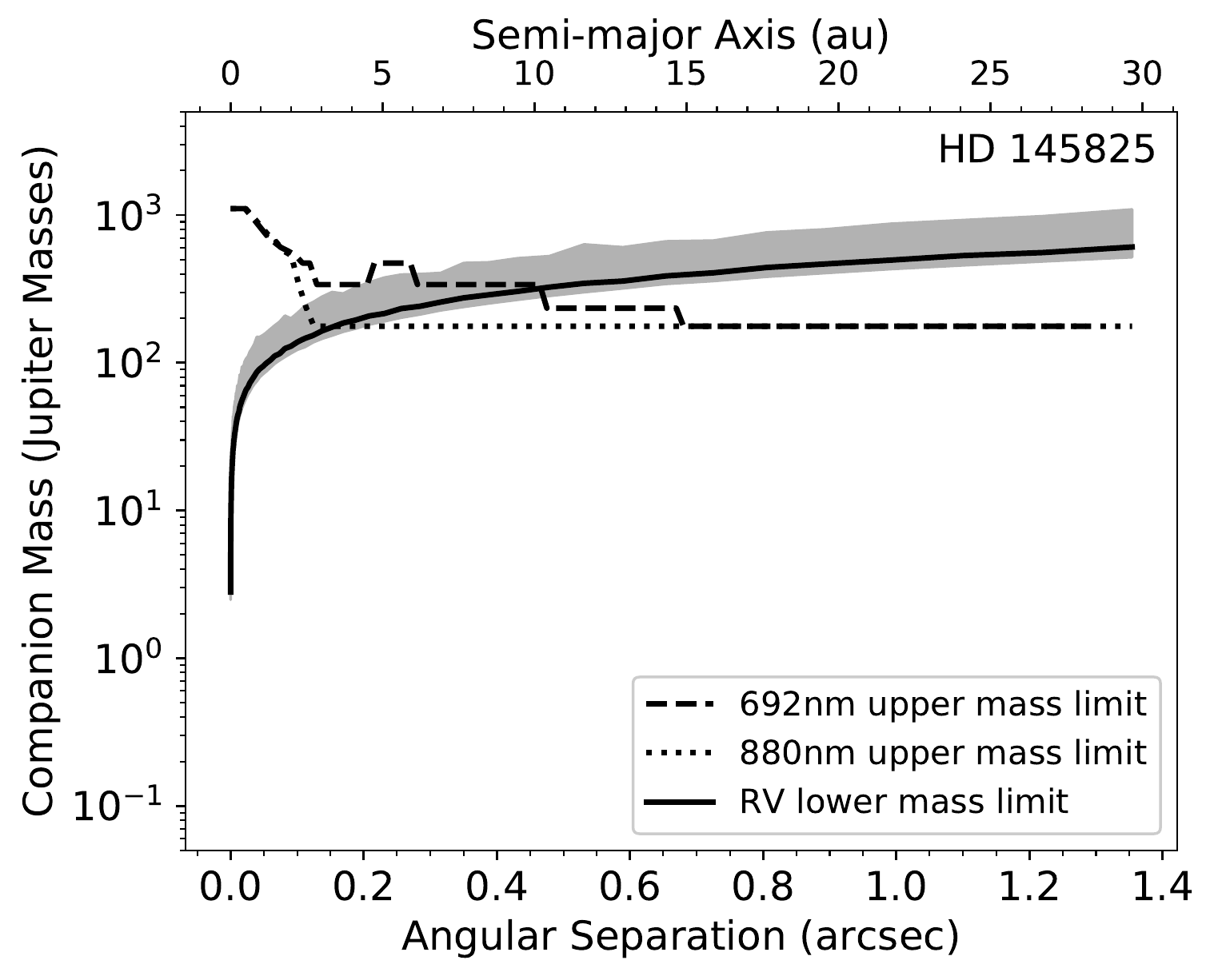} &
      \includegraphics[width=5.6cm]{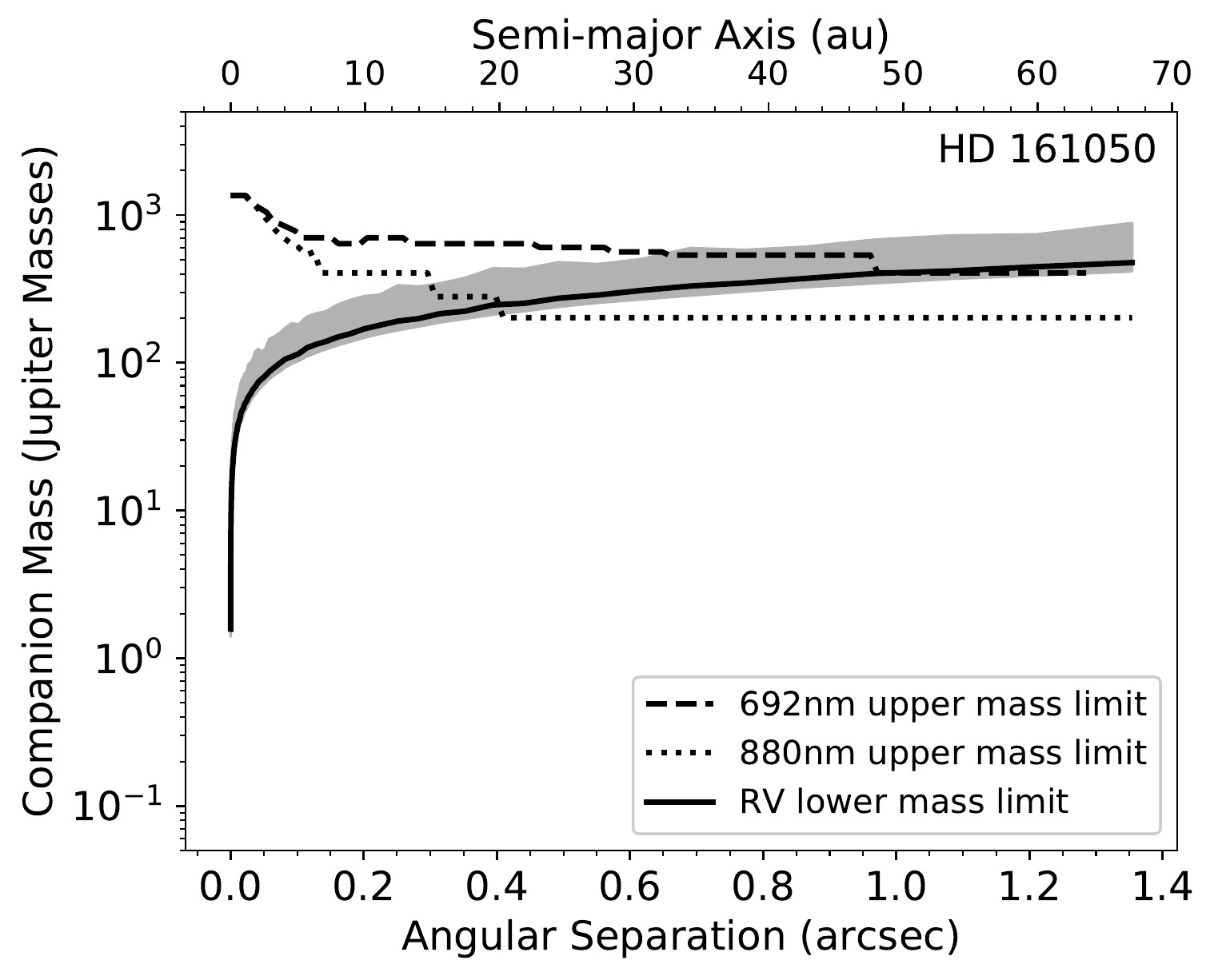} \\
      \includegraphics[width=5.6cm]{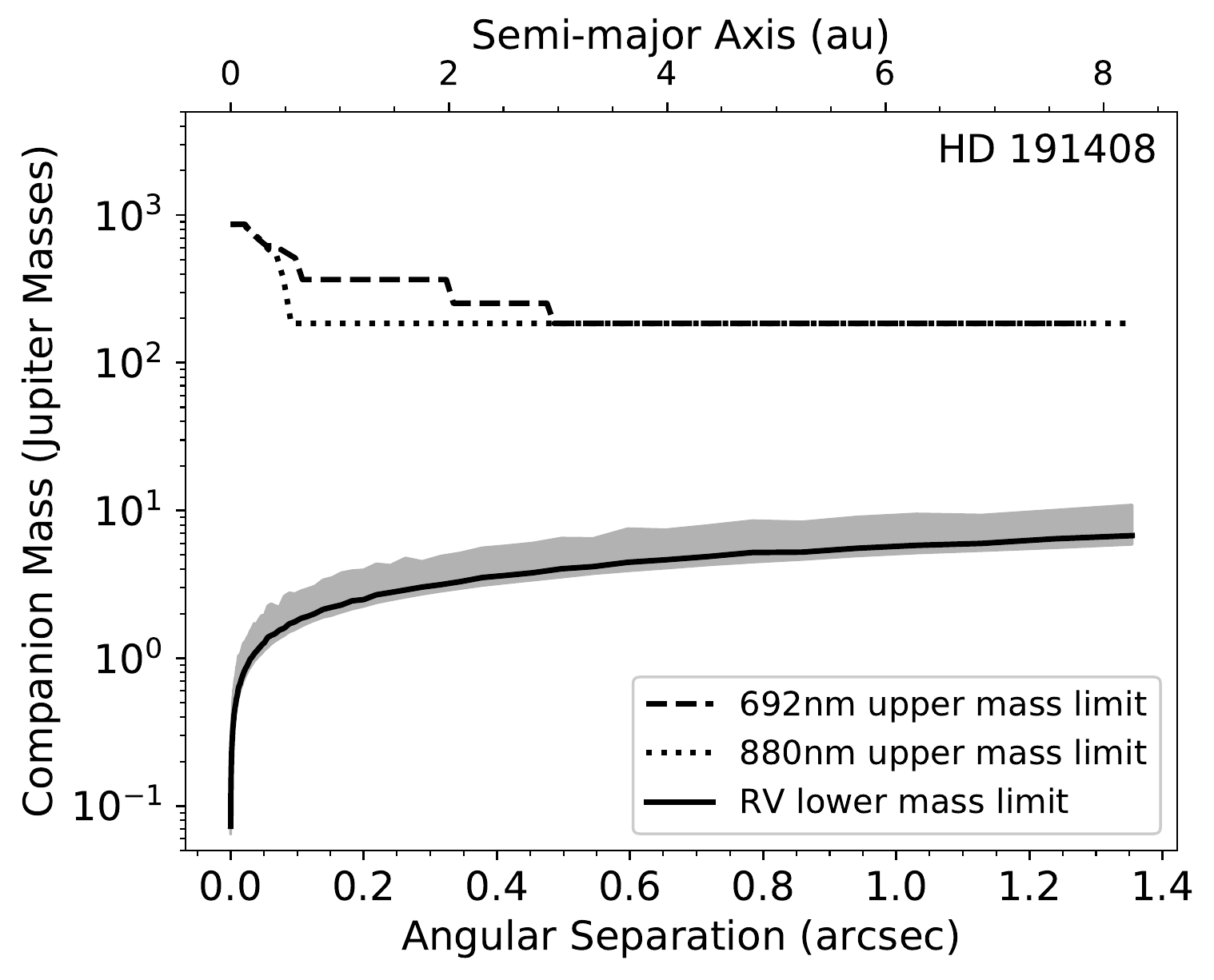} &
      \includegraphics[width=5.6cm]{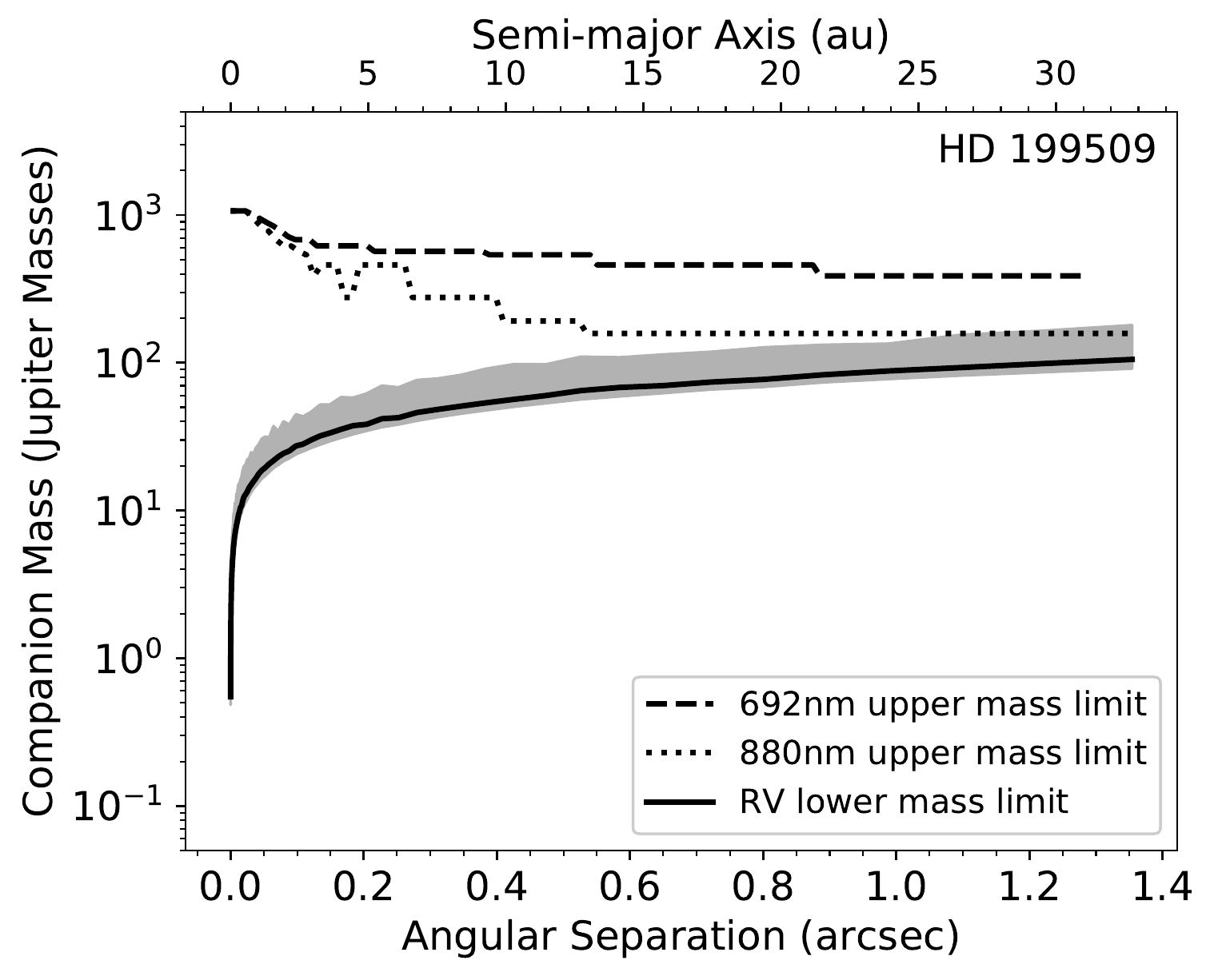} &
      \includegraphics[width=5.6cm]{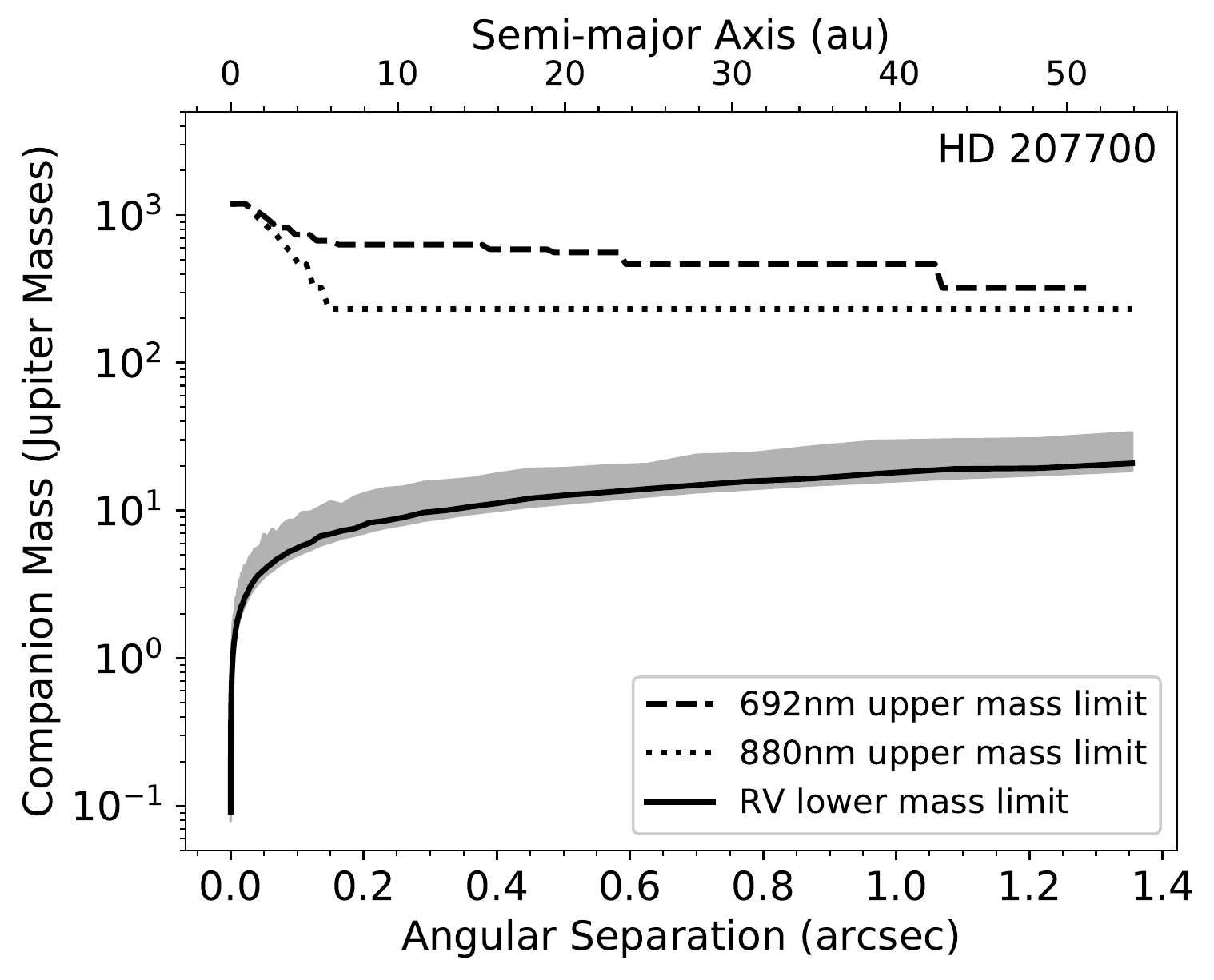} \\
      \includegraphics[width=5.6cm]{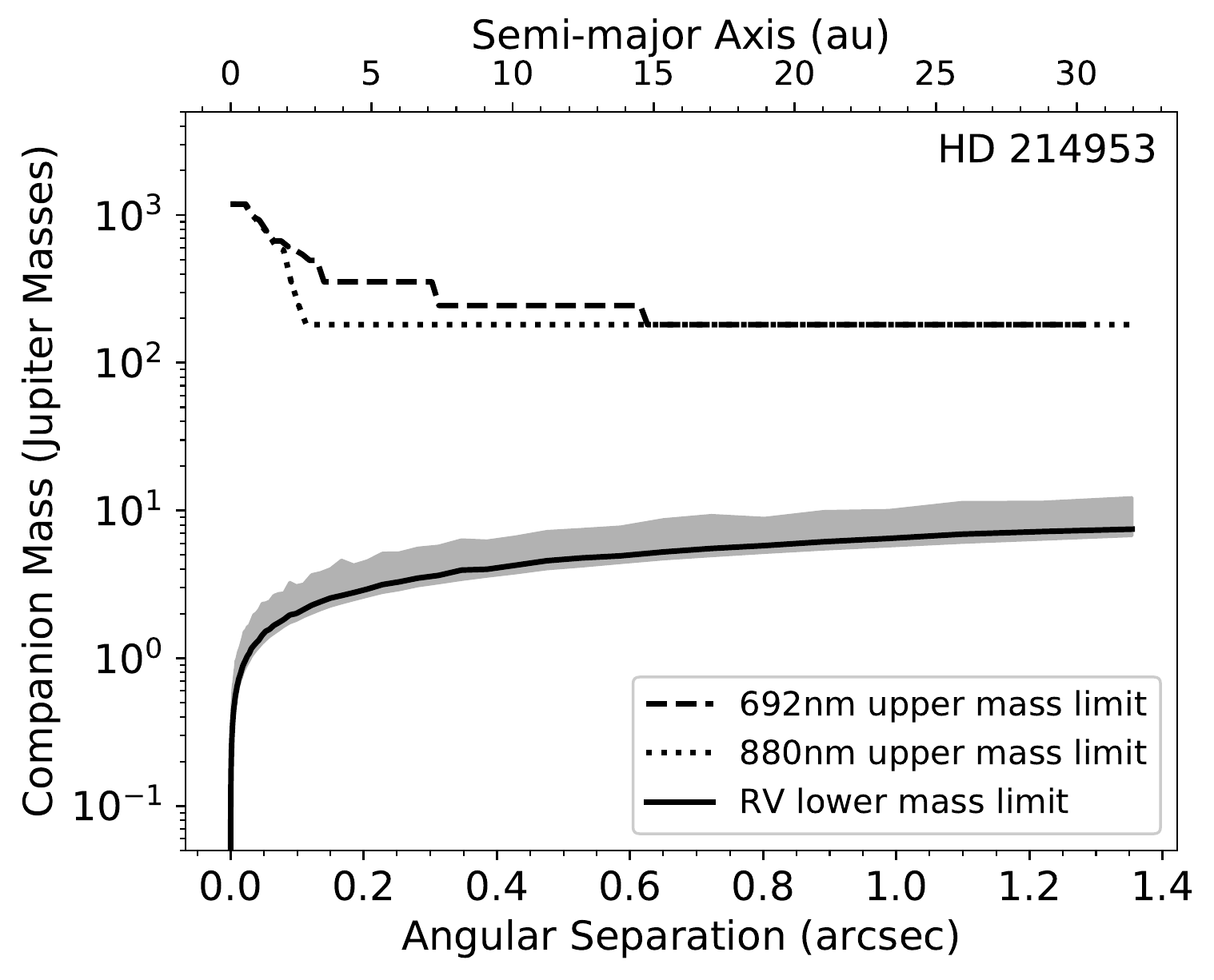} &
      \includegraphics[width=5.6cm]{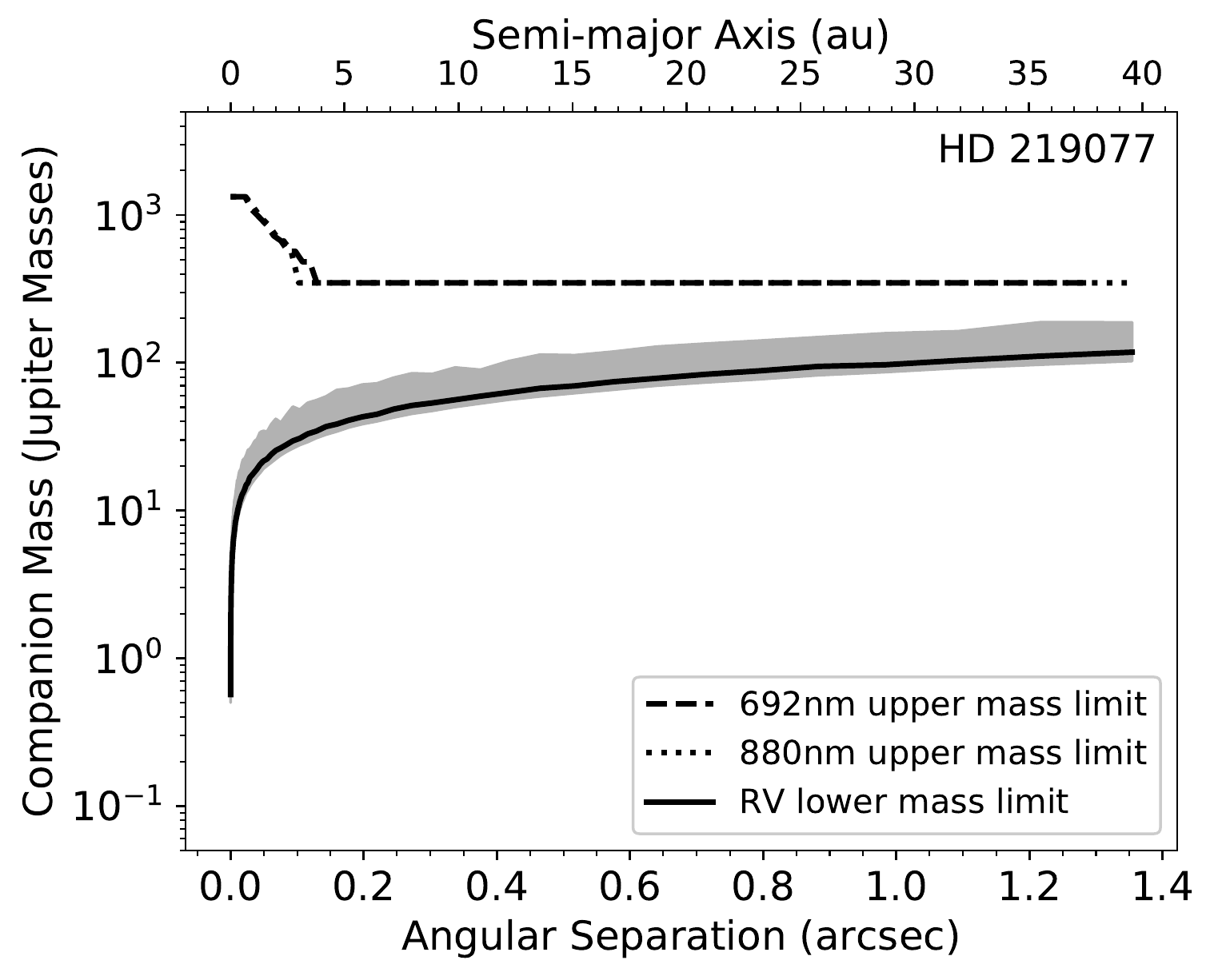} &
      \includegraphics[width=5.6cm]{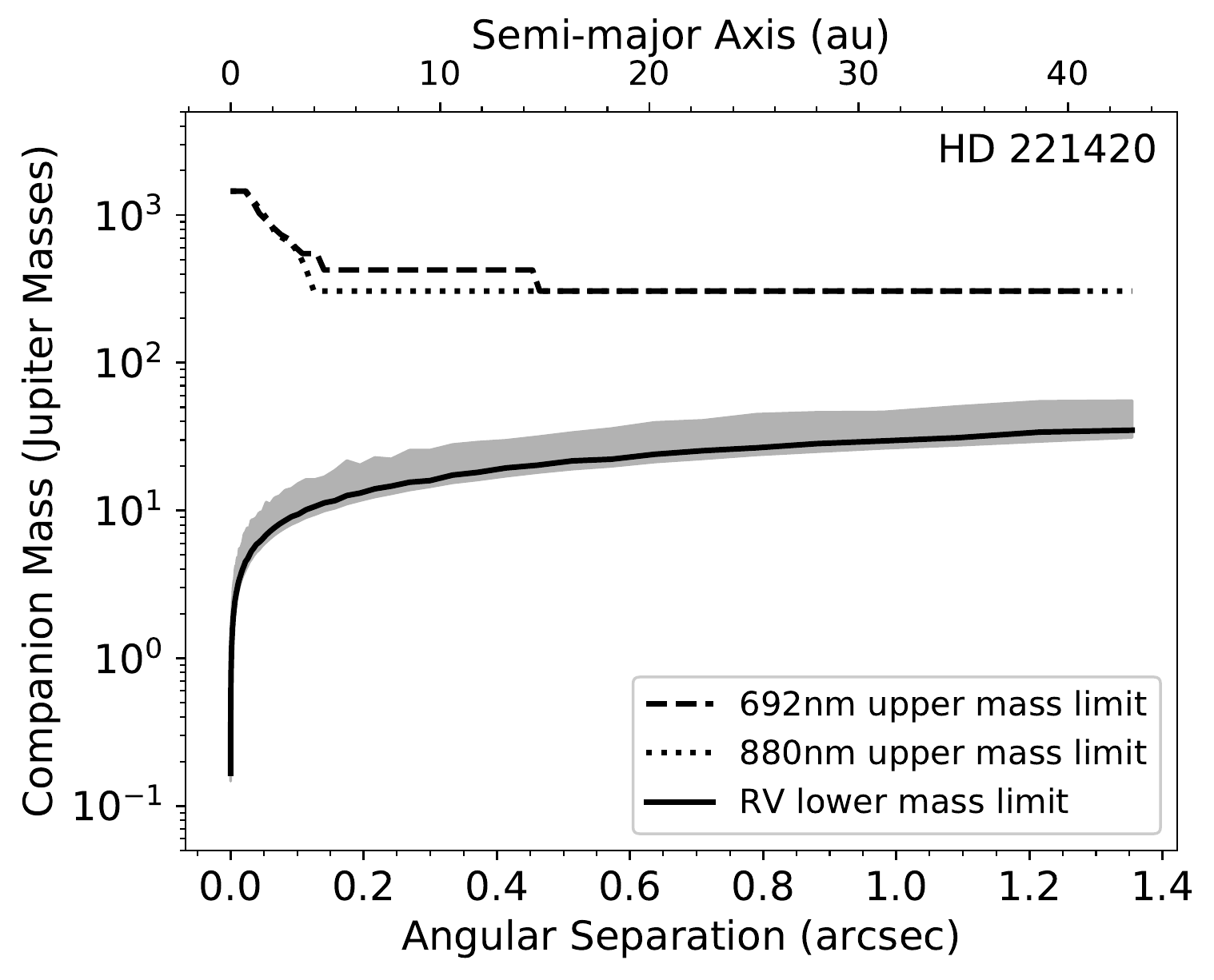}
    \end{tabular}
  \end{center}
  \caption{Companion mass limits for the fifteen targets included in
    this survey for which no stellar companion was imaged. For the
    lower limits, the gray region spans the 68\% confidence interval
    and the black line denotes the median.}
  \label{masslimplots}
\end{figure*}

Upper and lower limits for the fifteen targets without directly imaged
stellar companions are shown in Figure~\ref{masslimplots}. The lower
limits include the 68\% confidence region accounting for the unknown
inclination and eccentricity. Other than the three cases where stellar
companions have been identified from a full Keplerian orbit (see
Table~\ref{keptab}), the combination of RV and imaging observations
generally rules out the presence of stellar companions more massive
than several hundred Jupiter masses at most orbital separations.

\begin{figure}
  \begin{center}
    \includegraphics[width=\columnwidth]{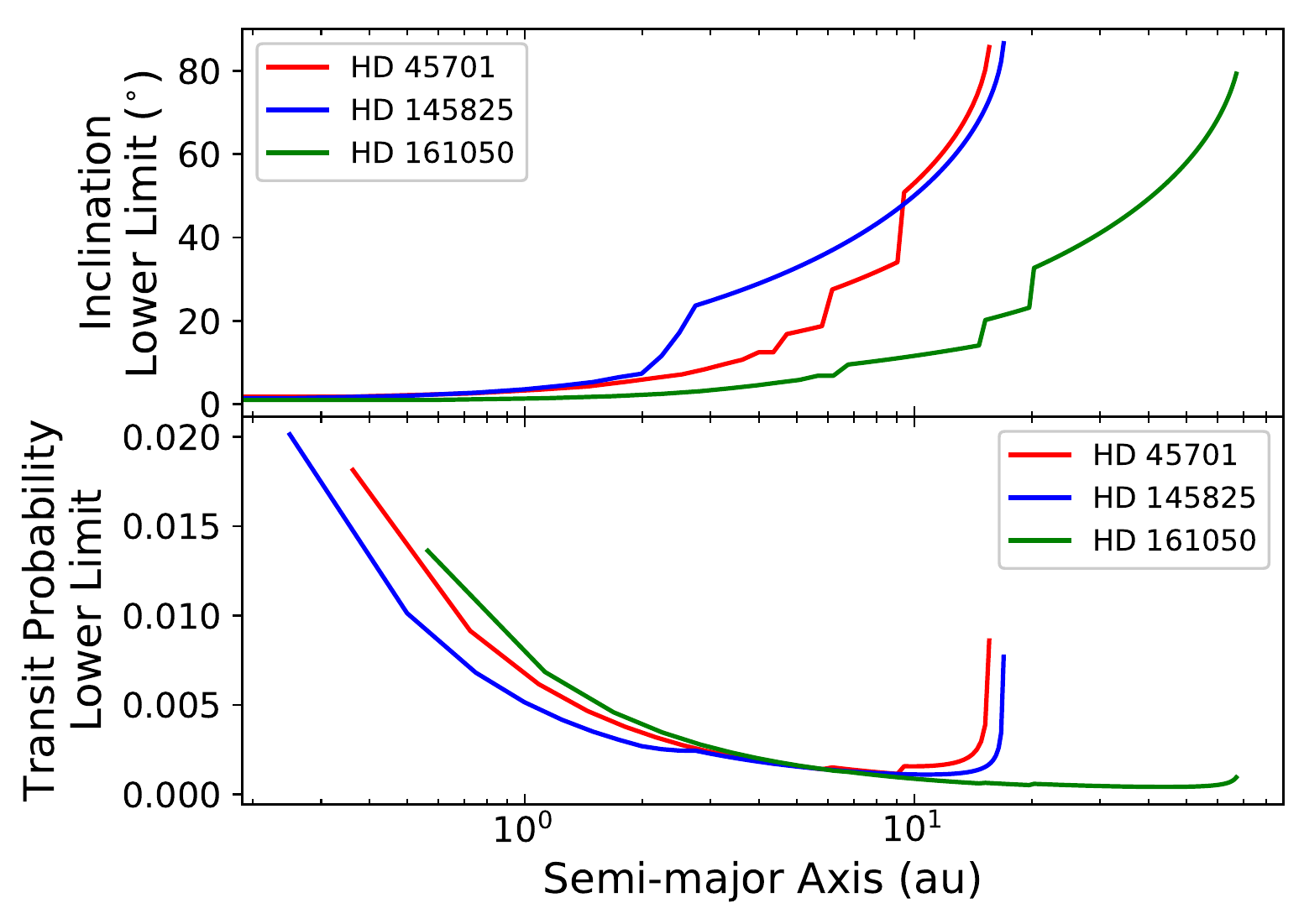}
  \end{center}
  \caption{Lower limits on inclination and transit probability placed
    on a subset of targets in the sample. In all cases, the
    inclination limits cause the transit probability to begin
    increasing as a function of semi-major axis.}
  \label{ptransit}
\end{figure}

For several targets, the combination of RV and imaging observations
dramatically reduce the size of parameter space where a companion
could exist, thereby enabling lower limits on the orbital inclination
of a companion. We demonstrate this for HD~45701, HD~145825, and
HD~161050, which have $\Delta$RV values of a few km\,s$^{-1}$. Using
the mass limits from the 880 nm images, we invert
Equation~\ref{eq:rvsemiamp} to solve for the lower limit on
inclination. At small semi-major axes, these limits are weak. However,
at wider separations the mass constraints force the inclinations to
high values to maintain consistency with the $\sim$km\,s$^{-1}$ RV
signals (see Figure~\ref{ptransit}).

Limits on orbital inclination are highly valuable for considerations
of transit probability. Again considering HD~45701, HD~145825, and
HD~161050, we approximate the lower limit transit probabilities
corresponding to the inclination limits in two steps. First, we
approximate the \textit{a priori} geometric transit probability as
$(R_p + R_\star)/a$, where $R_{\star}$ is the host star's radius and
$R_p$ is the companion radius\footnote{We note that the calculation of
  \textit{a posteriori} transit probabilities for these RV-detected
  companions would require a more thorough assessment of underlying
  mass distribution of giant planet and sub-stellar objects
  \citep[e.g.,][]{stevens2013}.}. We estimate $R_\star$ by applying
the known properties of the targets (Table~\ref{sumtab}) to the
relations of \citet{torres2010}. We also assume that all companions
have a Jupiter radius. Second, we increase the geometric transit
probability by a factor matching the relative decrease in allowed
inclination values set by our lower limits. As shown in
Figure~\ref{ptransit}, the inclination limits are strong enough that
the transit probability actually begins increasing as a function of
$a$ for companions at wide separations.

The severe short-orbit bias of the transit method of exoplanet
discovery largely prohibits the known sample of transiting exoplanets
to those with semi-major axes of several tenths of an AU. For
reference, the geometric transit probability of a Jupiter-size
companion at 10~AU around a Sun-like star is roughly 0.0005. The
detection of a transiting companion to HD~45701 or HD~145825, for
which we estimate lower limits on transit probability that are
$\sim$0.005--0.01 at $10<a<20$~AU, would be unprecedented. Such a
discovery would enable novel characterization efforts, including
atmospheric studies via transmission spectroscopy
\citep[e.g.,][]{dalba2015} and dynamical or photometric exomoon
searches \citep[e.g.,][]{kipping2012b}. Previously, only a small
number of known RV exoplanets have been thoroughly observed in search
of transits \citep[see][and references therein]{dalba2019} largely due
to limitations in photometric follow-up resources. The potentially
optimistic transit probabilities of the companions to HD~45701,
HD~145825, and HD~161050 make them ideal cases for follow-up transit
ephemeris refinement \citep[e.g.,][]{kane2008b}. Furthermore, this
work demonstrates how combining RV and imaging data to constrain
inclination may be used to identify such potentially transiting
systems.


\section{Conclusions}
\label{conclusions}

Over the past few decades, the number of known exoplanets has grown at
a dramatic rate. However, the most prolific techniques of RVs and
transits contain an intrinsic detection bias toward relatively small
star--planet separation that continues to dominate the semi-major axis
parameter space that has been explored. It is therefore critical that
these and other techniques be pushed toward larger semi-major axis
sensitivity to gain deeper insights into overall planetary system
architectures.

In this work, we presented the results of an extensive study of twenty
stars that show evidence of long-period companions in order to
ascertain the possibility that the observed signatures are planetary
in origin. As described in Section~\ref{intro}, the technique of
pairing RV with high-resolution imaging is widely used to detect
potential stellar origins of RV signatures. Of our twenty targets, six
have sufficient RV phase coverage to produce Keplerian orbital
solutions, of which three are stellar and three are planetary in
nature. Five of the twenty targets are revealed via DSSI data to have
stellar objects at relatively small angular separations, one of which
is amongst those with a Keplerian orbital solution. Thus, half of the
sample have either confirmed planetary companions or evidence of bound
stellar companions. The remaining ten cases consist of RV linear
trends with no directly imaged stellar companion. The analysis of
Section~\ref{planets} utilizes the available data to place constraints
on the companion mass, of which the most likely explanation is that a
planet with a presently unresolved orbit is the cause of the RV
signature. The caveat to the planetary explanation is that the targets
may have been observed with DSSI at times when the angular separation
between the primary and secondary was too small to detect a stellar
signature from the secondary. The ambiguous nature of the orbits in
these ten cases make it difficult to plan effective direct imaging
observations when the companion would be located at an optimal angular
separation from the primary. It is hence important that these targets
continue to be monitored with precision RVs to characterize the orbits
to enable the confirmation of the planetary hypothesis behind the
observed signatures.


\section*{Acknowledgements}

This work is based on observations obtained at the Gemini South
Observatory, which is operated by the Association of Universities for
Research in Astronomy, Inc., under a cooperative agreement with the
NSF on behalf of the Gemini partnership: the National Science
Foundation (United States), National Research Council (Canada),
CONICYT (Chile), Ministerio de Ciencia, Tecnolog\'{i}a e
Innovaci\'{o}n Productiva (Argentina), Minist\'{e}rio da Ci\^{e}ncia,
Tecnologia e Inova\c{c}\~{a}o (Brazil), and Korea Astronomy and Space
Science Institute (Republic of Korea). We acknowledge the traditional
owners of the land on which the AAT stands, the Gamilaraay people, and
pay our respects to elders past and present. The results reported
herein benefited from collaborations and/or information exchange
within NASA's Nexus for Exoplanet System Science (NExSS) research
coordination network sponsored by NASA's Science Mission Directorate.




\end{document}